\documentclass[twoside,ukrainian]{article}
\usepackage[T1]{fontenc}
\usepackage[utf8]{inputenc}
\usepackage{geometry}
\usepackage{verbatim}
\usepackage{amsmath}
\usepackage{color}
\usepackage{setspace}
\usepackage{amssymb}

\makeatletter

\newcommand{\noun}[1]{\textsc{#1}}
\newcommand{\mathcircumflex}[0]{\mbox{\^{}}}

\newcommand{\lyxaddress}[1]{
\par {\raggedright #1
\vspace{1.4em}
\noindent\par}
}
\newenvironment{lyxlist}[1]
{\begin{list}{}
{\settowidth{\labelwidth}{#1}
 \setlength{\leftmargin}{\labelwidth}
 \addtolength{\leftmargin}{\labelsep}
 }}
{\end{list}}

\usepackage[english,ukrainian]{babel}
\usepackage{amsfonts,amsmath,amsthm,amssymb, amscd,latexsym,mathrsfs}
\usepackage{ifthen,cite}
\usepackage[active]{srcltx}
\usepackage[dvips]{graphicx} 
\date{\empty}

\newtheorem{theorem}{Theorem}
\@addtoreset{theorem}{section}
\newtheorem{lemma}{Lemma}
\@addtoreset{lemma}{section}
\newtheorem{assertion}{Assertion}
\@addtoreset{assertion}{section}
\newtheorem{nasl}{Corollary}
\@addtoreset{nasl}{section}
\newtheorem{definition}{Definition}
\@addtoreset{definition}{section}
\newtheorem{problem}{Problem}
\@addtoreset{problem}{section}
\newtheorem{props}{Properties}
\@addtoreset{props}{section}
\theoremstyle{remark}
\newtheorem{rmk}{Remark}
\@addtoreset{rmk}{section}
\newtheorem{example}{{\bf\em Example}}
\@addtoreset{example}{section}

\newcommand{\BeginDef}{\begin{definition}}
\newcommand{\EndDef}  {\end{definition}}
\newcommand{\BeginLem}{\begin{lemma}}
\newcommand{\EndLem}  {\end{lemma}}
\newcommand{\BeginThm}{\begin{theorem}}
\newcommand{\EndThm}  {\end{theorem}}
\newcommand{\BeginAs}{\begin{assertion}}
\newcommand{\EndAs}{\end{assertion}}
\newcommand{\BeginProp}{\begin{props}}
\newcommand{\EndProp}{\end{props}}
\newcommand{\BeginNasl}{\begin{nasl}}
\newcommand{\EndNasl}  {\end{nasl}}
\newcommand{\BeginRmk}{\begin{rmk}}
\newcommand{\EndRmk}  {\end{rmk}}
\newcommand{\BeginProof}{\begin{proof}[Proof]}
\newcommand{\EndProof}{\end{proof}}
\newcommand{\BeginEx}{\begin{example}}
\newcommand{\EndEx}{\end{example}}
\newcommand{\BeginProbl}{\begin{problem}}
\newcommand{\EndProbl}{\end{problem}}

\newcommand{\evenhead}{Author \ name}
\newcommand{\oddhead}{Article \ name}
\newcommand{\ShortArticleName}[1]{\renewcommand{\oddhead}{#1}}
\newcommand{\AuthorNameForHeading}[1]{\renewcommand{\evenhead}{#1}}

\renewcommand{\@evenhead}{
\hspace*{-3pt}\raisebox{-7pt}[\headheight][0pt]{\vbox{\hbox to 
\textwidth {\thepage \hfil \evenhead}\vskip4pt \hrule}}}

\renewcommand{\@oddhead}{
\hspace*{-3pt}\raisebox{-7pt}[\headheight][0pt]{\vbox{\hbox to 
\textwidth {\oddhead \hfil \thepage}\vskip4pt\hrule}}}

\allowdisplaybreaks

\usepackage{babel}
\makeatother

\begin{document}

\newcommand{\N}{\mathbb{N}}
\newcommand{\Z}{\mathbb{Z}}
\newcommand{\R}{\mathbb{R}}
\newcommand{\CC}{\mathbb{C}}
\newcommand{\Q}{\mathbb{Q}}
\newcommand{\card}{\mathbf{card}}
\newcommand{\nlm}{\nolimits}\newcommand{\Dm}{\mathfrak{D}}
\newcommand{\Rg}{\mathfrak{R}}
\newcommand{\NN}{\mathfrak{N}}
\newcommand{\Nxv}{\widetilde{N}}
\newcommand{\diremb}{\underline{\hookrightarrow}}
\newcommand{\Lxv}{\widetilde{L}}
\newcommand{\LL}{\mathfrak{L}}
\newcommand{\T}{\mathbf{T}}
\newcommand{\Txv}{\tilde{\T}}
\newcommand{\TT}{\mathbb{T}}
\newcommand{\bS}{\mathbf{S}}
\newcommand{\les}{\trianglelefteq}
\newcommand{\tn}{\tilde{t}}
\newcommand{\ls}{\vartriangleleft}
\newcommand{\MM}{\mathcal{M}}
\newcommand{\xxx}{\mathop{\ls\negthickspace\negthickspace\negthickspace\negmedspace-\negmedspace-}\nlm}
\newcommand{\diag}{\mathsf{diag}}
\newcommand{\nxxx}{\mathop{\;\:\not\hspace{-3.5mm}\xxx}}
\newcommand{\ag}{g^{[-1]}}
\newcommand{\af}{f^{[-1]}}
\newcommand{\arc}[1]{#1^{[-1]}}
\newcommand{\cyeq}{\mathop{\equiv}\limits ^{\mathbf{o}}}
\newcommand{\fff}{\mathop{\leftarrow}}
\newcommand{\nff}{\mathop{\:\not\not\negmedspace\negmedspace\fff}}
\newcommand{\tff}{\mathop{\twoheadleftarrow}}
\newcommand{\ntff}{\mathop{\:\not\not\negmedspace\negmedspace\tff}}
\newcommand{\QQQ}{\overleftarrow{\mathcal{Q}}}
\newcommand{\AAA}{\mathcal{A}}
\newcommand{\BBB}{\mathcal{B}}
\newcommand{\Bs}{\mathfrak{Bs}}
\newcommand{\bs}[1]{\mathsf{bs}\left(#1\right)}
\newcommand{\Ind}[1]{\mathcal{I}nd\left(#1\right)}
\newcommand{\ind}[1]{\mathtt{ind}\left(#1\right)}
\newcommand{\Bz}{\mathfrak{B}z}
\newcommand{\supp}{\mathbf{supp}}
\newcommand{\Lk}[1]{\mathcal{L}k\left(#1\right)}
\newcommand{\YY}{\mathbf{Y}}
\newcommand{\ffm}{\fff\limits _{\MM}}
\newcommand{\heq}{\upuparrows}
\newcommand{\BsM}{\Bs(\MM)}
\newcommand{\ffs}{\mathop{\fff\left(s\right)}}
\newcommand{\tfs}{\mathop{\tff(s)}}
\newcommand{\mff}{\mathop{\fff\negmedspace\textnormal{{\small(m)}}}}
\newcommand{\mtf}{\mathop{\tff\negmedspace\textnormal{{\small(m)}}}}
\newcommand{\pps}{(s)\rightarrow}
\newcommand{\tps}{\mathop{\left(s\right)\twoheadrightarrow}}
\newcommand{\mtp}{\mathop{\textnormal{{\small(m)}\negmedspace}\twoheadrightarrow}}
\newcommand{\cH}{\mathcal{H}}
\newcommand{\HH}{\mathbf{H}}
\newcommand{\hh}{\mathbf{h}}
\newcommand{\cS}{\mathfrak{S}}
\newcommand{\cP}{\mathcal{P}}
\newcommand{\cQ}{\mathcal{Q}}
\newcommand{\cL}{\mathcal{L}}
\newcommand{\cR}{\mathcal{R}}
\newcommand{\BsP}{\Bs(\cP)}
\newcommand{\Tm}{\mathbf{Tm}}
\newcommand{\TM}{\mathbb{T}\mathbf{m}}
\newcommand{\TmP}{\Tm(\cP)}
\newcommand{\tm}[1]{\mathsf{tm}\left(#1\right)}
\newcommand{\synh}{\Uparrow}
\newcommand{\BS}{\mathbb{B}\mathfrak{s}}
\newcommand{\BSP}{\BS(\cP)}
\newcommand{\BsB}{\Bs(\BBB)}
\newcommand{\BSB}{\BS(\BBB)}
\newcommand{\TmB}{\Tm(\BBB)}
\newcommand{\TMB}{\TM(\BBB)}
\newcommand{\ttm}{^{\vee}}
\newcommand{\BSBB}[1]{\BS\left(\BBB_{#1}\right)}
\newcommand{\psx}{\tilde{\psi}}
\newcommand{\Ll}{\mathbb{L}l}
\newcommand{\Ld}{\mathbb{L}d}
\newcommand{\dl}{\mathsf{dl}}
\newcommand{\LlB}{\Ll(\BBB)}
\newcommand{\LdB}{\Ld(\BBB)}
\newcommand{\At}{\mathcal{A}t}
\newcommand{\Atp}{\At p}
\newcommand{\Sxv}{S^{\sim}}
\newcommand{\Uxv}{U^{\sim}}
\newcommand{\Bxv}{\tilde{B}}
\newcommand{\inn}[1]{\in\negthinspace\negthinspace[#1]\,\,}
\newcommand{\cU}{\mathcal{U}}
\newcommand{\fU}{\mathfrak{U}}
\newcommand{\vcB}{\overleftarrow{\BBB}}
\newcommand{\vfU}{\overleftarrow{\fU}}
\newcommand{\cZ}{\mathcal{Z}}
\newcommand{\EE}{\mathbb{E}}
\newcommand{\Znv}[1]{\cZ\mathsf{nv}\left(#1\right)}
\newcommand{\Zpv}[1]{\cZ\mathsf{pv}\left(#1\right)}
\newcommand{\Zim}[1]{\cZ\mathsf{im}\left(#1\right)}
\newcommand{\vW}{\overleftarrow{W}}
\newcommand{\vU}{\overleftarrow{U}}
\newcommand{\bU}{\mathbf{U}}
\newcommand{\bbU}{\mathbb{U}}
\newcommand{\IndZ}{\Ind{\cZ}}
\newcommand{\LkZ}{\Lk{\cZ}}
\newcommand{\un}[2]{\left\langle #2\fff#1\right\rangle }
\newcommand{\unn}[3]{\un{#1}{!\:#2}#3}
\newcommand{\uni}[2]{\left\langle #1\rightarrow#2\right\rangle }
\newcommand{\ffff}{\mathop{\fff\textnormal{{\small(f)}}}}
\newcommand{\nfff}{\mathop{\nff\textnormal{{\small(f)}}}}
\newcommand{\ol}[1]{#1\mathcircumflex}
\newcommand{\oll}{\ol{l}}
\newcommand{\w}{\mathrm{w}}
\newcommand{\vi}{\succ}
\newcommand{\nvi}{\mathop{\,\vi!\,}}
\newcommand{\fvi}{\mathop{\,\vi!!\,}}
\newcommand{\fvieq}{\equiv!\:}
\newcommand{\vicnt}{\prec\hspace{-0.3mm}\vi}
\newcommand{\vieq}{\widehat{\equiv}}
\newcommand{\hdpz}[1]{\uparrow_{#1}^{+}}
\newcommand{\hdppz}[1]{\uparrow_{#1}^{++}}
\newcommand{\hdmz}[1]{\uparrow_{#1}^{-}}
\newcommand{\hdmmz}[1]{\uparrow_{#1}^{--}}
\newcommand{\hdp}{\hdpz{\,}}
\newcommand{\hdpp}{\hdppz{\,}}
\newcommand{\hdm}{\hdmz{\,}}
\newcommand{\hdmm}{\hdmmz{\,}}
\newcommand{\hdspz}[1]{\upuparrows_{#1}}
\newcommand{\hdsmz}[1]{\uparrow\negmedspace\downarrow_{#1}}
\newcommand{\hdsp}{\hdspz{\,}}
\newcommand{\hdsm}{\hdsmz{\,}}
\newcommand{\QQ}{\,\boldsymbol{Q}\,}
\newcommand{\QQm}{\,\boldsymbol{Q}_{-}\,}
\newcommand{\QQq}{\,\boldsymbol{Q}'\,}
\newcommand{\QQqm}{\,\boldsymbol{Q}'_{-}\,}

\selectlanguage{english} 
\large\sloppy
\ShortArticleName{Abstract Concept of Changeable Set}
\AuthorNameForHeading{Grushka Ya.I.}

\author{Grushka Ya.I. }

\title{\textbf{Abstract Concept of Changeable Set}}

\maketitle

\lyxaddress{\emph{Institute of Mathematics NAS of Ukraine}\textbf{\emph{.}}\emph{
}\\
\emph{3, Tereschenkivska st., }\\
\emph{Kyiv(Kiev)-4, }\\
\emph{01601 Ukraine }\\
\textbf{\emph{e-mail}}\emph{: grushka@imath.kiev.ua} }

\begin{flushright}
\begin{minipage}[t][1\totalheight]{0.93\columnwidth}%
\textbf{Abstract}. The work lays the foundations of the theory of
changeable sets. In author opinion, this theory, in the process of
it's development and improvement, can become one of the tools of solving
the sixth Hilbert problem least for physics of macrocosm. 

From a formal point of view, changeable sets are sets of objects which,
unlike the elements of ordinary (static) sets may be in the process
of continuous transformations, and which may change properties depending
on the point of view on them (the area of observation or reference
frame). From the philosophical and intuitive point of view the changeable
sets can look like as {}``worlds'' in which changes obey arbitrary
laws.\bigskip{}


\begin{flushright}
\emph{\bigskip{}
}
\par\end{flushright}

\end{minipage}%

\par\end{flushright}

\newpage{}

\tableofcontents{}

\newpage{}

\section{Introduction}

In spite of huge success of modern theoretical physics and the power
of mathematical tools, which it applies, its foundations remain unclear.
Well-known sixth Hilbert's problem of mathematically strict formulation
of the foundations of theoretical physics, delivered in 1900 \cite{Hilbert6_1},
completely is not solved to this day \cite{Hilbert6_2,Hilbert6_3}.
Some attempts to formalize certain physical theories was done in the
papers  {[}\ref{bibl:McKinsey+Suppes}-\ref{bibl:Pimenov2}]. The
main defect of these works is the absence of a single abstract and
systematic approach, and, consequently, insufficiency of flexibility
of the mathematical apparatus of these works, excessive its adaptability
to the specific physical theories under consideration. Moreover trying
in \cite{daCosta+Doria} to immediately formalize the maximum number
of known physical objects, without creating a hierarchy of elementary
abstract mathematical concepts has led to not very easy for the analysis
mathematical object \cite[page. 177, definition 4.1]{daCosta+Doria}.
 In general, it should be noted, that the main feature of existing
mathematically strict models of theoretical physics is that the investigators
try to find intuitively the mathematical tools to describe physical
phenomena under consideration, and only then they try to formalize
the description of this phenomena, identifying physical objects with
some constructs, generated by these mathematical tools, for example,
with solutions of some differential equations on some space or manifold.
As a result, quite complicated mathematical structures appear, whereas
most elementary physical concepts and postulates, obtained by a help
of experiments, life experience or common sense (which led to the
appearance of this mathematical model), remain not formulated mathematically
strictly. In works \cite{Levich01,Levich03} it is expressed the view
that, in the general case, it is impossible to solve this problem
by means of existing mathematical theories. Also in \cite{Levich01,Levich03}
it is posed the problem of constructing the theory of {}``dynamic
sets'', that is the theory of new abstract mathematical structures
for modeling various processes in physical, biological and other complex
systems. 

In the present work the foundations of the theory of changeable sets
are laid and the basic properties of these sets are established. The
theory of changeable sets can be considered as attempt to give a solution
of the problem, posed in \cite{Levich01,Levich03}. And author hopes,
that the apparatus of the theory of changeable sets can generate the
necessary mathematical structures at least for physics and some other
natural sciences in macrocosm. 

From a formal point of view, changeable sets are sets of objects which,
unlike the elements of ordinary (static) sets may be in the process
of continuous transformations, and which may change properties depending
on the point of view on them (the area of observation or reference
frame). From the philosophical and intuitive point of view the changeable
sets can look like as {}``worlds'' in which changes obey arbitrary
laws. Note that the main statements of the theory of changeable sets
has been announced in \cite{MyTmm01}. Fundamentals of the theory
of primitive changeable sets (which is contained in the sections \ref{sec:OrientedSets}--\ref{sec:Atp(R)}
of this paper) also has been presented in \cite{MyTmm02}.

\section{Oriented Sets and their Properties \label{sec:OrientedSets}}

When we try to see on any picture of reality (area of reality) from
the most abstract point of view, we can only say that this picture
in every moment of its existence consists of certain things (objects).
During the research of this area of reality the objects of which it
consists can be divided into smaller, elementary, objects that we
call the elementary states. Method of division a given area of reality
into elementary states depends on our knowledge of this area, of practice
required level of detailing of research, or of the level of physical
and mathematical idealization of the studied system. Depending on
these factors as elementary states may be, for example, the position
of a material point or an elementary particle in a given time, the
value of scalar, vector or tensor field at a given point in space-time,
state of individuals of a species in a given time (in mathematical
models of biology) and others. And if a picture of reality will not
changed over time, then to describe this picture of reality (in the
most abstract form) it is sufficient the classical set theory, when
elementary states are interpreted as elements of a certain set. However,
the reality is changeable. Elementary states in the process of evolution
can change their properties (and thus lose its formal mathematical
self-identity), also elementary states may born or disappear, decompose
into several elementary states, or, conversely, several elementary
states may merge into one. But it is obvious that whenever it is possible
to trace the {}``evolution lines'' of the studied system, we can
clearly answer the question whether the elementary state \char`\"{}$y$\char`\"{}
is the result of transformations (ie, \char`\"{}transformation descendant\char`\"{})
of elementary state \char`\"{}$x$\char`\"{}. Therefore, the next
definition may be considered as the simplest (starting) model of a
set of changing objects. 

\BeginDef  \label{Def:OrientedSet} 

Let, $M$ be any \noun{non-empty} set.

Arbitrary ref\hspace{0.1mm}lexive binary relation $\xxx$ on $M$
(that is a relation satisfying $\forall x\in M\; x\xxx x$) we will
name an \textbf{orientation}, and the pair $\MM=\left(M,\xxx\right)$
will be called an \textbf{oriented set}. In this case the set $M$
we will name a \textbf{basic} set or a set of all \textbf{elementary
states} of oriented set $\MM$ and we will denote it by $\BsM$. The
relation $\xxx$ we will name a \textbf{directing relation of changes
(transformations)} of $\MM$, and we will denote it by $\ffm$. 

\EndDef 

In the case where the oriented set $\MM$ is known in advance, the
char $\MM$ in the denotation $\ffm$ will be released, and we will
use denotation $\fff$ instead. For the elements $x,y\in\BsM$ the
record $y\fff x$ should be read as {}``the elementary state $y$
is the result of transformations (or the transformation prolongation)
of the elementary state $x$''

\BeginRmk 

Some attempts to construct abstract mathematical structures for modeling
physical systems made in \cite{Pimenov1,Pimenov2}. In these works
as a basic abstract model it is proposed to consider a pair of kind
$\left(M,\prec\right)$, where $M$ is some set and $\prec$ is the
local sequence relation (that is asymmetric and locally transitive
(in the sense of \cite[page 28]{Pimenov2}) relation), which satisfies
the additional axioms $\text{TK}_{1}$-$\text{TK}_{3}$ \cite{Pimenov2}.
The main deficiency of this approach is, that it is not motivated
by abstract philosophical arguments, while the main motivation is
provided by the specific example of order relation, generated by the
{}``light cone'' in Minkowski space-time. Due to these factors,
the model, suggested in \cite{Pimenov1,Pimenov2}, is not enough flexible.
In particular, this model is unusable for the description of discrete
processes. Also, this model is not enough comfortable for consideration
(at the abstract level) of complex branched processes, where different
{}``branches'' of the process can {}``intersect'' or {}``merge''
during transformations. Moreover the construction of mathematical
model of the special relativity theory, based on the order relation
of {}``light cone'' makes it impossible the mathematically strict
study of tachyons under this model.

Note, that the directing relation of changes in the definition \ref{Def:OrientedSet}
displays only real transformations, of the elementary states which
have appeared in the oriented set, while the the local sequence relation
\cite{Pimenov1,Pimenov2} (in particular {}``light cone'' order
relation), display all potentially possible transformations. 

\EndRmk 

Let $\MM$ be an oriented set. 

\BeginDef  

The subset $N\subseteq\BsM$ will be referred to as \textbf{transitive}
in $\MM$ if for any $x,y,z\in N$ such, that $z\fff y$ and $y\fff x$
we have $z\fff x$. 

The transitive subset $N\subseteq\BsM$ will be called \textbf{maximum
transitive} if there not exist a transitive set $N_{1}\subseteq\BsM$,
such, that $N\subset N_{1}$ (where the symbol $\subset$ denotes
the strict inclusion, that is $N\neq N_{1}$). 

The transitive subset $L\subseteq\BsM$ will be referred to as \textbf{chain}
in $\MM$ if for any $x,y\in L$ at least one of the relations $y\fff x$
or $x\fff y$ is true. The chain $L\subseteq\BsM$ we will name the
\textbf{maximum chain} if there not exist a chain $L_{1}\subseteq\BsM$,
such, that $L\subset L_{1}$. 

\EndDef 

\BeginAs \label{As:2elem-transitive-chain}

Let $\MM$ be an oriented set. 

\begin{enumerate}
\item Any non-empty subset $N\subseteq\BsM$, containing not more than,
two elements is transitive.
\item Any non-empty subset $L=\left\{ x,y\right\} \subseteq\BsM$, containing
not more than, two elements is chain if and only if $y\fff x$ or
$x\fff y$. In particular, any singleton $L=\left\{ x\right\} \subseteq\BsM$
is chain. 
\end{enumerate}
\EndAs 

The proof of the assertion \ref{As:2elem-transitive-chain} is reduced
to trivial verification.

\BeginLem \label{Lem:TransitiveChainUnion}

Let $\MM$ be an oriented set. 

\begin{enumerate}
\item Union of an arbitrary family of transitive sets of $\MM$, linearly
ordered by the inclusion relation, is a transitive set in $\MM$. 
\item Union of an arbitrary family of chains of $\MM$, linearly ordered
by the inclusion relation, is a chain in $\MM$. 
\end{enumerate}
\EndLem 

\BeginProof 

1. Let $\NN\subseteq2^{\BsM}$ be a family of transitive sets of $\MM$,
linearly ordered by the inclusion relation. Denote: 

\[
\Nxv:=\bigcup_{N\in\NN}N.\]
 Consider any elementary states $x,y,z\in\Nxv$ such, that $z\fff y$
and $y\fff x$. Since $x,y,z\in\Nxv=\bigcup_{N\in\NN}N$ then there
exist $N_{x},N_{y},N_{z}\in\NN$ such, that $x\in N_{x}$, $y\in N_{y}$,
$z\in N_{z}$. Since the family of sets $\NN$ is linearly ordered
by the inclusion relation, then there exists the set $N_{0}\in\left\{ N_{x},N_{y},N_{z}\right\} $
such, that $N_{x},N_{y},N_{z}\subseteq N_{0}$. So, we have $x,y,z\in N_{0}$.
Since $N_{0}\in\left\{ N_{x},N_{y},N_{z}\right\} \subseteq\NN$, then
$N_{0}$ is the transitive set. Therefore from conditions $z\fff y$
and $y\fff x$ it follows, that $z\fff x$. Thus $\Nxv$ is the transitive
set.

2. Let $\LL\subseteq2^{\BsM}$ be a family of chains of $\MM$, linearly
ordered by the inclusion relation. Denote: 

\[
\Lxv:=\bigcup_{L\in\LL}L.\]
By the post 1, $\Lxv$ is the transitive set. Consider any elementary
states $x,y\in\Lxv$. Since the family of sets $\LL$ is linearly
ordered by the inclusion relation, then, similarly as in the post
1, there exists a chain $L_{0}\in\LL$ such, that $x,y\in L_{0}$.
And, because $L_{0}$ is chain, at least one of the relations $y\fff x$
or $x\fff y$ is true. Thus $\Lxv$ is the chain of $\MM$. ~ ~
~ \EndProof 

Using the lemma \ref{Lem:TransitiveChainUnion} and the Zorn's lemma,
we obtain the following assertion.\\

\BeginAs  \label{As:MaxTransitiveChain}~

\begin{enumerate}
\item For any transitive set $N$ of oriented set $\MM$ there exists a
maximum transitive set $N_{\max}$ such, that $N\subseteq N_{\max}$. 
\item For any chain $L$ of oriented set $\MM$ there exists a maximum chain
$L_{\max}$ such, that $L\subseteq L_{\max}$.
\end{enumerate}
\EndAs 

It should be noted that the second post of the assertion \ref{As:MaxTransitiveChain}
can be referred to as the generalization of the Hausdorff maximal
principle under this theory. 

The following corollaries results from the assertions \ref{As:MaxTransitiveChain}
and \ref{As:2elem-transitive-chain}. 

\BeginNasl 

For any two elements $x,y\in\BsM$ in the oriented set $\MM$ there
exists a maximum transitive set $N\subseteq\BsM$ such that $x,y\in N$. 

\EndNasl 

\BeginNasl  \label{Nasl:MaxChainExist}

For any two elements $x,y\in\BsM$, such that $y\fff x$, in the oriented
set $\MM$ there exists a maximum chain $L$ such that $x,y\in L$. 

\EndNasl 

If we put $x=y\in\BsM$ (by definition $\BsM\neq\emptyset$), we obtain,
that in any oriented set $\MM$ necessarily there exist maximum transitive
sets and maximum chains.

\section{Definition of the Time. Primitive Changeable Sets }

In theoretical physics, scientists tend to think, that the moments
of time are real numbers. But the abstract mathematics deal with objects
of a arbitrarily large cardinality. Therefore in out abstract theory
we will not restricted to the real moments of time. In the next definition
moments of time are elements of any linearly ordered set. Such definition
of time is close to the philosophy conception of time as some {}``chronological
order'', agreed upon the processes of transformations. 

\BeginDef \label{Def:ChronoMain}

Let $\MM$ be an oriented set and $\TT=\left(\T,\leq\right)$ be a
linearly ordered set. A map $\psi:\T\mapsto2^{\BsM}$ is referred
to as \textbf{time} on $\MM$ iff the following conditions are satisfied: 

1) For any elementary state $x\in\BsM$ there exists a element $t\in\T$
such that $x\in\psi(t)$.

2) If $x_{1},x_{2}\in\MM$, $x_{2}\fff x_{1}$ and $x_{1}\neq x_{2}$,
then there exist elements $t_{1},t_{2}\in\T$ such that $x_{1}\in\psi\left(t_{1}\right)$,
$x_{2}\in\psi\left(t_{2}\right)$ and $t_{1}<t_{2}$ (this means that
there is a temporal separateness of successive unequal elementary
states). 

In this case the elements $t\in\T$ we will call the moments of time,
the pair \[
\cH=\left(\TT,\psi\right)=\left(\left(\T,\leq\right),\psi\right)\]
 will be named \textbf{chronologization} of $\MM$ and the triple
\[
\cP=\left(\MM,\TT,\psi\right)=\left(\MM,\left(\T,\leq\right),\psi\right)\]
 we will call the \textbf{primitive changeable set}.

\EndDef 

\BeginRmk 

In \cite{Pimenov1,Pimenov2} linearly ordered sets has been used as
time-scales also. But the conception of time in the definition \ref{Def:ChronoMain}
is significantly different from \cite{Pimenov1,Pimenov2}. Note, that
the definition of time in \cite{Pimenov1,Pimenov2} is less general,
then the definition \ref{Def:ChronoMain} due to less generality of
the model, suggested in \cite{Pimenov1,Pimenov2}. 

\EndRmk 

We say that an oriented set \textbf{\emph{can be chronologized}} if
there exists at least one chronologization of $\MM$. It turns out
that any oriented set can be chronologized. To make sure this we may
consider any linearly ordered set $\TT=\left(\T,\leq\right)$, which
contains at least two elements and put:\[
\psi(t):=\BsM,\quad t\in\T.\]
The conditions of the definition \ref{Def:ChronoMain} for the function
$\psi(\cdot)$ apparently are satisfied. More non-trivial methods
to chronologize an oriented set we will consider in the section \ref{sec:ChronoThms}.

The following two assertions (\ref{As:SetEmbedChrono} and \ref{As:DeadHistoryDelete})
are trivial consequences of the definition \ref{Def:ChronoMain}.

\BeginAs  \label{As:SetEmbedChrono} Let $\MM$ and $\MM_{1}$ be
oriented sets, and while $\BsM\subseteq\Bs\left(\MM_{1}\right)$ and
$\fff\limits _{\MM}\subseteq\fff\limits _{\MM_{1}}$ (last inclusion
means that for $x,y\in\BsM$ the condition $y\fff\limits _{\MM}x$
implies $y\fff\limits _{\MM_{1}}x$). 

If a mapping $\psi_{1}:\T\mapsto2^{\Bs\left(\MM_{1}\right)}$ (where
$\TT=\left(\T,\leq\right)$ is a linearly ordered set) is a time on
$\MM_{1}$ then the mapping: 

\[
\psi(t)=\psi_{1}(t)\cap\BsM\]
is the time om $\MM$. 

\EndAs

\BeginAs  \label{As:DeadHistoryDelete}

Let $\MM$ be an oriented set and and $\psi:\T:\mapsto2^{\BsM}$ be
a time om $\MM$. 

(1) If $\T_{1}\subseteq\T$, $\T_{1}\neq\emptyset$ and $\psi(t)=\emptyset$
for $t\in\T\setminus\T_{1}$, then the mapping $\psi_{1}=\psi\upharpoonright\T_{1}$,
which is the restriction of $\psi$ on the set $\T_{1}$ also is time
on $\MM$. 

(2) If the ordered set $\T$ is embedded in a linearly ordered set
$\left(\Txv,\leq_{1}\right)$ (preserving order), then the mapping
$\tilde{\psi}:\Txv\mapsto2^{\BsM}$: \[
\tilde{\psi}(t)=\begin{cases}
\psi(t), & t\in\T\\
\emptyset, & t\in\Txv\setminus\T\end{cases}\]
also is time on $\MM$. 

\EndAs

The assertion \ref{As:DeadHistoryDelete} affirms, that {}``moments
of full death'' may be erased from or added to {}``chronological
history'' of primitive changeable set.

\section{One-point and Monotone Time. Chronologization Theorems \label{sec:ChronoThms}}

\BeginDef \label{Def:Mono_Point_Time}

Let $\left(\MM,\TT,\psi\right)=\left(\MM,\left(\T,\leq\right),\psi\right)$
be a primitive changeable set. 

1) The time $\psi$ will be called \textbf{quasi one-point} if for
any $t\in\T$ the set $\psi(t)$ is a singleton. 

2) The time $\psi$ will be called \textbf{one-point} if the following
conditions are satisfied: 

\quad{}(a) The time $\psi$ is quasi one-point; 

\quad{}(b) If $x_{1}\in\psi(t_{1})$, $x_{2}\in\psi(t_{2})$ and
$t_{1}\leq t_{2}$ then $x_{2}\fff x_{1}$. 

3) The time $\psi$ will be called \textbf{monotone} if for any elementary
states $x_{1}\in\psi\left(t_{1}\right)$, $x_{2}\in\psi\left(t_{2}\right)$
the conditions $x_{2}\fff x_{1}$ and $x_{1}\nff x_{2}$ imply $t_{1}<t_{2}$. 

In the case, when the time $\psi$ is quasi one-point (one-point/monotone)
the chronologization $\left(\TT,\psi\right)$ of the oriented set
$\MM$ will be called quasi one-point (one-point/monotone) correspondingly. 

\EndDef 

\BeginEx \label{Ex:PointTime}

Let us consider an arbitrary mapping $f:\R\mapsto\R^{d}$ ($d\in\N$).
This mapping can be interpreted as equation of motion of single material
point in the space $\R^{d}$. This mapping generates the oriented
set $\MM=\left(\BsM,\ffm\right)$, where $\BsM=\Rg(f)=\left\{ f(t)\,|\, t\in\R\right\} \subseteq\R^{d}$
and for $x,y\in\BsM$ the correlation $y\ffm x$ is true if and only
if there exist $t_{1},t_{2}\in\R$ such, that $x=f\left(t_{1}\right)$,
$y=f\left(t_{2}\right)$ and $t_{1}\leq t_{2}$. It is easy to verify,
that the mapping: 

\[
\psi(t)=\left\{ f(t)\right\} \subseteq\BsM,\quad t\in\R.\]
 is a one-point time on $\MM$. 

\EndEx 

The example \ref{Ex:PointTime} makes clear the definition of one-point
time. It is evident, that \emph{any one-point time is quasi one-point
and monotone}. It turns out that a quasi one-point time need not be
monotone (and thus one-point), and monotone time need not be quasi
one-point (and thus one-point). The next examples prove what is written
above. 

\BeginEx \label{Ex:QuaziPointNoMono}

Let us consider any two element set $M=\left\{ x_{1},x_{2}\right\} $.
We construct the oriented set $\MM=\left(\BsM,\ffm\right)$ by the
following way: 

\begin{gather*}
\BsM=M=\left\{ x_{1},x_{2}\right\} ;\\
\ffm=\left\{ \left(x_{2},x_{1}\right),\;\left(x_{1},x_{1}\right),\left(x_{2},x_{2}\right)\right\} \end{gather*}
(or, in other words, $x_{2}\ffm x_{1}$, $x_{1}\ffm x_{1}$, $x_{2}\ffm x_{2}$).
Note that the directing relation of changes $\ffm$ can be represented
in more laconic form: $\ffm=\left\{ \left(x_{2},x_{1}\right)\right\} \cup\diag(M)$,
where $\diag(M)=\left\{ \left(x,x\right)\,|\, x\in M\right\} $. As
a linearly ordered set we take $\TT=\left(\R,\leq\right)$ (with the
usual order on the real field). On the oriented set $\MM$ we define
time by the following way: \[
\psi(t):=\begin{cases}
\left\{ x_{1}\right\} , & t\notin\Q;\\
\left\{ x_{2}\right\} , & t\in\Q,\end{cases}\]
where $\Q$ is the field of rational numbers. It is easy to verify,
that the mapping $\psi$ really is time on $\MM$ in the sense of
definition \ref{Def:ChronoMain}. Since $\psi(t)$ is a singleton
for any $t\in\R$, the time $\psi$ is quasi one-point. If we put
$t_{1}=\sqrt{2}$, $t_{2}=1$, we obtain $x_{1}\in\psi\left(t_{1}\right)$,
$x_{2}\in\psi\left(t_{2}\right)$, $x_{2}\fff x_{1}$, $x_{1}\nff x_{2}$,
but $t_{1}>t_{2}$. Thus time $\psi$ is not monotone. 

\EndEx 

\BeginEx \label{Ex:MonoNoQuasiPopint}

Let us consider an arbitrary four-element set $M=\left\{ x_{1},x_{2},x_{3},x_{4}\right\} $
and construct the oriented set $\MM=\left(\BsM,\ffm\right)$ by the
following way: \begin{gather*}
\BsM=M=\left\{ x_{1},x_{2},x_{3},x_{4}\right\} ;\\
\ffm=\left\{ \left(x_{2},x_{1}\right),\left(x_{4},x_{3}\right)\right\} \cup\diag(M).\end{gather*}
As a linearly ordered set we take $\TT=\left(\left\{ 1,2\right\} ,\leq\right)$
(with the usual ordering on the real axis). Time on $\MM$ is defined
by the following way: \[
\psi(t):=\begin{cases}
\left\{ x_{1},x_{3}\right\} , & t=1\\
\left\{ x_{2},x_{4}\right\} , & t=2.\end{cases}\]
It is not hard to prove, that the mapping $\psi$ is monotone time
on $\MM$. But this time, obviously, is not quasi one-point. 

\EndEx 

It appears that quasi one-point and, simultaneously, monotone time
need not be one-point. This fact is illustrated by the following example. 

\BeginEx 

Let the oriented set $\MM$ be same as in the example \ref{Ex:MonoNoQuasiPopint}.
We consider the ordered set $\TT=\left(\left\{ 1,2,3,4\right\} ,\leq\right)$
(with the usual real number ordering). Time on $\MM$ we define by
the following:

\[
\psi(t):=\left\{ x_{t}\right\} ,\quad t\in\left\{ 1,2,3,4\right\} .\]
It is not hard to verify, that $\psi(\cdot)$ is quasi one-point and
monotone time on $\MM$. Although, if we put $t_{1}:=2$, $t_{2}:=3$,
we receive, $x_{2}\in\psi\left(t_{1}\right)$, $x_{3}\in\psi\left(t_{2}\right)$,
$t_{1}\leq t_{2}$, but $x_{3}\nff x_{2}$. Thus, the time $\psi$
is not one-point. 

\EndEx 

\BeginDef 

Oriented set $\MM$ will be called a \textbf{chain oriented set} if
the set $\BsM$ is the chain of $\MM$, that is if the relation $\fff$
if transitive on $\BsM$ and for any $x,y\in\BsM$ at least one of
the conditions $x\fff y$ or $y\fff x$ is satisfied. 

Oriented set $\MM$ will be called a \textbf{cyclic} if for any $x,y\in\BsM$
both of the relations $x\fff y$ and $y\fff x$ are true. 

\EndDef

It is evident, that any cyclic oriented set is a chain. 

\BeginLem \label{Lem:CycleChrono}

Any cyclic oriented set can be one-point chronologized. 

\EndLem

\BeginProof 

Let $\MM$ be a cyclic oriented set. By definition of oriented set,
$\BsM\neq\emptyset$. Choose any two disjoint sets $\T_{1},\T_{2}$
equipotent to the set $\BsM$ ($\T_{1}\cap\T_{2}=\emptyset$). (Such
sets must exist, because we can put $\T_{1}:=\BsM$ and construct
the set $\T_{2}$ from the elements of set $\Txv=2^{\T_{1}}\setminus\T_{1}$,
cardinality of which is not smaller the cardinality of $\T_{1}$.)
Let $\les_{i}$ ($i=1,2$) be any linear order relation on $\T_{i}$
(by Zermelo's theorem, such linear order relations necessarily exist).
~ Denote: \[
\T:=\T_{1}\cup\T_{2}.\]
On the set $\T$ we construct the relation:\[
\leq=\les_{1}\cup\Delta_{2}\cup\left\{ \left(t,\tau\right)\,|\: t\in\T_{1},\:\tau\in\T_{2})\right\} ,\]
 or, in the other words, for $t,\tau\in\T$ relation $t\leq\tau$
holds if and only if one of the following conditions is satisfied: 

\medskip{}

(O1) $t,\tau\in\T_{i}$ and $t\les_{i}\tau$ for some $i\in\left\{ 1,2\right\} $;

(O2) $t\in\T_{1}$, $\tau\in\T_{2}$. \medskip{}

The pair $\left(\T,\leq\right)$ is the ordered union of the linear
ordered sets $\left(\T_{1},\les_{1}\right)$ and $\left(\T_{2},\les_{2}\right)$.
Thus, by \cite[p. 208]{Kuratovskii}, $\left(\T,\leq\right)$ is
a linear ordered set. Let $f:\T_{2}\mapsto\T_{1}$ be any bijection
(one-to-one correspondence) between the (equipotent) sets $\T_{1}$
and $\T_{2}$. And let $g:\T_{1}\mapsto\BsM$ be any bijection between
the (equipotent) sets $\T_{1}$ and $\BsM$. 

Let us consider the following mapping $\psi:\T\mapsto2^{\BsM}$: 

\begin{equation}
\psi(t):=\begin{cases}
\left\{ g(t)\right\} , & t\in\T_{1};\\
\left\{ g(f(t))\right\} , & t\in\T_{2}.\end{cases}\label{eq:TimeDef01}\end{equation}
We are going to prove, that $\psi$ is a time on the oriented set
$\MM$. 

1) Let $x\in\BsM$. Since the mapping $g:\T_{1}\mapsto\BsM$ is bijection
between the sets $\T_{1}$ and $\BsM$, there exists the inverse mapping
$\ag:\BsM\mapsto\T_{1}$. Let us consider the element $t_{x}=\ag(x)\in\T_{1}\subseteq\T$.
According to (\ref{eq:TimeDef01}): 

\begin{gather*}
\psi(t_{x})=\left\{ g(t_{x})\right\} =\left\{ g(\ag(x))\right\} =\left\{ x\right\} .\end{gather*}
Therefore, $x\in\psi(t_{x})$. Thus the first condition of the time
definition \ref{Def:ChronoMain} is satisfied. 

2) Let $x,y\in\BsM$ be elements of $\BsM$ such, that $y\fff x$
and $x\neq y$. Denote: 

\begin{gather*}
t_{x}:=\ag(x)\in\T_{1};\\
t_{y}:=\af\left(\ag(y)\right)\in\T_{2}.\end{gather*}
By (O2), $t_{x}\leq t_{y}$. Since $\T_{1}\cap\T_{2}=\emptyset$,
we have $t_{x}\neq t_{y}$. Thus $t_{x}<t_{y}$. In accordance with
(\ref{eq:TimeDef01}), we obtain: 

\begin{gather*}
\psi(t_{x})=\left\{ g(t_{x})\right\} =\left\{ g\left(\ag(x)\right)\right\} =\left\{ x\right\} ;\\
\psi(t_{y})=\left\{ g\left(f\left(t_{y}\right)\right)\right\} =\left\{ g\left(f\left(\af\left(\ag(y)\right)\right)\right)\right\} =\left\{ y\right\} .\end{gather*}
 Consequently, $x\in\psi(t_{x})$, $y\in\psi(t_{y})$. That is the
second condition of the time definition \ref{Def:ChronoMain} also
is satisfied. 

Thus $\psi$ is a time on $\MM$. It remains to prove that the time
$\psi$ is one-point. 

According to (\ref{eq:TimeDef01}), for any $t\in\T$ the set $\psi(t)$
consists of one element (is a singleton). Thus the condition (a) of
the one-point time definition \ref{Def:Mono_Point_Time} is satisfied.
Since the oriented set $\MM$ is a cyclic, the condition (b) of the
definition \ref{Def:Mono_Point_Time} is also satisfied. Thus time
$\psi$ is one-point. ~ ~ \EndProof

\BeginThm \label{Thm:ChainChrono}

Any chain oriented set can be one-point chronologized. 

\EndThm

\BeginProof  

Let $\MM$ be a chain oriented set. Then the set $\BsM$ is a chain
of oriented set $\MM$, ie the relation $\fff=\ffm$ is quasi order
on $\BsM$. 

We will say that elements $x,y\in\BsM$ are \emph{cyclic equivalent}
(denotation $x\cyeq y$) if $x\fff y$ and $y\fff x$. In accordance
with \cite[page 21]{Birkhoff}, relation $\cyeq$ is equivalence relation
on $\BsM$. Let $F_{1}$ and $F_{2}$ be any two classes of equivalence,
generated by the relation $\cyeq$. We will denote $F_{2}\xxx F_{1}$
if for any $x_{1}\in F_{1}$, $x_{2}\in F_{2}$ it is true $x_{2}\fff x_{1}$.
According to \cite[page 21]{Birkhoff}, the relation $\xxx$ is ordering
on the quotient set $\BsM/\cyeq$ of all equivalence classes, generated
by $\cyeq$. We aim to prove, that this ordering is linear. Chose
any equivalence classes $F_{1},F_{2}\in\BsM/\cyeq$. Because $F_{1}$,$F_{2}$
are equivalence classes, they are nonempty, therefore there exists
the elements $x_{1}\in F_{1}$, $x_{2}\in F_{2}$. Since the oriented
set $\MM$ is a chain, at least one from the relations $x_{2}\fff x_{1}$
or $x_{1}\fff x_{2}$ is true. But, because any two elements, belonging
to the same class of equivalence, are cyclic equivalent, in the case
$x_{2}\fff x_{1}$ we will have $F_{2}\xxx F_{1}$, and in the case
$x_{1}\fff x_{2}$ we obtain $F_{1}\xxx F_{2}$. Thus $(\BsM/\cyeq,\xxx)$
is a linear ordered set. 

Any equivalence class $F\in\BsM/\cyeq$ is a cyclic oriented set relatively
the relation $\fff$ (restricted to this class). Consequently, by
lemma \ref{Lem:CycleChrono}, any such equivalence class can be one-point
chronologized. Let $\left(\TT_{F},\psi_{F}\right)=\left(\left(\T_{F},\leq_{F}\right),\psi_{F}\right)$
be a one-point chronologization of the class of equivalence $F\in\BsM/\cyeq$.
Without loss of generality we can assume that $\T_{F}\cap\T_{G}=\emptyset$
for $F\neq G$. Indeed, otherwise we may use the sets: \[
\Txv_{F}=\left\{ \left(t,F\right):t\in\T_{F}\right\} ,\quad F\in\BsM/\cyeq,\]
 with ordering: \[
\left(t_{1},F\right)\lesssim_{F}\left(t_{2},F\right)\Longleftrightarrow t_{1}\leq_{F}t_{2},\quad t_{1},t_{2}\in\T_{F}\qquad(F\in\BsM/\cyeq)\]
 and times: \[
\tilde{\psi}_{F}\left((t,F)\right)=\psi_{F}(t),\quad t\in\T_{F}\qquad(F\in\BsM/\cyeq),\]
it is evident, that these times are one-point. 

Thus, we will assume that $\T_{F}\cap\T_{G}=\emptyset$, $F\neq G$.
Denote: \[
\T:=\bigcup_{F\in\BsM/\cyeq}\T_{F}.\]

According to this denotation, for any element $t\in\T$ there exists
an equivalence class $F(t)\in\BsM/\cyeq$ such, that $t\in\T_{F(t)}$.
Since $\T_{F}\cap\T_{G}=\emptyset$, $F\neq G$, such equivalence
class $F(t)$ is for an element $t\in\T$ unique, ie the following
assertion is true: 

\begin{lyxlist}{00.00.0000}
\item [{\textbf{~~~~(F)}}] \emph{For any element $t\in\T$ the condition
$t\in\T_{F}$ ($F\in\BsM/\cyeq$) results in $F=F(t)$. }
\end{lyxlist}
For arbitrary elements $t,\tau\in\T$ we will denote $t\leq\tau$
if and only if at least one of the following conditions is true: 

(O1) $F(t)\neq F(\tau)$ and $F(\tau)\xxx F(t)$.

(O2) $F(t)=F(\tau)$ and $t\leq_{F(t)}\tau$.

The pair $\left(\T,\leq\right)$ is the ordered union of the (linear
ordered) family of linear ordered sets $\left(\T_{F}\right)_{F\in\BsM/\cyeq}$.
Thus, by \cite[p. 208]{Kuratovskii}, $\leq$ is a linear ordering
on $\T$. 

Denote: 

\begin{equation}
\psi(t):=\psi_{F(t)}(t),\qquad t\in\T.\label{eq:TimeDef02}\end{equation}
Since $\psi_{F(t)}(t)\subseteq F(t)\subseteq\BsM$, $t\in\T$, the
mapping $\psi$ reflects $\T$ into $2^{\BsM}$. Now we are going
to prove, that $\psi$ is one-point time. 

(a) Let $x\in\BsM$. Then there exists an equivalence class $\Phi\in M/\cyeq$,
such, that $x\in\Phi$. Since the mapping $\psi_{\Phi}:\T_{\Phi}\mapsto2^{\Phi}$
is a time on the oriented set $\left(\Phi,\fff\right)$, there exists
a time moment $t\in\T_{\Phi}$, such, that $x\in\psi_{\Phi}(t)$.
Since $t\in\T_{\Phi}$, then by virtue of assertion (F) we have $\Phi=F(t)$.
Therefore: 

\[
\psi(t)=\psi_{F(t)}(t)=\psi_{\Phi}(t)\ni x.\]
 Thus, the first condition of the time definition \ref{Def:ChronoMain}
is satisfied. 

(b) Let $x,y\in\BsM$, $y\fff x$ and $y\neq x$. According to the
item (a), there exist $t,\tau\in\T$ such, that $x\in\psi(t)$, $y\in\psi(\tau)$.
And, using (\ref{eq:TimeDef02}), we obtain, $x\in\psi(t)=\psi_{F(t)}(t)\subseteq F(t)$,
$y\in F(\tau)$. Hence, since $y\fff x$, for any $x'\in F(t)$, $y'\in F(\tau)$
(taking into account, that $x'\cyeq x$, $y'\cyeq y$), we obtain
$y'\fff x'$. This means, that $F(\tau)\xxx F(t)$. And, in the case
$F(t)\neq F(\tau)$, using (O1), we obtain, $t\leq\tau$, so, taking
into account, that $F(t)\neq F(\tau)$ causes $t\neq\tau$, we have
$t<\tau$. Thus it remains to consider the case $F(t)=F(\tau)$. In
this case we have $x,y\in F(t)$. And since $y\fff x$, $y\neq x$
and $\psi_{F(t)}$ is a time on $\left(F(t),\fff\right)$, there exist
the elements $t',\tau'\in\T_{F(t)}$ such, that $x\in\psi_{F(t)}(t')$,
$y\in\psi_{F(t)}(\tau')$ and $t'<_{F(t)}\tau'$. Therefore, since
$t',\tau'\in\T_{F(t)}$, using assertion (F), we obtain $x\in\psi_{F(t)}(t')=\psi_{F(t')}(t')=\psi(t')$
and $y\in\psi(\tau')$. Hence $x\in\psi(t')$, $y\in\psi(\tau')$,
where, $t'<_{F(t)}\tau'$ (that is $t'\leq_{F(t)}\tau'$ and $t'\neq\tau'$).
So, by (F) and (O2), we obtain $t'<\tau'$. 

Thus $\psi$ is a time on $\MM$. 

(c) It remains to prove, that the time $\psi$ is one-point. Since
for any equivalence class $G\in\BsM/\cyeq$ the mapping $\psi_{G}$
is a one-point time, by (\ref{eq:TimeDef02}), set $\psi(t)$ is a
singleton for any $t\in\T$. Thus, the first condition of the one-point
time definition is satisfied. 

Let $x\in\psi(t)$, $y\in\psi(\tau)$, where $t\leq\tau$. Then by
(\ref{eq:TimeDef02}) $x\in\psi(t)=\psi_{F(t)}(t)\subseteq F(t)$,
$y\in F(\tau)$. And in the case $F(t)=F(\tau)$ the relation $y\fff x$
follows from the relation $x\cyeq y$. Concerning the case $F(t)\neq F(\tau)$,
in this case, by (O1),(O2), we obtain $F(\tau)\xxx F(t)$, which involves
$y\fff x$. Thus, the second condition of the one-point time definition
also is satisfied. 

Therefore, the time $\psi$ is one-point. ~ ~ ~ ~ ~ \EndProof

\BeginThm  \label{Thm:MainQuasipointCrono} 

Any oriented set can be quasi one-point chronologized. 

\EndThm

\BeginProof 

\textbf{1.} Let $\MM$ be an oriented set. Denote by $\LL$ the set
of all maximum chains of the $\MM$. In accordance with theorem \ref{Thm:ChainChrono},
for any chain $L\in\LL$ there exists an one-point chronologization
$\left((\T_{L},\leq_{L}),\psi_{L}\right)$ of the oriented set $(L,\fff)$.
Similarly to the proof of the theorem \ref{Thm:ChainChrono}, without
loss of generality we can assume, that $\T_{L}\cap\T_{M}=\emptyset$,
$L\neq M$. Denote: 

\begin{equation}
\T:=\bigcup_{L\in\LL}\T_{L}.\label{eq:ChainTUnionDef}\end{equation}
Let $\les$ be any linear order relation on $\LL$ (by Zermelo's theorem,
such linear order relation necessarily exists). By virtue of (\ref{eq:ChainTUnionDef}),
for any element $t\in\T$ chain $L(t)\in\LL$ exists such, that $t\in\T_{L(t)}$.
Since $\T_{F}\cap\T_{G}=\emptyset$ ($F\neq G$), this chain $L(t)$
is unique. This means, that the following assertion is true: 

\begin{lyxlist}{00.00.0000}
\item [{\textbf{~~~~(L)}}] \emph{For any element $t\in\T$ the condition
$t\in\T_{L}$ ($L\in\LL$) causes $L=L(t)$.}
\end{lyxlist}
Let $t,\tau\in\T$. We shall put $t\leq\tau$ if and only if one of
the following conditions is satisfied: 

(O1) $L(t)\neq L(\tau)$ and $L(t)\les L(\tau)$.

(O2) $L(t)=L(\tau)$ and $t\leq_{L(t)}\tau$.

The pair $\left(\T,\leq\right)$ is the ordered union of the (linear
ordrered) family of linear ordered sets $\left(\T_{L}\right)_{L\in\LL}$.
Thus, by \cite[p. 208]{Kuratovskii}, $\left(\T,\leq\right)$ is a
linear ordered set. Denote: 

\begin{equation}
\psi(t):=\psi_{L(t)}\left(t\right)\qquad t\in\T.\label{eq:TimeDef03}\end{equation}
Note, that $\psi(t)=\psi_{L(t)}\left(t\right)\subseteq L(t)\subseteq\BsM$,
$t\in\T$. 

\textbf{2.} We intend to prove, that the mapping $\psi:\T\mapsto2^{\BsM}$
is a time. 

2.a) Let $x\in\BsM$. In accordance with corollary \ref{Nasl:MaxChainExist},
there exists the maximum chain $N_{x}\in\LL$ such, that $x\in N_{x}$.
And, since the mapping $\psi_{N_{x}}:\T_{N_{x}}\mapsto2^{N_{x}}$
is a time, there exists an element $t_{x}\in\T_{N_{x}}\subseteq\T$,
such, that $x\in\psi_{N_{x}}\left(t_{x}\right)$. Since $t_{x}\in\T_{N_{x}}$,
by assertion (L) (see above) we have $N_{x}=L\left(t_{x}\right)$.
Therefore: 

\[
\psi\left(t_{x}\right)=\psi_{L\left(t_{x}\right)}\left(t_{x}\right)=\psi_{N_{x}}\left(t_{x}\right)\ni x.\]
Thus, for any element $x\in\BsM$ always an element $t_{x}\in\T$
exists, such, that $x\in\psi\left(t_{x}\right)$. 

2.b) Let $x,y\in\BsM$, $y\fff x$, $x\neq y$. According to the corollary
\ref{Nasl:MaxChainExist}, a maximum chain $N_{xy}\in\LL$ exists,
such, that $x,y\in N_{xy}$. Since $y\fff x$, $x\neq y$ and mapping
$\psi_{N_{xy}}:\T_{N_{xy}}\mapsto2^{N_{xy}}$ is a time, there exist
elements $t_{x},t_{y}\in\T_{N_{xy}}$ such, that $t_{x}<_{N_{xy}}t_{y}$
(ie $t_{x}\leq_{N_{xy}}t_{y}$, $t_{x}\neq t_{y}$) and $x\in\psi_{N_{xy}}\left(t_{x}\right)$,
$y\in\psi_{N_{xy}}\left(t_{y}\right)$. Since $t_{x},t_{y}\in\T_{N_{xy}}$,
in accordance with assertion (L), we obtain $L\left(t_{x}\right)=L\left(t_{y}\right)=N_{xy}$.
Therefore: \begin{gather*}
\psi\left(t_{x}\right)=\psi_{L\left(t_{x}\right)}\left(t_{x}\right)=\psi_{N_{xy}}\left(t_{x}\right)\ni x;\\
\psi\left(t_{y}\right)=\psi_{L\left(t_{y}\right)}\left(t_{y}\right)=\psi_{N_{xy}}\left(t_{y}\right)\ni y.\end{gather*}
Since $L\left(t_{x}\right)=L\left(t_{y}\right)=N_{xy}$, $t_{x}\leq_{N_{xy}}t_{y}$
and $t_{x}\neq t_{y}$, by (O2) we obtain $t_{x}\leq t_{y}$ and $t_{x}\neq t_{y}$,
that is $t_{x}<t_{y}$. 

Consequently for any elements $x,y\in\BsM$ such, that $y\fff x$,
$x\neq y$ there exists elements $t_{x},t_{y}\in\T$, such, that $t_{x}<t_{y}$,
$x\in\psi(t_{x})$, $y\in\psi(t_{y})$. 

Thus, the mapping $\psi:\T\mapsto2^{\BsM}$ really is a time on $\MM$.

\textbf{3.} Since the times $\left\{ \psi_{L}|L\in\LL\right\} $ are
one point, from (\ref{eq:TimeDef03}) it follows, that for any $t\in\T$
the set $\psi(t)$ is a singleton. Thus, the time $\psi$ is quasi
one-point. ~ ~ ~\EndProof

It is clear, that any oriented set $\MM$, containing elementary states
$x_{1},x_{2}\in\BsM$ such, that $x_{2}\nff x_{1}$ and $x_{1}\nff x_{2}$,
can not be one-point chronologized. Thus, not any oriented set can
be one-point chronologized. The next assertion shows, that not any
oriented set can be monotone chronologized. 

\BeginAs

If oriented set $\MM$ contains elementary states $x_{1},x_{2},x_{3}\in\BsM$
such, that $x_{2}\fff x_{1}$,~$x_{1}\nff x_{2}$,~ $x_{3}\fff x_{2}$,~$x_{2}\nff x_{3}$,~
$x_{1}\fff x_{3}$,~$x_{1}\neq x_{3}$, then this oriented set can
not be monotone chronologized.

\EndAs

\BeginProof Let oriented set $\MM$ contains elementary states $x_{1},x_{2},x_{3}\in\BsM$,
satisfying the conditions of assertion. Suppose, that the monotone
chronologization $\left(\left(\T,\leq\right),\psi\right)$ of the
oriented set $\MM$ exists. This means, that the mapping $\psi:\T:\mapsto2^{\BsM}$
is a monotone time on $\MM$. Since $x_{1}\fff x_{3}$ and $x_{1}\neq x_{3}$,
by time definition \ref{Def:ChronoMain}, there exist time points
$t_{1},t_{3}\in\T$ such, that $x_{1}\in\psi(t_{1})$,~$x_{3}\in\psi(t_{3})$
and $t_{3}<t_{1}$. Also, by time definition \ref{Def:ChronoMain},
there exists time point $t_{2}\in\T$, such, that $x_{2}\in\psi(t_{2})$.
Then, by definition of monotone time \ref{Def:Mono_Point_Time}, from
conditions $x_{2}\fff x_{1}$,~$x_{1}\nff x_{2}$,~ $x_{3}\fff x_{2}$,~$x_{2}\nff x_{3}$
it follows, that $t_{1}<t_{2}$, $t_{2}<t_{3}$. Hence $t_{1}<t_{3}$,
which contradicts inequality above ($t_{3}<t_{1}$). Thus, the assumption
about the existence of monotone chronologization of $\MM$ is wrong.
~ ~ ~ \EndProof

\BeginProbl Find necessary and sufficient conditions of existence
of monotone chronologization for oriented set. 

\EndProbl

\section{Time and Simultaneity. Internal Time }

\BeginDef

Let $\left(\MM,\TT,\psi\right)=\left(\MM,\left(\T,\leq\right),\psi\right)$
be a primitive changeable set. The set 

\[
Y_{\psi}=\left\{ \psi(t)\:|\; t\in\T\right\} \]
will be referred to as the\textbf{ set of simultaneous states}, generated
by the time $\psi$. 

\EndDef

Directly from the time definition (definition \ref{Def:ChronoMain})
it follows the next assertion.

\BeginAs \label{As:SynhrDefMotivation}

Let $\left(\MM,\TT,\psi\right)=\left(\MM,\left(\T,\leq\right),\psi\right)$
be a primitive changeable set, and $Y_{\psi}$ be a set of simultaneous
states, generated by the time $\psi$. Then:

\[
\bigcup_{A\in Y_{\psi}}A=\BsM.\]

\EndAs  

\BeginDef 

Let $\MM$ be an oriented set. Any family of sets $\YY\subseteq2^{\BsM}$,
which possesses the property $\bigcup_{A\in\YY}\, A=\BsM$ we will
call the \textbf{simultaneity} on $\MM$.

\EndDef 

According to the assertion \ref{As:SynhrDefMotivation}, any set of
simultaneous states, generated by the time $\psi$ of a primitive
changeable set $\left(\MM,\TT,\psi\right)$ is a simultaneity. 

Let $\YY$ be a simultaneity on an oriented set $\MM$ and $A,B\in\YY$.
We will denote $B\fff A$ (or $B\ffm A$) if and only if: 

\[
A=B=\emptyset,\;\textrm{or}\quad\exists\, x\in A\:\exists\, y\in B\;\left(y\fff x\right).\]

The next lemma is trivial.

\BeginLem \label{Lem:SyncrOrient}

Let $\YY$ be a simultaneity on an oriented set $\MM$. Then the pair
$\left(\YY,\fff\right)$ itself is an oriented set. 

\EndLem

\BeginThm \label{Thm:SyncronizationCrono}

Let $\MM$ be an oriented set and $\YY\subseteq2^{\BsM}$ be a simultaneity
on $\MM$. Then there exists time $\psi$ on the oriented set $\MM$,
such, that: 

\[
\YY=Y_{\psi},\]
where $Y_{\psi}$ is set of simultaneous states, generated by the
time $\psi$. 

\EndThm 

\BeginProof  \newcommand{\xxv}{\tilde{x}}
\newcommand{\Mxv}{\tilde{M}}
\newcommand{\Yxv}{\tilde{\YY}}

Let $\MM$ be an oriented set and $\YY\subseteq2^{\BsM}$ be a simultaneity
on $\MM$. 

\textbf{a)} First we prove the theorem in the case, where the simultaneity
$\YY$ {}``separates'' sequential unequal elementary states, that
is where the following condition holds: 

\begin{description}
\item [{(Rp)}] For any $x,y\in\BsM$ such, that $y\fff x$ and $x\neq y$
there exists sets $A,B\in\YY$ such, that $x\in A$, $y\in B$ and
$A\neq B$. 
\end{description}
By lemma \ref{Lem:SyncrOrient}, $\left(\YY,\fff\right)$ is an oriented
set. According to theorem \ref{Thm:MainQuasipointCrono}, oriented
set $\left(\YY,\fff\right)$ can be quasi one-point chronologized.
Let $\Psi:\T\mapsto2^{\YY}$ be quasi one-point time on $\left(\YY,\fff\right)$.
By definition \ref{Def:Mono_Point_Time} of quasi one-point time,
for any $t\in\T$ the set $\Psi(t)$ is a singleton. This means, that: 

\[
\forall\, t\in\T\;\exists A_{t}\in\YY\quad\Psi(t)=\left\{ A_{t}\right\} .\]
Denote: \[
\psi(t):=A_{t},\quad t\in\T.\]
The next aim is to prove, that $\psi$ is time on $\MM$. Since $\psi$
is time on $\YY$, then $\bigcup_{t\in\T}\Psi(t)=\YY$. And, taking
into account, that $\Psi(t)=\left\{ A_{t}\right\} $, $t\in\T$, we
obtain $\left\{ A_{t}\,|\: t\in\T\right\} =\YY$. Therefore, since
the family of sets $\YY$ is simultaneity on $\MM$, we have, $\bigcup_{t\in\T}\,\psi(t)=\bigcup_{t\in\T}\, A_{t}=\bigcup_{A\in\YY}A=\BsM$.
Hence, for any $x\in\BsM$ there exists a time moment $t\in\T$ such,
that $x\in\psi(t)$. Thus, the first condition of the time definition
\ref{Def:ChronoMain} is satisfied. Now, we are going to prove, that
the second condition of the definition \ref{Def:ChronoMain} also
is satisfied. Let $x,y\in M$, $y\fff x$ and $x\neq y$. By condition
(Rp), there exist sets $A,B\in\YY$, such, that $x\in A$, $y\in B$
and $A\neq B$. Taking into account, that $x\in A$, $y\in B$ and
$y\fff x$, we obtain $B\fff A$. Since $B\fff A$, $A\neq B$ and
$\Psi$ --- time on $\left(\YY,\fff\right)$, there exist time moments
$t,\tau\in\T$ such, that $A\in\Psi(t)$, $B\in\Psi(\tau)$ and $t<\tau$.
And, taking into account $\Psi(t)=\left\{ A_{t}\right\} $, $\Psi(\tau)=\left\{ A_{\tau}\right\} $,
we obtain $A=A_{t}$, $B=A_{\tau}$, that is $A=\psi(t)$, $B=\psi(\tau)$.
Since $x\in A$, $y\in B$, then $x\in\psi(t)$, $y\in\psi(\tau)$,
where $t<\tau$. 

Thus, $\psi$ is a time on $\MM$. Moreover, taking into account what
has been proven before, we get:

\[
Y_{\psi}=\left\{ \psi(t)\,|\: t\in\T\right\} =\left\{ A_{t}\,|\: t\in\T\right\} =\YY.\]

Hence, in the case, when (Rp) is true, the theorem is proved.

\textbf{b)} Now we consider the case, when the condition (Rp) is not
satisfied. Chose any element $\xxv$, such, that $x\notin\BsM$. Denote: 

\[
\Mxv:=\BsM\cup\left\{ \xxv\right\} .\]
 For elements $x,y\in\Mxv$ we put $y\widetilde{\fff}x$ if and only
if one of the following conditions is satisfied:

\begin{center}
(a) $x,y\in\BsM$ and $y\fff x$; ~ ~ (b) $x=y=\xxv$. \smallskip{}

\par\end{center}

\noindent That is the relation $\widetilde{\fff}$ can be represented
by formula $\widetilde{\fff}=\fff\cup\left\{ (\xxv,\xxv)\right\} $.
Taking into account, that for $x,y\in\BsM$ the condition $y\widetilde{\fff}x$
is is equivalent to the condition $y\fff x$, further for relations
$\widetilde{\fff}$ and $\fff$ we will use the same denotation $\fff$,
assuming, that the relation $\fff$ is expanded to the set $\Mxv$.
It is obvious, that $\left(\Mxv,\fff\right)$ is an oriented set.
Denote: 

\noindent \[
\YY_{0}:=\left\{ B\in\YY\,|\:\exists x,y\in B:\, x\neq y,\, y\fff x\right\} .\]
 Since the condition (Rp) is not satisfied, then $\YY_{0}\neq\emptyset$.
For $B\in\YY_{0}$ we put: 

\noindent \[
\Bxv:=B\cup\{\xxv\}.\]
Also we put: \begin{gather*}
\Yxv_{0}:=\left\{ \Bxv\,|\: B\in\YY_{0}\right\} \\
\Yxv:=\YY\cup\Yxv_{0}.\end{gather*}
Since $\YY$ is a simultaneity on $\MM$, and $\xxv\in\Bxv$ for any
set $\Bxv\in\Yxv_{0}$, then $\Yxv$ is a simultaneity on $\left(\Mxv,\fff\right)$.
The simultaneity $\Yxv$ readily satisfies the condition (Rp). Therefore,
according to proven in paragraph a), there exists the time $\psi_{1}\,:\,\T\mapsto2^{\Mxv}$
on $\left(\Mxv,\fff\right)$, such, that $Y_{\psi_{1}}=\left\{ \psi_{1}(t)\,|\: t\in\T\right\} =\Yxv$.
Now, we denote: 

\noindent \[
\psi(t):=\psi_{1}(t)\cap\BsM,\quad t\in\T.\]
In accordance with the assertion \ref{As:SetEmbedChrono}, $\psi$
is a time on $\MM$. Moreover we obtain: \begin{multline*}
Y_{\psi}=\left\{ \psi(t)\,|\: t\in\T\right\} =\left\{ \psi_{1}(t)\cap\BsM\,|\: t\in\T\right\} =\left\{ A\cap\BsM\,|\: A\in\Yxv\right\} =\\
=\left\{ A\cap\BsM\,|\: A\in\YY\right\} \cup\left\{ A\cap\BsM\,|\: A\in\Yxv_{0}\right\} =\\
=\left\{ A\,|\: A\in\YY\right\} \cup\left\{ \Bxv\cap\BsM\,|\: B\in\YY_{0}\right\} =\YY\cup\left\{ B\,|\: B\in\YY_{0}\right\} =\\
=\YY\cup\YY_{0}=\YY.\end{multline*}

\EndProof 

\BeginDef 

Let $\YY\subseteq2^{\BsM}$ be any simultaneity on the oriented set
$\MM$. Time $\psi$ on $\MM$ will be named the \textbf{generating
time} of the simultaneity $\YY$ if and only if $\YY=Y_{\psi}$, where
$Y_{\psi}$ is the set of simultaneous states, generated by the time
$\psi$.

\EndDef 

Thus, the theorem \ref{Thm:SyncronizationCrono} asserts, that any
simultaneity always has a generating time. Below we consider the question
about uniqueness of a generating time for a simultaneity (under the
certain conditions). To ensure the correctness of staging this question,
first of all, we need to introduce the concept of equivalence of two
chronologizations. 

\BeginDef \label{Def:ChronoEq}

Let $\MM$ be an oriented set and $\psi_{1}:\T_{1}\mapsto2^{\BsM}$,
$\psi_{2}:\T_{2}\mapsto2^{\BsM}$ be times for $\MM$, defined on
the linear ordered sets $\left(\T_{1},\leq_{1}\right)$, $\left(\T_{2},\leq_{2}\right)$.
We say, that the chronologizations $\cH_{1}=\left(\left(\T_{1},\leq_{1}\right),\psi_{1}\right)$
and $\cH_{2}=\left(\left(\T_{2},\leq_{2}\right),\psi_{2}\right)$
are\textbf{ equivalent} (using the denotation $\cH_{1}\heq\cH_{2}$)
if and only if there exist an one-to-one correspondence $\xi:\T_{1}\mapsto\T_{2}$
such, that: 

1) $\xi$ is order isomorphism between the linearly ordered sets $\left(\T_{1},\leq_{1}\right)$,
$\left(\T_{2},\leq_{2}\right)$, that is for any $t,\tau\in\T_{1}$
the inequality $t\leq_{1}\tau$ is equivalent to the inequality $\xi\left(t\right)\leq_{2}\xi\left(\tau\right)$. 

2) For any $t\in\T_{1}$ it is valid the equality $\psi_{1}(t)=\psi_{2}(\xi(t))$. 

\EndDef

\newcommand{\VV}{\mathcal{V}}
\BeginAs Let $\MM$ be any oriented set and $\VV$ is any set, which
consists of chronologizations of $\MM$. Then the relation $\heq$
is an equivalence relation on $\VV$. 

\EndAs

\BeginProof 

Throughout in this proof $\cH_{i}=\left(\left(\T_{i},\leq_{i}\right),\psi_{i}\right)\in\VV$
($i=1,2,3$) mean any three chronologizations of the oriented set
$\MM$. 

\textbf{1)} Reflexivity.  Denote $\xi_{11}(t):=t$, $t\in\T_{1}$.
It is obvious that $\xi_{11}$ is a order isomorphism between $\left(\T_{1},\leq_{1}\right)$
and $\left(\T_{1},\leq_{1}\right)$. Besides we have $\psi(t)=\psi(\xi(t))$,
$t\in\T$. Thus $\cH_{1}\heq\cH_{1}$. 

\textbf{2)} Symmetry. Let $\cH_{1}\heq\cH_{2}$. Then, by definition
\ref{Def:ChronoEq}, there exist an one-to-one correspondence $\xi_{12}:\T_{1}\mapsto\T_{2}$
such, that: 

1) $\xi_{12}$ is order isomorphism between the linearly ordered sets
$\left(\T_{1},\leq_{1}\right)$, $\left(\T_{2},\leq_{2}\right)$. 

2) $\psi_{1}(t)=\psi_{2}(\xi_{12}(t))$, for any $t\in\T_{1}$. 

Since the mapping $\xi_{12}$ is bijection, there exists the inverse
mapping $\xi_{21}(t):=\arc{\xi_{12}}(t)\in\T_{1}$, $t\in\T_{2}$.
Since $\xi_{12}$ is order isomorphism between the linearly ordered
sets $\left(\T_{1},\leq_{1}\right)$, $\left(\T_{2},\leq_{2}\right)$,
then $\xi_{21}$ is order isomorphism between $\left(\T_{2},\leq_{2}\right)$
and $\left(\T_{1},\leq_{1}\right)$. Moreover, for any $t\in\T_{2}$
we obtain: 

\[
\psi_{2}(t)=\psi_{2}\left(\xi_{12}\left(\arc{\xi_{12}}(t)\right)\right)=\psi_{1}\left(\xi_{21}(t)\right).\]
 Thus, $\cH_{2}\heq\cH_{1}$.

\textbf{3)} Transitivity. Let $\cH_{1}\heq\cH_{2}$, $\cH_{2}\heq\cH_{3}$.
Then there exist order isomorphisms $\xi_{12}:\T_{1}\mapsto\T_{2}$
and $\xi_{23}:\T_{2}\mapsto\T_{3}$ such, that $\psi_{1}(t)=\psi_{2}\left(\xi_{12}(t)\right)$,
$t\in\T_{1}$ and $\psi_{2}(t)=\psi_{3}\left(\xi_{23}(t)\right)$,
$t\in\T_{2}$. Denote, $\xi_{13}(t):=\xi_{23}\left(\xi_{12}(t)\right)$,
$t\in\T_{1}$. It is easy to verify, that $\xi_{13}$ is an order
isomorphism between $\left(\T_{1},\leq_{1}\right)$ and $\left(\T_{3},\leq_{3}\right)$.
Moreover, for any $t\in\T_{1}$ we obtain: 

\begin{gather*}
\psi_{1}(t)=\psi_{2}\left(\xi_{12}(t)\right)=\psi_{3}\left(\xi_{23}\left(\xi_{12}(t)\right)\right)=\psi_{3}\left(\xi_{13}(t)\right).\end{gather*}
 Therefore, $\cH_{1}\heq\cH_{3}$. ~ ~ ~ \EndProof

Now, if we consider the question about uniqueness of a generating
time for a simultaneity up to equivalence of corresponding chronologizations,
the answer is still negative. For example we can consider a linearly
ordered sets $\left(\T,\leq\right)$ and $\left(\T_{1},\leq\right)$
such, that $\emptyset\neq\T_{1}\subset\T$ (more accurately linear
order relation on $\T_{1}$ is a restriction of order relation on
$\T$, and both relations are denoted by the same symbol {}``$\leq$'').
If $\psi_{1}:\T_{1}\mapsto2^{\BsM}$ is a time on the oriented set
$\MM$, then for any element $t_{1}\in\T_{1}$ we can define the time: 

\[
\psi(t):=\begin{cases}
\psi_{1}(t), & t\in\T_{1}\:;\\
\psi_{1}\left(t_{1}\right), & t\in\T\setminus\T_{1}\:,\end{cases}\]
This time is such, that $Y_{\psi}=Y_{\psi_{1}}$, although, in the
case, when the ordered sets $\left(\T,\leq\right)$ and $\left(\T_{1},\leq\right)$
are not isomorphic, the chronologizations $\left(\left(\T,\leq\right),\psi\right)$
and $\left(\left(\T_{1},\leq\right),\psi_{1}\right)$ are not equivalent.
That is why, to obtain the positive answer for the above question,
further we will impose additional conditions on simultaneity and generating
time. 

\BeginDef \label{Def:mff_mtf}

Let $\MM$ be an oriented set. 

1) We will say, that a set $B\subseteq\BsM$ is \textbf{monotonously
sequential} to the set $A\in\BsM$ in the oriented set $\MM$ if and
only if there exist elements $x\in A$ and $y\in B$ such, that $y\ffm x$
and $x\nff\limits _{\MM}y$. In this case we will use the denotation
$B\mff\limits _{\MM}A$. 

2) Let $\mathcal{Q}\subseteq2^{\BsM}$ be any system of subsets of
$\BsM$. We will say, that a set $A\in\mathcal{Q}$ is \textbf{transitively
monotonously sequential} to a set $B\in\mathcal{Q}$ relative to the
system $\cQ$ if and only if there exist a finite sequence of sets
$C_{0},C_{1},\cdots,C_{n}\in\mathcal{Q}$ ($n\in\N$) such, that $C_{0}=A$,
$C_{n}=B$ and $C_{k}\mff\limits _{\MM}C_{k-1}$, for any $k\in\overline{1,n}$
(where $\overline{1,n}=\left\{ 1,\dots,n\right\} $). In this case
we will use the denotation $B\mtf\limits _{\MM}^{\mathcal{Q}}A$. 

\EndDef 

In the case where the oriented set $\MM$ is known in advance, the
char $\MM$ in the denotations $\mff\limits _{\MM}$ and $\mtf\limits _{\MM}^{\mathcal{Q}}$
will be released, and we will use denotations $\mff$ and $\mtf\limits ^{\mathcal{Q}}$
instead.

\BeginRmk \label{Rmk:mtfTransitive}

It is easy to prove, that for any system of sets $\mathcal{Q}\subseteq2^{\BsM}$
(in any oriented set $\MM$) the binary relation $\mtf\limits ^{\mathcal{Q}}$
is transitive on the set $\mathcal{Q}$. 

\EndRmk

\BeginAs \label{As:SubsetTrMFollow}

Let $\MM$ be an oriented set, and $\cS,\cS'\subseteq2^{\BsM}$ are
systems of subsets of $\BsM$, moreover $\cS\sqsubseteq\cS'$ (this
means, that for any set $A\in\cS$ there exist a set $A'\in\cS'$
such, that $A\subseteq A'$). 

Then for any $A,B\in\cS$ and $A',B'\in\cS'$ such, that $A\subseteq A'$,
$B\subseteq B'$ correlation $B\mtf\limits ^{\cS}A$ involves the
correlation $B'\mtf\limits ^{\cS'}A'$.

\EndAs 

\BeginProof 

Suppose that the conditions of the assertion are performed. Let $A,B\in\cS$,
$A',B'\in\cS'$, $A\subseteq A'$, $B\subseteq B'$ and $B\mtf\limits ^{\cS}A$.
Then, there exists a finite sequence of sets $C_{0},\cdots,C_{n}\in\cS$
($n\in\N$) such, that $C_{0}=A$, $C_{n}=B$ and $C_{k}\mff C_{k-1}$
(for any $k\in\overline{1,n}$). Since $\cS\sqsubseteq\cS'$, there
exist sets $C_{0}',\cdots,C_{n}'\in\cS'$ such, that $C_{k}\subseteq C_{k}'$
($k\in\overline{0,n}$). Moreover, since $C_{0}=A\subseteq A'\in\cS'$,
$C_{n}=B\subseteq B'\in\cS'$, we can believe that $C_{0}'=A'$, $C_{n}'=B'$.
Taking into account that $C_{k}\subseteq C_{k}'$ ($k\in\overline{0,n}$),
and $C_{k}\mff C_{k-1}$ ($k\in\overline{1,n}$), we obtain $C_{k}'\mff C_{k-1}'$,
$k\in\overline{1,n}$ (where $C_{0}'=A'$, $C_{n}'=B'$). Thus $B'\mtf\limits ^{\cS'}A'$.
~ ~ ~ \EndProof 

\BeginDef \label{Def:NorepSynh}

Let $\MM$ be an oriented set. 

1) System of sets $\cS\subseteq2^{\BsM}$ will be referred to as \textbf{unrepeatable}
if and only if there not exist sets $A,B\in\cS$ such, that $A\mtf\limits ^{\cS}B$
and $B\mtf\limits ^{\cS}A$. In particular, in the case, where a simultaneity
$\YY\subseteq2^{\BsM}$ is unrepeatable system of sets, this simultaneity
we will call an \textbf{unrepeatable simultaneity}. 

2) Simultaneity $\YY\subseteq2^{\BsM}$ will be referred to as \textbf{precise}
if and only if for any $x,y\in\BsM$ such, that $y\fff x$ and $x\neq y$
there exist sets $A,B\in\YY$ such, that $x\in A$, $y\in B$, $A\neq B$
and $B\mtf\limits ^{\YY}A$ (this means, that this simultaneity {}``fixes''
all changes on the oriented set $\MM$). 

3) Simultaneity $\YY$ will be called \textbf{precisely-unrepeatable}
if and only if it is precise and, at the same time, unrepeatable. 

\EndDef 

\BeginRmk \label{Rmk:mtfStrOrder(unrep_sim)}

From the remark \ref{Rmk:mtfTransitive} it readily follows, that
in the case, where a simultaneity $\YY\subseteq2^{\BsM}$ is unrepeatable,
the relation $\mtf\limits ^{\YY}$ is a strict order on $\YY$ (ie
$\mtf\limits ^{\YY}$ is anti-reflexive and transitive relation).

\EndRmk 

\BeginAs \label{As:NorepSysProp} 

Let $\MM$ be an oriented set and $\cS\subseteq2^{\BsM}$ is unrepeatable
system of sets. Then:

1) For any $A,B\in\cS$, such, that $B\mtf\limits ^{\cS}A$ is true
$A\neq B$. 

2) If $\cS_{1}\subseteq2^{\BsM}$ and $\cS_{1}\sqsubseteq\cS$, then
$\cS_{1}$ also is unrepeatable system of sets. 

\EndAs 

\BeginProof 

1) Let $\cS\subseteq2^{\BsM}$ be unrepeatable system of sets. If
we suppose, that $B\mtf\limits ^{\cS}A$ and $A=B$ (for some $A,B\in\cS$),
then we obtain $A\mtf\limits ^{\cS}B$ and $B\mtf\limits ^{\cS}A$,
which is impossible, since the system of sets $\cS$ is unrepeatable. 

2) Let $\cS_{1}\sqsubseteq\cS$. Suppose, that the system of sets
$\cS_{1}$ is not unrepeatable. Then, there exists sets $A_{1},B_{1}\in\cS_{1}$
such, that $A_{1}\mtf\limits ^{\cS_{1}}B_{1}$ and $B_{1}\mtf\limits ^{\cS_{1}}A_{1}$.
Since $\cS_{1}\sqsubseteq\cS$, there exist sets $A,B\in\cS$ such,
that $A_{1}\subseteq A$, $B_{1}\subseteq B$. Hence, by assertion
\ref{As:SubsetTrMFollow}, we obtain $A\mtf\limits ^{\cS}B$ and $B\mtf\limits ^{\cS}A$,
which is impossible, since the system of sets $\cS$ is unrepeatable.
Thus, the system of sets $\cS_{1}$ is unrepeatable, because the opposite
assumption is wrong. ~ ~ ~ ~ \EndProof 

\BeginLem \label{Lem:ForSingleMNCrono1} 

Let $\psi:\T\mapsto2^{\BsM}$ be a monotone time on an oriented set
$\MM$, and $Y_{\psi}$ be a simultaneity, generated by the time $\psi$.
Then for any $t_{1},t_{2}\in\T$ the condition $\psi\left(t_{2}\right)\mtf\limits ^{Y_{\psi}}\psi\left(t_{1}\right)$
leads to $t_{1}<t_{2}$. 

\EndLem 

\BeginProof 

1) First we consider the case, where $\psi\left(t_{2}\right)\mff\psi\left(t_{1}\right)$.
In this case, by definition \ref{Def:mff_mtf}, there exist elements
$x_{1}\in\psi\left(t_{1}\right)$, $x_{2}\in\psi\left(t_{2}\right)$
such, that $x_{2}\fff x_{1}$ and $x_{1}\nff x_{2}$. Hence, since
the time $\psi$ is monotone, we obtain $t_{1}<t_{2}$ (by the definition
\ref{Def:Mono_Point_Time}). 

Now, we consider the general case, $\psi\left(t_{2}\right)\mtf\limits ^{Y_{\psi}}\psi\left(t_{1}\right)$.
In this case, by definition \ref{Def:mff_mtf}, there exist time points
$\tau_{0},\tau_{1},\cdots,\tau_{n}\in\T$ such, that $\tau_{0}=t_{1}$,
$\tau_{n}=t_{2}$ and $\psi\left(\tau_{k}\right)\mff\psi\left(\tau_{k-1}\right)$
for any $k\in\overline{1,n}$. By statement 1), $\tau_{k-1}<\tau_{k}$,
$k\in\overline{1,n}$. Thus, $t_{1}=\tau_{0}<\tau_{1}<\cdots<\tau_{n}=t_{2}$.
~ ~ ~ \EndProof 

\BeginDef \label{Def:SnConnectedSynh}

We will say, that a simultaneity $\YY$ on an oriented set is \textbf{monotone-connected}
if and only if for any sets $A,B\in\YY$ such, that $A\neq B$ it
holds one of the conditions $A\mtf\limits ^{\YY}B$ or $B\mtf\limits ^{\YY}A$.

\EndDef 

\BeginRmk \label{Rmk:mtfLinOrder}

Directly from the definition \ref{Def:SnConnectedSynh} and remark
\ref{Rmk:mtfStrOrder(unrep_sim)} it follows, that if a simultaneity
$\YY\subseteq2^{\BsM}$ is unrepeatable and monotone-connected, then
the relation $\mtf\limits ^{\YY}$ is a strict linear order on $\YY$. 

\EndRmk 

\BeginDef 

Let $\MM$ be an oriented set and $\left(\T,\leq\right)$ be a linear
ordered set. Time $\psi:\T\mapsto2^{\BsM}$ will be called \textbf{incessant}
if and only if there not exist time points $t_{1},t_{2}\in\T$ such,
that $t_{1}<t_{2}$ and for any $t\in\T$, $t_{1}\leq t\leq t_{2}$
it is true the equality $\psi(t)=\psi\left(t_{1}\right)$. In the
case, where the time $\psi$ is both monotone and incessant it will
be called \textbf{strictly monotone}.

\EndDef

\BeginLem  \label{Lem:ForSingleMNCrono2}

Let $\YY$ be precisely-unrepeatable and monotone-connected simultaneity
on the oriented set $\MM$ and $\psi\,:\:\T\mapsto2^{\BsM}$ is the
time, generating this simultaneity. 

1) If the time $\psi$ is strictly monotone, then it is \textbf{unrepeatable}
(this means, that for any $t_{1},t_{2}\in\T$ such, that $t_{1}\neq t_{2}$
the correlation $\psi\left(t_{1}\right)\neq\psi\left(t_{2}\right)$
is valid). 

2) The time $\psi$ is strictly monotone if and only if for any $t_{1},t_{2}\in\T$
inequality $t_{1}<t_{2}$ implies the correlation $\psi\left(t_{2}\right)\mtf\limits ^{\YY}\psi\left(t_{1}\right)$. 

3) If the time $\psi$ is strictly monotone, then the strictly linearly
ordered sets $\left(\T,>\right)$ and $\left(\YY,\mtf\limits ^{\YY}\right)$
are isomorphic relative the order, and the mapping $\psi:\T\mapsto\YY$
is the order isomorphism between them. 

\EndLem 

\BeginProof 

1) Let, under conditions of the lemma, time $\psi\,:\:\T\mapsto2^{\BsM}$
be strictly monotone. Suppose, there exist time points $t_{1,}t_{2}\in\T$
such, that $t_{1}<t_{2}$ and $\psi\left(t_{1}\right)=\psi\left(t_{2}\right)$.
Since the time $\psi$ (being strictly monotone) is incessant, there
exists a time point $t_{3}\in\T$ such, that $t_{1}<t_{3}<t_{2}$
and $\psi\left(t_{3}\right)\neq\psi\left(t_{1}\right)=\psi\left(t_{2}\right)$.
So far as $\psi\left(t_{3}\right)\neq\psi\left(t_{1}\right)$ and
the simultaneity $\YY$ is monotone-connected, one of the conditions
$\psi\left(t_{3}\right)\mtf\limits ^{\YY}\psi\left(t_{1}\right)$
or $\psi\left(t_{1}\right)\mtf\limits ^{\YY}\psi\left(t_{3}\right)$
is performed. But since $t_{1}<t_{3}$ the correlation $\psi\left(t_{1}\right)\mtf\limits ^{\YY}\psi\left(t_{3}\right)$
is impossible by lemma \ref{Lem:ForSingleMNCrono1}. Therefore, $\psi\left(t_{3}\right)\mtf\limits ^{\YY}\psi\left(t_{1}\right)$.
Similarly, since $t_{3}<t_{2}$ and $\psi\left(t_{3}\right)\neq\psi\left(t_{2}\right)$,
we obtain $\psi\left(t_{2}\right)\mtf\limits ^{\YY}\psi\left(t_{3}\right)$.
Hence, taking into account, that $\psi\left(t_{1}\right)=\psi\left(t_{2}\right)$,
we have $\psi\left(t_{3}\right)\mtf\limits ^{\YY}\psi\left(t_{1}\right)$
and $\psi\left(t_{1}\right)\mtf\limits ^{\YY}\psi\left(t_{3}\right)$,
which is impossible, because the simultaneity $\YY=Y_{\psi}$ is unrepeatable. 

2.a) Suppose, that the time $\psi\,:\:\T\mapsto2^{\BsM}$ is strictly
monotone. Chose any time points $t_{1},t_{2}\in\T$ such, that $t_{1}<t_{2}$.
By the first statement of this lemma, $\psi\left(t_{1}\right)\neq\psi\left(t_{2}\right)$.
Since the simultaneity $\YY$ is monotone-connected, one of the conditions
$\psi\left(t_{2}\right)\mtf\limits ^{\YY}\psi\left(t_{1}\right)$
or $\psi\left(t_{1}\right)\mtf\limits ^{\YY}\psi\left(t_{2}\right)$
is performed. But, so far as $t_{1}<t_{2}$, the condition $\psi\left(t_{1}\right)\mtf\limits ^{\YY}\psi\left(t_{2}\right)$
is impossible by lemma \ref{Lem:ForSingleMNCrono1}. Thus: 

\begin{equation}
\forall t_{1},t_{2}\in\T\; t_{1}<t_{2}\Rightarrow\psi\left(t_{2}\right)\mtf\limits ^{\YY}\psi\left(t_{1}\right).\label{eq:t1_lt_t2then}\end{equation}

Now we suppose, that the condition (\ref{eq:t1_lt_t2then}) holds.
The first aim is to prove, that the time $\psi$ is monotone. Consider
any elementary states $x_{1},x_{2}\in\BsM$ such, that $x_{1}\in\psi\left(t_{1}\right)$,
$x_{2}\in\psi\left(t_{2}\right)$, $x_{2}\fff x_{1}$ and $x_{1}\nff x_{2}$
(where $t_{1},t_{2}\in\T$). By definition \ref{Def:mff_mtf}, $\psi\left(t_{2}\right)\mff\psi\left(t_{1}\right)$.
Consequently, 

\begin{equation}
\psi\left(t_{2}\right)\mtf\limits ^{\YY}\psi\left(t_{1}\right).\label{eq:Psi(t2)tfsPsi(t1)}\end{equation}
If we suppose $t_{1}\geq t_{2}$, we must obtain:\begin{equation}
\psi\left(t_{1}\right)\mtf\limits ^{\YY}\psi\left(t_{2}\right)\label{eq:Psi(t1)tfsPsi(t2)}\end{equation}
(indeed, in the case $t_{1}=t_{2}$ the correlation (\ref{eq:Psi(t1)tfsPsi(t2)})
follows from (\ref{eq:Psi(t2)tfsPsi(t1)}), and in the case $t_{1}>t_{2}$
the correlation (\ref{eq:Psi(t1)tfsPsi(t2)}) is caused by the condition
(\ref{eq:t1_lt_t2then})). Thus, in the case $t_{1}\geq t_{2}$, both
of the conditions (\ref{eq:Psi(t2)tfsPsi(t1)}) and (\ref{eq:Psi(t1)tfsPsi(t2)})
must be performed, which is impossible (since the simultaneity $\YY$
is unrepeatable). Consequently, only the inequality $t_{1}<t_{2}$
is possible. This proves that the time $\psi$ is monotone.

Thus, it remains to prove, that the time $\psi$ is incessant. Suppose,
there exist time points $t_{1},t_{2}\in\T$ such, that $t_{1}<t_{2}$,
and $\psi(t)=\psi\left(t_{1}\right)$ for any $t\in\T$, satisfying
$t_{1}\leq t\leq t_{2}$. Then, in particular, $\psi\left(t_{1}\right)=\psi\left(t_{2}\right)$
(where $t_{1}<t_{2}$). Since $t_{1}<t_{2}$, by condition (\ref{eq:t1_lt_t2then}),
the correlation (\ref{eq:Psi(t2)tfsPsi(t1)}) must be performed. But
since $\psi\left(t_{1}\right)=\psi\left(t_{2}\right)$, the correlation
(\ref{eq:Psi(t1)tfsPsi(t2)}) also is performed, which is impossible
(since the simultaneity $\YY$ is unrepeatable). Therefore, the time
$\psi$ is incessant. 

Thus, the time $\psi$ is strictly monotone. 

3) Let the time $\psi\,:\:\T\mapsto2^{\BsM}$ be strictly monotone.
According to the first statement of the lemma, the mapping $\psi:\T\mapsto\YY=Y_{\psi}$
is one-to-one correspondence between $\T$ and $\YY=Y_{\psi}$. According
to the second statement of the lemma, for any $t_{1},t_{2}\in\T$
the inequality $t_{2}>t_{1}$ implies the correlation $\psi\left(t_{2}\right)\mtf\limits ^{\YY}\psi\left(t_{1}\right)$.
Hence, taking into account, that by remark \ref{Rmk:mtfLinOrder},
$\left(\YY,\mtf\limits ^{\YY}\right)$ is a linear ordered set (with
strict order), we conclude, that the mapping $\psi$ is isomorphism
between the strictly linear ordered sets $\left(\T,>\right)$ and
$\left(\YY,\mtf\limits ^{\YY}\right)$. ~ ~ ~ \EndProof 

\BeginRmk

It turns out, that for any precisely-unrepeatable and monotone-connected
simultaneity $\YY\subseteq2^{\BsM}$ the assertion, inverse to the
first statement, of the lemma \ref{Lem:ForSingleMNCrono2}, in general,
is not true. To demonstrate this we present the following example.
\smallskip{}

\EndRmk 

\BeginEx  Let us consider the following oriented set: 

\begin{gather*}
\BsM:=\left\{ 1,2,3,4\right\} ;\\
\ffm:=\left\{ (3,1),(4,2)\right\} \cup\diag\left(\BsM\right),\end{gather*}
that is, in the other words, $3\fff1$, $4\fff2$, $1\fff1$, $2\fff2$,
$3\fff3$, $4\fff4$. In this oriented set we consider the following
simultaneity: \[
\YY:=\left\{ \left\{ 1,2\right\} ,\:\left\{ 3,4\right\} ,\:\left\{ 2,3\right\} \right\} .\]
Then, we have $\left\{ 2,3\right\} \mff\left\{ 1,2\right\} $, $\left\{ 3,4\right\} \mff\left\{ 2,3\right\} $,
$\left\{ 3,4\right\} \mff\left\{ 1,2\right\} $, and $\left\{ 2,3\right\} \nff(m)\left\{ 3,4\right\} $,
$\left\{ 1,2\right\} \nff(m)\left\{ 2,3\right\} $, $\left\{ 1,2\right\} \nff(m)\left\{ 3,4\right\} $,
moreover, any set of simultaneity $\YY$ is not monotonously sequential
by the itself. That is, schematically: 

\begin{gather*}
\begin{array}{ccccc}
\left\{ 3,4\right\}  & \mff & \left\{ 2,3\right\}  & \mff & \left\{ 1,2\right\} \\
\nwarrow & <-- & \mff & <-- & \swarrow\end{array},\end{gather*}
and, moreover, the relation {}``$\mff$'' on the simultaneity $\YY$
is fully generated by the last scheme. And from this scheme it is
evident, that the simultaneity $\YY$ is unrepeatable and monotone-connected.
Moreover, it is easy to verify, that this simultaneity is precise. 

Also we consider the following linear ordered set: 

\[
\T:=\left\{ 1,2,3\right\} ,\]
 with the standard linear order relation on the natural numbers ($\leq$).
The simultaneity $\YY$ can be generated by the following times: 

\begin{gather*}
\psi_{1}:\quad\psi_{1}(1):=\left\{ 1,2\right\} ,\:\psi_{1}(2):=\left\{ 2,3\right\} ,\:\psi_{1}(3):=\left\{ 3,4\right\} ;\\
\psi_{2}:\quad\psi_{2}(1):=\left\{ 1,2\right\} ,\:\psi_{2}(2):=\left\{ 3,4\right\} ,\:\psi_{2}(3):=\left\{ 2,3\right\} .\end{gather*}
Both of times $\psi_{1}$ and $\psi_{2}$ are, evidently, unrepeatable,
but the time $\psi_{2}$ is not monotone, because of: 

\begin{gather*}
2\in\psi_{2}(3),\quad4\in\psi_{2}(2),\\
4\fff2,\:2\nff4,\quad\textrm{but}\;3\not<2.\end{gather*}

\EndEx

\BeginThm \label{Thm:SingleMNCrono1}

For any precisely-unrepeatable and monotone-connected simultaneity
$\YY$ an unique up to equivalence of chronologizations strictly monotone
time $\psi$ exists, such, that $\YY=Y_{\psi}$.

\EndThm 

It should be noted, that the uniqueness up to equivalence of chronologizations
in the theorem \ref{Thm:SingleMNCrono1} is understood as follows:

{}``if strictly monotone times $\psi_{1}$ and $\psi_{2}$, defined
on linear ordered sets $\TT_{1}$ and $\TT_{2}$ are such, that $\YY=Y_{\psi_{1}}=Y_{\psi_{2}}$,
then $\cH_{1}\heq\cH_{2}$, where $\cH_{1}$ and $\cH_{2}$ are corresponding
chronologizations ($\cH_{i}=\left(\TT_{i},\psi_{i}\right)$, $i\in\left\{ 1,2\right\} $)''. 

\BeginProof 

\textbf{1}. Let $\YY$ be precisely-unrepeatable and monotone-connected
simultaneity on an oriented set $\MM$. Then, by remark \ref{Rmk:mtfLinOrder},
$\mtf\limits ^{\YY}$ is a strict linear order on $\YY$. Hence, the
relation $\mtp\limits ^{\YY}$, being inverse to the relation $\mtf\limits ^{\YY}$,
also is a strict linear order on $\YY$. Denote: 

\[
\T:=\YY.\]
 For $t,\tau\in\T=\YY$ we will assume, that $t\leq\tau$ if and only
if: \[
t=\tau\;\textrm{or}\; t\mtp\limits ^{\YY}\tau.\]
 That is $\leq$ is (non-strict) linear order, generated by the strict
order $\mtp\limits ^{\YY}$. Therefore, for $t,\tau\in\T$ the following
logical equivalence is true: 

\begin{equation}
t<\tau\quad\textrm{if and only if}\quad t\mtp\limits ^{\YY}\tau,\label{eq:MainEquvivilence01}\end{equation}
where record means, that $t\leq\tau$ and $t\neq\tau$. Thus, $\left(\T,\leq\right)$
is a linear ordered set. Denote: 

\[
\psi(t):=t,\quad t\in\T=\YY.\]
 Since $\T=\YY$, then $\psi(t)=t\in\YY\subseteq2^{\BsM}$ for $t\in\T$
. 

\textbf{2}. The next aim is to prove, that $\psi$ is a time on $\MM$. 

(a) Since $\YY$ is a simultaneity, then for any $x\in\BsM$ there
exists set $t_{x}\in\YY=\T$, such, that $x\in t_{x}$. Therefore,
we obtain $\psi\left(t_{x}\right)=t_{x}\ni x$. Thus, the first condition
of the time definition \ref{Def:ChronoMain} is performed. 

(b) Suppose, that $x,y\in\BsM$, $y\fff x$ and $x\neq y$. Since
the simultaneity $\YY$ is precise, there exist $t_{x},t_{y}\in\YY=\T$
such, that $x\in t_{x}$, $y\in t_{y}$ and $t_{y}\mtf\limits ^{\YY}t_{x}$.
Then, by (\ref{eq:MainEquvivilence01}), $t_{x}<t_{y}$. Moreover,
since $\psi(t)=t$, $t\in\T$, we have: 

\[
x\in t_{x}=\psi\left(t_{x}\right);\quad y\in t_{y}=\psi\left(t_{y}\right).\]
Consequently, the second condition of the definition \ref{Def:ChronoMain}
also is satisfied. 

Thus, the mapping $\psi$ is a time. 

\textbf{3}. Now we are aim to prove, that the time $\psi$ is strictly
monotone. 

(a) Let $x\in\psi\left(t_{x}\right)$, $y\in\psi\left(t_{y}\right)$,
$y\fff x$ and $x\nff y$. Then $t_{y}=\psi\left(t_{y}\right)\mff\psi\left(t_{x}\right)=t_{x}$.
Consequently, $t_{y}\mtf\limits ^{\YY}t_{x}$, ie, by (\ref{eq:MainEquvivilence01}),
$t_{x}<t_{y}$. Thus, the time $\psi$ is monotone. 

(b) Suppose, that this time is not incessant. Then there exist $t_{1},t_{2}\in\T$
such, that $t_{1}<t_{2}$ and $\psi\left(t\right)=\psi\left(t_{1}\right)$
for any $t\in\T$, satisfying the inequality $t_{1}\leq t\leq t_{2}$.
In particular this means, that $\psi\left(t_{2}\right)=\psi\left(t_{1}\right)$.
And, since $\psi(\tau)=\tau$, $\tau\in\T$, we obtain $t_{2}=t_{1}$,
which contradicts the inequality $t_{1}<t_{2}$. Therefore, the assumption
is wrong, and the time $\psi$ is incessant. 

Thus, the time $\psi$ is strictly monotone. 

\textbf{4}. It remains to prove, that the time $\psi$ is unique up
to equivalence of chronologizations. Let $\psi_{1}:\T_{1}\mapsto2^{\BsM}$
be an other strictly monotone time such, that $Y_{\psi_{1}}=\YY$
(where $\left(\T_{1},\leq_{1}\right)$ is a linear ordered set. Then,
by lemma \ref{Lem:ForSingleMNCrono2}, the linear ordered (by strict
order) sets $\left(\T_{1},>_{1}\right)$ and $\left(\YY,\mtf\limits ^{\YY}\right)$
are isomorphic relative the order, with the mapping $\psi_{1}:\T_{1}\mapsto\YY$
being isomorphism, where $>_{1}$ is relation, inverse to the relation
$<_{1}$, and $<_{1}$ is strict order, generated by non-strict order
$\leq_{1}$. Thus, the ordered sets $\left(\T_{1},\leq_{1}\right)$
and $\left(\YY,\leq\right)=\left(\T,\leq\right)$ also are isomorphic
with the isomorphism $\psi_{1}$. Moreover, for any $t\in\T_{1}$,
we have: 

\[
\psi_{1}(t)=\psi\left(\psi_{1}(t)\right),\]
 ie, by definition \ref{Def:ChronoEq}, $\left(\left(\T_{1},\leq_{1}\right),\psi_{1}\right)\heq\left(\left(\T,\leq\right),\psi\right)$.
~ ~ ~ \EndProof

\BeginDef \label{Def:ChronoProc} 

Let $\MM$ be an oriented set, and $\psi:\T\mapsto2^{\BsM}$ is a
time on $\MM$. 

\medskip{}

A mapping $\hh:\T\mapsto2^{\BsM}$ will be referred to as \textbf{chronometric
process} (for the time $\psi$), if and only if: 

1) $\hh(t)\subseteq\psi(t)$ for any $t\in\T$. \vspace{-1.7mm}

2) For arbitrary $t,\tau\in\T$ inequality $t<\tau$ is valid if and
only if $\hh(\tau)\mtf\limits ^{\hh(\T)}\hh(t)$ and $\hh(t)\neq\hh(\tau)$,
where $\hh(\T)=\left\{ \hh(t)\,|\: t\in\T\right\} $;

\medskip{}

The time $\psi$ on $\MM$ will be referred to as \textbf{internal}
if and only if there exists at least one chronometric process for
this time. 

\EndDef  

Sense of the term {}``internal time'' lies in the fact that in the
case, where a time on a primitive changeable set is internal, this
time can be measured within this primitive changeable set, using the
chronometric process as a {}``clock'' and states of this process
as {}``indicators of time points''. The next aim is to establish
the sufficient condition of existence and uniquness of internal time
for given simultaneity. 

\BeginLem \label{Lem:InternalTime1}

The generating time for precisely-unrepeatable and monotone-connected
simultaneity is internal if and only if it is strictly monotone. 

\EndLem 

\BeginProof 

Let $\MM$ be an oriented set, $\YY$ is precisely-unrepeatable and
monotone-connected simultaneity and $\psi:\T\mapsto2^{\BsM}$ is a
time on $\MM$, which generates $\YY$ (ie $\YY=Y_{\psi}$). 

\textbf{1)} Suppose, that the time $\psi$ is internal. Then there
exists a chronometric process $\hh:\T\mapsto2^{\BsM}$ for the time
$\psi$. 

1.a) First we are going to prove, that the time $\psi$ is monotone.
Let $x_{1}\in\psi\left(t_{1}\right)$, $x_{2}\in\psi\left(t_{2}\right)$,
$x_{2}\fff x_{1}$ and $x_{1}\nff x_{2}$. Then $\psi\left(t_{2}\right)\mff\psi\left(t_{1}\right)$,
ie $\psi\left(t_{2}\right)\mtf\limits ^{\YY}\psi\left(t_{1}\right)$.
Hence, since the simultaneity $\YY$ is unrepeatable, using the assertion
\ref{As:NorepSysProp}, we obtain $\psi\left(t_{1}\right)\neq\psi\left(t_{2}\right)$,
ie $t_{1}\neq t_{2}$. Thus, the possible cases are $t_{1}<t_{2}$
or $t_{2}<t_{1}$. Let us suppose, that $t_{2}<t_{1}$. Then, since
$\hh$ is chronometric process, we have, $\hh\left(t_{1}\right)\mtf\limits ^{\hh(\T)}\hh\left(t_{2}\right)$.
From definition \ref{Def:ChronoProc} it follows, that $\hh(\T)\sqsubseteq\YY$,
consequently, using the assertion \ref{As:SubsetTrMFollow}, we obtain
$\psi\left(t_{1}\right)\mtf\limits ^{\YY}\psi\left(t_{2}\right)$,
which is impossible, because the simultaneity $\YY$ is unrepeatable,
and, according to the above proved, $\psi\left(t_{2}\right)\mtf\limits ^{\YY}\psi\left(t_{1}\right)$.
So only possible it remains the inequality $t_{1}<t_{2}$, which proves,
that the time $\psi$ is monotone. 

1.b) Now, we are going to prove, that the time $\psi$ is incessant.
Assume the contrary. Then there exist the time points $t_{1},t_{2}\in\T$
such, that $t_{1}<t_{2}$, and for any $t\in\T$, satisfying $t_{1}\leq t\leq t_{2}$,
the equality $\psi\left(t\right)=\psi\left(t_{1}\right)$ is true.
Then, in particular, $\psi\left(t_{2}\right)=\psi\left(t_{1}\right)$.
But, since $\hh$ is chronometric process, then $\hh\left(t_{2}\right)\mtf\limits ^{\hh(\T)}\hh\left(t_{1}\right)$,
and, by assertion \ref{As:SubsetTrMFollow}, $\psi\left(t_{2}\right)\mtf\limits ^{\YY}\psi\left(t_{1}\right)$.
Therefore, by assertion \ref{As:NorepSysProp}, $\psi\left(t_{2}\right)\neq\psi\left(t_{1}\right)$,
which contradicts the above written. Thus, the time $\psi$ is incessant.
And, taking into account what has been proved in paragraph 1.a), we
have, that time $\psi$ is strictly monotone. 

\textbf{2)} Now we suppose, that the time $\psi$ is strictly monotone.
Then, by lemma \ref{Lem:ForSingleMNCrono2}, the strictly linear ordered
sets $\left(\T,>\right)$ and $\left(\YY,\mtf\limits ^{\YY}\right)=\left(\YY,\mtf\limits ^{Y_{\psi}}\right)$
are isomorphic relative the order, and the mapping $\psi:\T\mapsto\YY$
is the order isomorphism between them. That is why, for ant $t_{1},t_{2}\in\T$
the conditions $t_{1}<t_{2}$ and $\psi\left(t_{2}\right)\mtf\limits ^{Y_{\psi}}\psi\left(t_{1}\right)$
are logically equivalent (where $Y_{\psi}=\YY=\psi\left(\T\right)$).
Thus, taking into account statement 1 of the assertion \ref{As:NorepSysProp},
we conclude, that the mapping $\hh(t)=\psi(t)$, $t\in\T$ is a chronometric
process for the time $\psi$. Consequently, the time $\psi$ is internal.
~ ~ ~ ~ \EndProof 

The next theorem follows from the lemma \ref{Lem:InternalTime1} and
theorem \ref{Thm:SingleMNCrono1}. 

\BeginThm \label{Thm:SingleInternalChrono1}

For any precisely-unrepeatable and monotone-connected simultaneity
$\YY$ an unique up to equivalence of chronologizations internal time
$\psi$ exists, such, that $\YY=Y_{\psi}$. 

\EndThm 

Philosophical content of the theorem \ref{Thm:SingleInternalChrono1}
is that the originality of pictures of reality, possibility to see
any changes in the sequential simultaneous states, and connectivity
of different pictures of reality by chains of changes are uniquely
generating the course of {}``internal'' time in {}``our'' world. 

\BeginRmk

Further we will denote primitive changeable sets by large calligraphic
letters. 

Let $\cP=(\MM,\TT,\phi)$ be a primitive changeable set, where $\TT=(\T,\trianglelefteq)$
is a linear ordered set. We introduce the following denotations: 

\begin{gather*}
\BsP:=\BsM;\qquad\quad\;\fff\limits _{\cP}:=\ffm;\\
\TmP:=\T;\quad\leq_{\cP}:=\trianglelefteq;\quad\psi_{\cP}:=\phi.\end{gather*}
Also we will use the records $\geq_{\cP}$,$<_{\cP}$,$>_{\cP}$ to
denote the inverse, strict and strict inverse orders, generated by
the order $\leq_{\cP}$. The set $\BsP$ we will name a \emph{basic}
set or a set of all \emph{elementary states} of the primitive changeable
set $\cP$ and we will denote it by $\BsM$. Elements of the set $\BsP$
will be named the \emph{elementary states }of $\cP$. The relation
$\fff\limits _{\cP}$ we will name a \emph{directing relation of changes}\textbf{
}of $\cP$. The set $\TmP$ will be named the \emph{set of time points}
of $\cP$. The relations $\leq_{\cP}$,$<_{\cP}$,$\geq_{\cP}$,$>_{\cP}$
will be referred to as the relations of non-strict, strict, non-strict
inverse and strict inverse time order on $\cP$. 

In the case, where the primitive changeable set $\cP$, in question
is clear, in the notations $\fff\limits _{\cP}$, $\leq_{\cP}$, $<_{\cP}$,
$\geq_{\cP}$, $>_{\cP}$, $\psi_{\cP}$ the symbol $\cP$ will be
omitted, and the notations $\fff$, $\leq$, $<$, $\geq$, $>$,
$\psi$ will be used instead. 

\EndRmk

\section{Systems of Abstract Trajectories and Primitive Changeable Sets, Generated
by them\label{sec:Atp(R)} }

\BeginDef \label{Def:SysAbstrTraekt}

Let $M$ be an arbitrary set and $\TT=\left(\T,\leq\right)$ is any
linear ordered set. 

\begin{enumerate}
\item Any mapping $r:\Dm(r)\mapsto M$, where $\Dm(r)\subseteq\T$, will
be referred to as an \textbf{abstract trajectory} from $\TT$ to $M$
(here $\Dm(r)$ is the domain of the abstract trajectory $r$). 
\item Any set $\cR$, which consists of abstract trajectories from $\TT$
to $M$ and satisfies: \[
\bigcup_{r\in\cR}\Rg\left(r\right)=M\]
 will be named \textbf{system of abstract trajectories }from $\TT$
to $M$ (here $\Rg(r)$ is the range of the abstract trajectory $r$). 
\end{enumerate}
\EndDef 

\BeginThm \label{Thm:AtpDefMotivation}

Let $\cR$ be a system of abstract trajectories from $\TT=\left(\T,\leq\right)$
to $M$. Then there exists a unique primitive changeable set $\cP$,
which satisfies the following conditions: 

\begin{description}
\item [{1)}] $\Bs(\cP)=M;\quad\Tm(\cP)=\T,\quad\leq_{\cP}=\leq$.   
\item [{2)}] For any $x,y\in\BsP$ the condition $y\fff x$ is satisfied
if and only if there exist an abstract trajectory $r=r_{x,y}\in\cR$
and elements $t,\tau\in\Dm(r)\subseteq\T$ such, that $x=r(t)$, $y=r(\tau)$
and $t\leq\tau$. 
\item [{3)}] For arbitrary $x\in\BsP$ and $t\in\TmP$ the condition $x\in\psi(t)$
is satisfied if and only if there exist an abstract trajectory $r=r_{x}\in\cR$
such, that $t\in\Dm(r)$ and $x=r(t)$.
\end{description}
\EndThm 

\BeginProof 

Let $\cR$ be any system of abstract trajectories from $\TT=\left(\T,\leq\right)$
to $M$. Define the following relation: 

\[
\xxx\limits _{\cR}=\left\{ (y,x)\in M\times M\,|\:\exists r\in\cR\,\exists t,\tau\in\Dm(r):\: x=r(t),\: y=r(\tau),\: t\leq\tau\right\} \]
 on the set $M$ (where the symbol $\times$ denotes the Cartesian
product of sets). Or, in other words, for $x,y\in M$ the correlation
$y\xxx\limits _{\cR}x$ is true if and only if there exist an abstract
trajectory $r=r_{x,y}\in\cR$ and elements $t,\tau\in\Dm(r)$ such,
that $x=r(t)$, $y=r(\tau)$ and $t\leq\tau$. Also we define the
following mapping $\varphi_{\cR}:\T\mapsto2^{M}$:

\[
\varphi_{\cR}(t)=\bigcup_{r\in\cR,\, t\in\Dm(r)}\left\{ r(t)\right\} =\left\{ r(t)\,|\: r\in\cR,\: t\in\Dm(r)\right\} .\]
 In particular, $\varphi_{\cR}(t)=\emptyset$ in the case, where there
not exist a trajectory $r\in\cR$ such, that $t\in\Dm(r)$. 

It is not hard to verify, that the pair $\MM=\left(M,\xxx\limits _{\cR}\right)$
is an oriented set and the mapping $\varphi_{\cR}$ is a time on $\MM$.
Therefore, the triple: 

\[
\cP=\left(\MM,\TT,\varphi_{\cR}\right)=\left(\left(M,\xxx\limits _{\cR}\right),\left(\T,\leq\right),\varphi_{\cR}\right)\]
is a primitive changeable set. And it is not hard to see, that this
primitive changeable set satisfies the conditions 1),2),3) of this
theorem. 

Inversely, if a primitive changeable set $\cP_{1}$ satisfies the
conditions 1),2),3) of this theorem, then from the first condition
it follows, that $\Bs\left(\cP_{1}\right)=M$, $\Tm\left(\cP_{1}\right)=\T$,
$\leq_{\cP_{1}}=\leq$. And the second and third conditions imply
the equalities $\fff\limits _{\cP_{1}}=\xxx\limits _{\cR}$, $\psi_{\cP_{1}}=\varphi_{\cR}$.
Thus, 

\[
\cP_{1}=\left(\left(\Bs\left(\cP_{1}\right),\fff\limits _{\cP_{1}}\right),\left(\Tm\left(\cP_{1}\right),\leq_{\cP_{1}}\right),\psi_{\cP_{1}}\right)=\left(\left(M,\xxx\limits _{\cR}\right),\left(\T,\leq\right),\varphi_{\cR}\right)=\cP.\]

\EndProof 

\BeginDef \label{Def:AtpR}

Let $\cR$ be any system of abstract trajectories from $\TT=\left(\T,\leq\right)$
to $M$. The primitive changeable set $\cP$, which satisfies the
conditions 1),2),3) of the theorem \ref{Thm:AtpDefMotivation}, will
be named a primitive changeable set, \textbf{generated by the system
of abstract trajectories} $\cR$, and it will be denoted by $\Atp(\TT,\cR)$:

\[
\Atp(\TT,\cR):=\cP.\]

\EndDef

Thus, systems of abstract trajectories provide the simple tool for
creation of primitive changeable sets.

\section{Elementary-time States and Basic Changeable Sets}

\subsection{Elementary-time States of Primitive Changeable Sets and their Properties
\label{sub:EChS!}}

\BeginDef \label{Def:EChS}

Let $\cP$ be a primitive changeable set. Any pair $(t,x)$ ($x\in\BsP$,
$t\in\TmP$) such, that $x\in\psi(t)$, will be named an \textbf{elementary-time
state}. 

The set of all elementary-time states of $\cP$ will be denoted by
$\BSP$: 

\[
\BSP:=\left\{ \omega\;|\:\omega=\left(t,x\right),\:\textrm{where}\; t\in\TmP,\; x\in\psi(t)\right\} .\]

\EndDef  

For any elementary-time state $\omega=(t,x)\in\BSP$ we introduce
the following denotations: 

\vspace{-1mm}
\[
\bs{\omega}:=x,\quad\tm{\omega}:=t.\]

\vspace{-2mm}

\BeginDef \label{Def:EChS_Synh,fff}  

We will say, that an elementary-time state $\omega_{2}\in\BSP$ is
\textbf{formally sequential} to an elementary-time state $\omega_{1}\in\BSP$
if and only if $\omega_{1}=\omega_{2}$ or $\bs{\omega_{2}}\fff\limits _{\cP}\bs{\omega_{1}}$
and $\tm{\omega_{1}}<_{\cP}\tm{\omega_{2}}$. For this case we will
use the denotation: 

\[
\omega_{2}\ffff\limits _{\cP}\omega_{1}.\]

\EndDef  

In the case, where the primitive changeable set $\cP$, in question
is known, in the denotation $\omega_{2}\ffff\limits _{\cP}\omega_{1}$
the symbol $\cP$ will be omitted, and the notation $\omega_{2}\ffff\omega_{1}$
will be used instead. 

\BeginAs \label{As:EChsProperties1}

1) If $\omega_{1},\omega_{2}\in\BSP$ and $\omega_{2}\ffff\omega_{1}$,
then $\tm{\omega_{1}}\leq\tm{\omega_{2}}$. If, in addition, $\omega_{1}\neq\omega_{2}$,
then $\tm{\omega_{1}}<\tm{\omega_{2}}$. 

2) The relation $\ffff=\ffff\limits _{\cP}$ is asymmetric on the
set $\BSP$, that is if $\omega_{1},\omega_{2}\in\BSP$, $\omega_{2}\ffff\omega_{1}$
and $\omega_{1}\ffff\omega_{2}$, then $\omega_{1}=\omega_{2}$. 

\EndAs 

\BeginProof 

The first statement follows by a trivial way from the definition \ref{Def:EChS_Synh,fff},
and the second statement derives from the first. ~ ~ ~ \EndProof 

\BeginDef 

The oriented set $\MM$ is named\textbf{ anti-cyclical} if for any
$x,y\in\BsM$ the conditions $x\fff y$ and $y\fff x$ involve the
equality $x=y$. 

\EndDef 

\BeginAs \label{As:EChsProperties2}

Let $\cP$ be a primitive changeable set. Then:

1) The pair $\cQ=\left(\BSP,\ffff\limits _{\cP}\right)=\left(\BSP,\ffff\right)$
is an anti-cyclical oriented set. 

2) The mapping: 

\begin{equation}
\psx(t)=\psx_{\cP}(t):=\left\{ \omega\in\BSP\:|\:\tm{\omega}=t\right\} \in2^{\BSP},\quad t\in\TmP\label{eq:TimebbBSPdef}\end{equation}
is a monotone time on $\cQ$.

3) For $t_{1}\neq t_{2}$ we have $\psx\left(t_{1}\right)\cap\psx\left(t_{2}\right)=\emptyset$. 

4) If, in addition, $\psi(t)\neq\emptyset$, $t\in\TmP$, then the
time $\psi$ is strictly monotone. 

\EndAs 

\BeginProof 

1) The first statement of the assertion \ref{As:EChsProperties2}
follows from the definition \ref{Def:EChS_Synh,fff} and second statement
of the assertion \ref{As:EChsProperties1}.

2) 2.1) Let $\omega\in\BSP$. Then, by (\ref{eq:TimebbBSPdef}), $\omega\in\psx(t)$,
where $t=\tm{\omega}$. 

2.2) Let $\omega_{1},\omega_{2}\in\BSP$, $\omega_{2}\ffff\omega_{1}$
and $\omega_{1}\neq\omega_{2}$. According to (\ref{eq:TimebbBSPdef}),
for $t_{1}=\tm{\omega_{1}}$, $t_{2}=\tm{\omega_{2}}$ we obtain: 

\[
\omega_{1}\in\psx\left(t_{1}\right),\quad\omega_{2}\in\psx\left(t_{2}\right).\]
Since $\omega_{2}\ffff\omega_{1}$ and $\omega_{2}\neq\omega_{1}$,
then, by assertion \ref{As:EChsProperties1} (statement 1), $t_{1}<t_{2}$. 

From 2.1),2.2) it follows, that $\psx$ is a time on $\cQ$. 

2.3) Let $\omega_{1}\in\psx\left(t_{1}\right)$, $\omega_{2}\in\psx\left(t_{2}\right)$,
$\omega_{2}\ffff\omega_{1}$ and $\omega_{1}\nfff\omega_{2}$. Then,
by definition of time $\psx$ (\ref{eq:TimebbBSPdef}), $\tm{\omega_{1}}=t_{1}$,
$\tm{\omega_{2}}=t_{2}$. Therefore, by assertion \ref{As:EChsProperties1},
statement 1, $t_{1}<t_{2}$. Thus, the time $\psx$ is monotone. 

3) Let $t_{1},t_{2}\in\TmP$. Suppose, that $\psx\left(t_{1}\right)\cap\psx\left(t_{2}\right)\neq\emptyset$.
Then there exists an elementary-time state $\omega\in\psx\left(t_{1}\right)\cap\psx\left(t_{2}\right)$.
Hence, by (\ref{eq:TimebbBSPdef}), we obtain $t_{1}=\tm{\omega}=t_{2}$. 

4) Assume, that, in addition, $\psi(t)\neq\emptyset,t\in\TmP$. Then
for an arbitrary $t\in\TmP$ there exists an elementary state $x_{t}\in\BsP$
such, that $x_{t}\in\psi(t)$. Consequently, the elementary-time state
$\omega_{t}=\left(t,x_{t}\right)\in\BSP$ satisfies the condition
$\tm{\omega_{t}}=t$, that is $\omega_{t}\in\psx(t)$. Thus, $\psx(t)\neq\emptyset$,
$t\in\TmP$. Hence, taking into account the statement 3) of this assertion,
we obtain, $\psx\left(t_{1}\right)\neq\psx\left(t_{2}\right)$ for
$t_{1},t_{2}\in\TmP$, $t_{1}\neq t_{2}$. Consequently, the time
$\psx$ is incessant, and, taking into account the statement 2) of
this assertion, we conclude, that the time $\psx$ is strictly monotone.
~ ~ ~ ~ \EndProof

\subsection{Base of Elementary Processes and Basic Changeable Sets }

As it had been proved in the assertion \ref{As:EChsProperties2},
for any primitive changeable set $\cP$ the pair $\left(\BSP,\ffff\right)$
is an oriented set, in which $\ffff$ is the directing relation of
changes. But, it turns out, that in the reality, the relation $\ffff$
may generate such {}``transformations'' of elementary-time states,
which never took place in the real physical system. To illustrate
this fact, we consider the following example.

\BeginEx \label{Ex:ParasiteBSPtransforms}

Let us consider the system of abstract trajectories, which describes
the uniform linear motion of the system of identical material points,
evenly distributed on the straight trajectory of their own motion.
The identity of the material points assumes, that all characteristics
of these points in a some time moment can be reduced to only their
coordinates. This means, that a material point, which has a certain
coordinates at a some time moment is completely mathematically identical
to the one point that have the same coordinates in another time. This
system of material points can be described by the following system
of abstract trajectories from $\R$ to $\R$: 

\begin{gather*}
\cR=\left\{ r_{\alpha}\,|\:\alpha\in\R\right\} ,\quad\textrm{where}\\
r_{\alpha}(t):=t+\alpha,\quad t\in\R,\enskip\alpha\in\R\qquad(\Dm\left(r_{\alpha}\right)=\R,\;\alpha\in\R).\end{gather*}
 Denote: \[
\cP:=\Atp(\left(\R,\leq\right),\cR),\]
where {}``$\leq$'' is the standard linear order relation on the
real numbers. By the definition \ref{Def:AtpR} and condition 1) of
the theorem \ref{Thm:AtpDefMotivation}, $\BsP=\TmP=\R.$ We are aim
to prove, that for the elements $x_{1},x_{2}\in\BsP=\R$ the condition
$x_{2}\fff x_{1}$ is equivalent to the inequality $x_{1}\leq x_{2}$.
Indeed, in the case $x_{1}\leq x_{2}$ for $t_{1}=x_{1}$, $t_{2}=x_{2}$
we obtain $x_{1}=r_{0}\left(t_{1}\right)$, $x_{2}=r_{0}\left(t_{2}\right)$,
where $t_{1}\leq t_{2}$. Therefore, by the condition 2) of the theorem
\ref{Thm:AtpDefMotivation}, we obtain $x_{2}\fff x_{1}$. Inversely,
if $x_{2}\fff x_{1}$, then, by condition 2) of the theorem \ref{Thm:AtpDefMotivation},
there exist numbers $\alpha,t_{1},t_{2}\in\R$ such, that $t_{1}\leq t_{2}$,
$x_{1}=r_{\alpha}\left(t_{1}\right)$, $x_{2}=r_{\alpha}\left(t_{2}\right)$,
that is $x_{1}=t_{1}+\alpha$, $x_{2}=t_{2}+\alpha$, where $t_{1}\leq t_{2}$.
Hence, $x_{1}\leq x_{2}$. 

The next aim is to prove, that $\BSP=\R\times\R$. Since $\BsP=\TmP=\R$,
we have $\BSP\subseteq\R\times\R$. Thus, it remains to prove, the
inverse inclusion. Let $\omega=(\tau,x)\in\R\times\R$. Denote $\alpha_{\omega}:=x-\tau$.
Then $r_{\alpha_{\omega}}(\tau)=\tau+(x-\tau)=x$. Therefore, by condition
3) of the theorem \ref{Thm:AtpDefMotivation}, $x\in\psi_{\cP}(\tau)$.
This means, that $\omega=(\tau,x)\in\BSP$. The equality $\BSP=\R\times\R$
has been proved. 

By definition \ref{Def:EChS_Synh,fff} of formally sequential elementary-time
states, for $\omega_{1}=\left(t_{1},x_{1}\right)$, $\omega_{2}=\left(t_{2},x_{2}\right)\in\BSP$
the condition $\omega_{2}\ffff\omega_{1}$ is performed if and only
if $\omega_{1}=\omega_{2}$ or $t_{1}<t_{2}$ and $x_{1}\leq x_{2}$.
Hence, if we choose any elementary-time states $\omega_{1}=\left(t_{1},x_{1}\right)$,
$\omega_{2}=\left(t_{2},x_{2}\right)\in\BSP=\R\times\R$, satisfying
$t_{1}<t_{2}$ and $x_{1}\leq x_{2}$, we obtain $\omega_{2}\ffff\omega_{1}$.
But in the case $x_{1}-t_{1}\neq x_{2}-t_{2}$ there not exist an
abstract trajectory $r_{\alpha}\in\cR$ such, that $\omega_{1},\omega_{2}\in r_{\alpha}$.
This means, that in this model of real physical process, the elementary-time
state $\omega_{2}$ may not be the result of transformations of the
elementary-time state $\omega_{1}$. Thus, in this example, the relation
$\ffff$ generates infinitely many {}``parasitic transformation relations'',
which never took place in the reality. 

\EndEx 

The above example shows, that to adequate describe real physical process,
the directing relation of changes should be defined not only on the
set of set of elementary states $\BsP$, but, also, on the set of
elementary-time states $\BSP$ of a primitive changeable set $\cP$.
Indeed, let us consider the primitive changeable set $\cP:=\Atp(\cR)$
from the example \ref{Ex:ParasiteBSPtransforms}. For $\omega_{1},\omega_{2}\in\BSP$
we can put $\omega_{2}\xxx\omega_{1}$ if and only if $\tm{\omega_{1}}\leq\tm{\omega_{2}}$
and there exist an abstract trajectory $r_{\alpha}\in\cR$ such, that
$\omega_{1},\omega_{2}\in r_{\alpha}$ (that is such, that $\bs{\omega_{1}}=r_{\alpha}\left(\tm{\omega_{1}}\right)$,
$\bs{\omega_{2}}=r_{\alpha}\left(\tm{\omega_{2}}\right)$). Thus,
we obtain the relation {}``$\xxx$'', which reflects only such transformations
of the elementary-time states, which actually take place in the reality. 

\BeginDef \label{Def:BazovaMM}

Let $\cP$ be a primitive changeable set. 

\textbf{1.} Relation $\xxx$ on $\BSP$ is named \textbf{base of elementary
processes} if and only if: 

\begin{description}
\item [{(1)}] $\forall\,\omega\in\BSP$ $\omega\xxx\omega$. 
\item [{(2)}] If $\omega_{1},\omega_{2}\in\BSP$ and $\omega_{2}\xxx\omega_{1}$,
then $\omega_{2}\ffff\omega_{1}$ (ie $\xxx\subseteq\ffff$). 
\item [{(3)}] For arbitrary $x_{1},x_{2}\in\BsP$ such, that $x_{2}\fff x_{1}$
there exist $\omega_{1},\omega_{2}\in\BSP$ such, that $\bs{\omega_{1}}=x_{1}$,
$\bs{\omega_{2}}=x_{2}$ and $\omega_{2}\xxx\omega_{1}$. 
\end{description}
\textbf{2.} In the case, where $\xxx$ is the base of elementary processes
on the primitive changeable set $\cP$, the pair: \[
\BBB=\left(\cP,\xxx\right)\]
 will be referred to as \textbf{basic changeable set}. 

\EndDef

\subsection{Remarks on the Denotations \label{sec:RemarkBMM}}

For further basic changeable sets will be denoted by large calligraphy
symbols. 

Let $\BBB=\left(\cP,\xxx\right)$ be a basic changeable set. We introduce
the following denotations: 

\[
\begin{array}{lll}
\BsB:=\BsP; & \BSB:=\BSP; & \fff\limits _{\BBB}:=\fff\limits _{\cP};\\
\ffff\limits _{\BBB}:=\ffff\limits _{\cP}; & \TmB:=\TmP; & \leq_{\BBB}:=\leq_{\cP};\\
<_{\BBB}:=<_{\cP}; & \geq_{\BBB}:=\geq_{\cP}; & >_{\BBB}:=>_{\cP};\\
\psi_{\BBB}:=\psi_{\cP}.\end{array}\]
 Also for elementary-time states $\omega_{1},\omega_{2}\in\BSB$
we will use the denotation $\omega_{2}\fff\limits _{\BBB}\omega_{1}$
instead of the denotation $\omega_{2}\xxx\omega_{1}$. 

In the case, where the basic changeable set $\cP$, is clear in the
denotations $\fff\limits _{\BBB}$, $\ffff\limits _{\BBB}$, $\leq_{\BBB}$,
$<_{\BBB}$, $\geq_{\BBB}$, $>_{\BBB}$, $\psi_{\BBB}$ the symbol
$\BBB$ will be omitted, and the denotations $\fff$, $\ffff$, $\leq$,
$<$, $\geq$, $>$, $\psi$ will be used instead. 

The next properties of basic changeable sets follow from the definitions
\ref{Def:BazovaMM} and \ref{Def:EChS_Synh,fff} (in the properties
1-5 the symbol $\BBB$ means a basic changeable set):

\BeginProp  ~ \label{Prop:BMM}

\begin{enumerate}
\item The pair $\BBB_{0}=\left(\BsB,\fff\right)$ is an oriented set. 
\item The mapping $\psi=\psi_{\BBB}$ is a time on $\BBB_{0}=\left(\BsB,\fff\right)$.
\item $\omega\fff\omega$ for any elementary-time state $\omega\in\BSB$. 
\item If $\omega_{1},\omega_{2}\in\BSB$ and $\omega_{2}\fff\omega_{1}$,
then $\omega_{2}\ffff\omega_{1}$,  \label{enu:BMMProp(fff,ffff)}
and therefore $\bs{\omega_{2}}\fff\bs{\omega_{1}}$ and $\tm{\omega_{1}}\leq\tm{\omega_{2}}$.
\label{enu:BMMProp(fff,bs,tm)} 
\item For arbitrary $x_{1},x_{2}\in\BsB$ the condition $x_{2}\fff x_{1}$
holds if and only if there exist elementary-time states $\omega_{1},\omega_{2}\in\BSB$
such, that $\bs{\omega_{1}}=x_{1}$, $\bs{\omega_{2}}=x_{2}$ and
$\omega_{2}\fff\omega_{1}$. \label{enu:BMMProp(fff,Bs)}
\item $\BsB=\left\{ \bs{\omega}\,|\,\omega\in\BSB\right\} $. \label{enu:BMMProp(Bs)}
\end{enumerate}
\EndProp

\subsection{Examples of Basic Changeable Sets }

\BeginEx 

Let $\cP$ be any primitive changeable set. Then the relation $\ffff=\ffff\limits _{\cP}$
is base of elementary processes on $\cP$. Indeed, the conditions
(1) and (2) of the definition \ref{Def:BazovaMM} for the relation
$\ffff$ are fulfilled by a trivial way. To verify the condition (3)
we consider arbitrary $x_{1},x_{2}\in\BsP$ such, that $x_{2}\fff x_{1}$.
In the case $x_{1}=x_{2}$ by the time definition \ref{Def:ChronoMain},
there exist a time point $t_{1}\in\TmP$ such, that $x_{1}\in\psi\left(t_{1}\right)$.
Hence, for $\omega_{1}=\omega_{2}=\left(t_{1},x_{1}\right)\in\BSP$
we obtain $\bs{\omega_{1}}=\bs{\omega_{2}}=x_{1}=x_{2}$ and $\omega_{2}\ffff\omega_{1}$.
Thus, in the case $x_{1}=x_{2}$ the condition (3) of the definition
\ref{Def:BazovaMM} is satisfied. In the case $x_{1}\neq x_{2}$,
by definition \ref{Def:ChronoMain}, there exist time points $t_{1},t_{2}\in\TmP$
such, that $x_{1}\in\psi\left(t_{1}\right)$, $x_{2}\in\psi\left(t_{2}\right)$
and $t_{1}<t_{2}$. Hence, for $\omega_{1}=\left(t_{1},x_{1}\right),\:\omega_{2}=\left(t_{1},x_{2}\right)\in\BSP$,
we obtain $\bs{\omega_{1}}=x_{1}$, $\bs{\omega_{2}}=x_{2}$ and $\omega_{2}\ffff\omega_{1}$.
Thus, in the case $x_{1}\neq x_{2}$ the condition (3) of the definition
\ref{Def:BazovaMM} also is satisfied.

Therefore any primitive changeable set can be interpreted as basic
changeable set $\cP_{(f)}=(\cP,\ffff)$ in with the relation $\ffff$
is the base of elementary processes. 

\EndEx 

\BeginEx \label{Ex:ATbmm}

Let $\cR$ be any system of abstract trajectories from $\TT=\left(\T,\leq\right)$
to $M$. Denote:

\[
\cP:=\Atp(\TT,\cR).\]
By theorem \ref{Thm:AtpDefMotivation}, $\BsP=M$, $\TmP=\T$. Moreover,
by third statement of the theorem for $\left(t,x\right)\in M\times\T$
the condition $\left(t,x\right)\in\BSP$ holds if and only if there
exist an abstract trajectory $r=r_{t,x}\in\cR$ such, that $t\in\Dm(r)$
and $x=r(t)$, ie such, that $\omega=\left(t,x\right)\in r$. Thus,

\begin{equation}
\BSP=\bigcup_{r\in\cR}r.\label{eq:BSAtpRepresentation}\end{equation}
Then, for $\omega_{1},\omega_{2}\in\BSP$ we put $\omega_{2}\xxx[\cR]\,\omega_{1}$
if and only if $\tm{\omega_{1}}\leq\tm{\omega_{2}}$ and there exists
an abstract trajectory $r\in\cR$ such, that $\omega_{1},\omega_{2}\in\cR$
(ie such, that $\bs{\omega_{1}}=r\left(\tm{\omega_{1}}\right)$, $\bs{\omega_{2}}=r\left(\tm{\omega_{2}}\right)$).
We are going to prove, that the relation $\xxx[\cR]\,$ provides base
of elementary processes. 

(a) Let $\omega\in\BSP$. Then, by (\ref{eq:BSAtpRepresentation}),
there exist an abstract trajectory $r\in\cR$ such, that $\omega\in r$.
Hence, by definition of the relation {}``$\xxx[\cR]$'', we have
$\omega\xxx[\cR]\omega$. 

(b) Let $\omega_{1}=\left(t_{1},x_{1}\right)$,~$\omega_{2}=\left(t_{2},x_{2}\right)\in\BSP$
and  $\omega_{2}\xxx[\cR]\omega_{1}$. Then, from definition of the
relation {}``$\xxx[\cR]$'', it follows, that $t_{1}\leq t_{2}$
and there exists an an abstract trajectory $r\in\cR$ such, that $\omega_{1},\omega_{2}\in\cR$
(ie such, that $x_{1}=r\left(t_{1}\right)$, $x_{2}=r\left(t_{2}\right)$).
Consequently, by statement 2) of the theorem \ref{Thm:AtpDefMotivation},
$x_{2}\fff\limits _{\Atp\left(\cR\right)}x_{1}$. Therefore, in the
case $t_{1}\neq t_{2}$ we have $t_{1}<t_{2}$ and $x_{2}\fff x_{1}$,
besides in the case $t_{1}=t_{2}$ we obtain $x_{1}=r\left(t_{1}\right)=r\left(t_{2}\right)=x_{2}$,
that is $\omega_{1}=\omega_{2}$. But, in the both cases the correlation
$\omega_{2}\ffff\omega_{1}$ is true. 

(c) Let $x_{1},x_{2}\in\BsP$, $x_{2}\fff x_{1}$ (ie $x_{2}\fff\limits _{\Atp\left(\cR\right)}x_{1}$).
Then, by statement 2) of the theorem \ref{Thm:AtpDefMotivation},
there exists an abstract trajectory $r\in\cR$ such, that $x_{1}=r\left(t_{1}\right)$,
$x_{2}=r\left(t_{2}\right)$ for some $t_{1},t_{2}\in\TmP$ such,
that $t_{1}\leq t_{2}$. Denote: 

\[
\omega_{i}:=\left(t_{i},x_{i}\right),\qquad i\in\left\{ 1,2\right\} .\]
Then, $\omega_{1},\omega_{2}\in r\subseteq\bigcup_{\rho\in\cR}\rho=\BSP$,
$\bs{\omega_{i}}=x_{i}$ ($i\in\left\{ 1,2\right\} $) and, by definition
of the relation {}``$\xxx[\cR]$'', $\omega_{2}\xxx[\cR]\omega_{1}$. 

From the items (a)-(c) it follows, that the relation $\xxx[\cR]\,$
is base of elementary processes on $\cP=\Atp(\TT,\cR)$. Thus, the
pair: 

\[
\At(\TT,\cR)=\left(\cP,\xxx[\cR]\,\right)=\left(\Atp(\TT,\cR),\xxx[\cR]\,\right)\]
 is a basic changeable set. 

 From the properties \ref{Prop:BMM}(\ref{enu:BMMProp(fff,Bs)},\ref{enu:BMMProp(Bs)})
it follows, that if for a some basic changeable set $\BBB$ we know
$\TmB$, $\leq_{\BBB}$, $\BSB$ and base of elementary processes
$\fff\limits _{\BBB}$ on $\BSB$, then we can we can recover the
set $\BsB$, the directing relation of changes $\fff\limits _{\BBB}$
on $\BsB$ and the time $\psi_{\BBB}(t)$ (using the formula $\psi_{\BBB}(t)=\left\{ x\in\BsB\,|\,\left(t,x\right)\in\BSB\right\} $,
$t\in\TmB$), and thus, we can recover the whole basic changable set
$\BBB$. Hence from the last example it follows the next theorem. 

\EndEx 

\BeginThm \label{Thm:AtDefProperties}

Let $\cR$ be a system of abstract trajectories from $\TT=\left(\T,\leq\right)$
to $M$. Then there exists a unique basic changeable set $\BBB=\At(\TT,\cR)$,
such, that: 

\begin{description}
\item [{1)}] $\left(\Tm(\At(\TT,\cR)),\leq_{\At\left(\TT,\cR\right)}\right)=\TT$; 
\item [{2)}] $\BS(\At(\TT,\cR))=\bigcup_{r\in\cR}r$; 
\item [{3)}] For arbitrary $\omega_{1},\omega_{2}\in\BS(\At(\TT,\cR))$
the condition $\omega_{2}\fff\limits _{\At(\TT,\cR)}\omega_{1}$ is
satisfied if and only if $\tm{\omega_{1}}\leq\tm{\omega_{2}}$ and
there exist an abstract trajectory $r\in\cR$ such, that $\omega_{1},\omega_{2}\in r$.
\end{description}
\EndThm 

\BeginRmk \label{Rmk:At(fff,time)Properties}

1. Since the construction of the basic changeable set $\At(\TT,\cR)$
is based on the primitive changeable set $\Atp\left(\TT,\cR\right)$,
for any basic changeable set of kind $\BBB=\At(\TT,\cR)$ the statements,
formulated in the items 2),3) of the theorem \ref{Thm:AtpDefMotivation}
remain true (with replacement the character $\cP$ by $\BBB$ or by
$\At(\TT,\cR)$). 

2. In the case, when the linear ordered set $\TT$ is given in advance,
we will use the denotation $\At(\cR)$ instead of $\At(\TT,\cR)$. 

\EndRmk

\section{Chains in the Set of Elementary-time States. Fate Lines and their
Properties}

Using the definition of basic changeable sets as well as the assertions
\ref{As:EChsProperties2} and \ref{As:EChsProperties1} (item 2) we
obtain the following assertion. 

\BeginAs \label{As:EChsProperties3}

Let $\BBB$ be a basic changeable set. Then: 

1) The pair $\cQ=\left(\BSB,\fff\limits _{\BBB}\right)=\left(\BSB,\fff\right)$
is an anti-cyclical oriented set. 

2) The mapping \begin{equation}
\psx(t)=\psx_{\BBB}(t):=\left\{ \omega\in\BSB\:|\:\tm{\omega}=t\right\} \in2^{\BSB},\quad t\in\TmB\label{eq:TimebbBSPdef1}\end{equation}
 is a monotone time on $\cQ$.

3) If, in addition, $\psi(t)\neq\emptyset$, $t\in\TmP$, then the
time $\psi$ is strictly monotone. 

\EndAs 

According to the assertion \ref{As:EChsProperties3}, for any basic
changeable set $\BBB$ the pair $\left(\BSB,\fff\right)$ is (anti-cyclical)
oriented set. As in an arbitrary oriented set, in $\left(\BSB,\fff\right)$
we may introduce transitive sets and chains. From anti-cyclicity of
the oriented set $\left(\BSB,\fff\right)$ it follows the following
assertion. 

\BeginAs \label{As:ChainEChsOrder}

Let $\BBB$ be a basic changeable set. 

1) Any transitive subset $\mathcal{N}\subseteq\BSB$ of the oriented
set $\left(\BSB,\fff\right)$ is a (partially) ordered set (relatively
the relation $\fff$). 

2) Any chain $\cL\subseteq\BSB$ of the oriented set $\left(\BSB,\fff\right)$
is a linearly ordered set (relatively the relation $\fff$). 

\EndAs 

\BeginDef \label{Def:LinDoli}

Let $\BBB$ be a basic changeable set. 

1) Any maximum chain $\cL\subseteq\BSB$ of the oriented set $\left(\BSB,\fff\right)$
will be named a \textbf{fate line} of $\BBB$. The set of all fate
lines of $\BBB$ will be denoted by $\LdB$: 

\[
\LdB=\left\{ \cL\subseteq\BSB\,|\:\cL\:\textrm{is a fate line of}\;\BBB\right\} .\]

2) Any fate line, which contains an elementary-time state $\omega\in\BSB$
will be named the (\textbf{eigen}) fate line of elementary-time state
$\omega$ (in $\BBB$). 

3) A fate line $\cL\in\LdB$ will be named the (\textbf{eigen}) fate
line of the elementary state $x\in\BsB$ if and only if there exists
the elementary-time state $\omega_{x}\in\BSB$ such, that $\bs{\omega_{x}}=x$
and $\cL$ is eigen fate line of $\omega_{x}$. 

\EndDef  

It is clear that, in the general case, an elementary (elementary-time)
state may have many fate lines. 

We will say, that elementary (elementary-time) states $x_{1},x_{2}\in\BsB$,
($\omega_{1},\omega_{2}\in\BSB$) are \textbf{\emph{united by fate}}
if and only if there exist at least one fate line $\cL\in\LdB$, which
is eigen fate line of both states $x_{1},x_{2}$ ($\omega_{1},\omega_{2}$). 

\BeginAs \label{As:LdProperties1}

1) Any elementary-time state $\omega\in\BSB$ must have at least one
eigen fate line. 

2) For elementary-time states $\omega_{1},\omega_{2}\in\BSB$ to be
united by fate it is necessary and sufficient satisfaction one of
the following conditions:

\begin{equation}
\omega_{2}\fff\omega_{1}\quad\textrm{or}\quad\omega_{1}\fff\omega_{2}.\label{eq:CommonLDcondition1}\end{equation}

\EndAs 

\BeginProof 

1) The first statement of this assertion follows from the corollary
\ref{Nasl:MaxChainExist}. 

2) 2.a) Suppose, that for the elementary-time states $\omega_{1},\omega_{2}\in\BSB$
there exist a common fate line $\cL\in\LdB$ ($\omega_{1},\omega_{2}\in\cL$).
Then, by assertion \ref{As:ChainEChsOrder}, item 2, the pair $(\cL,\fff)$
is a linearly ordered set. Thus at least one of the conditions (\ref{eq:CommonLDcondition1})
must be fulfilled. 

2.b) Let, $\omega_{1},\omega_{2}\in\BSB$ and $\omega_{2}\fff\omega_{1}$.
Then, by corollary \ref{Nasl:MaxChainExist}, there exist a maximum
chain (fate line) $\cL\subseteq\BSB$ ($\cL\in\LdB$) such, that $\omega_{1},\omega_{2}\in\cL$.
~ ~ ~ \EndProof 

\BeginAs \label{As:LdProperties2}

1) Any elementary state $x\in\BsB$ must have at least one eigen fate
line. 

2) For elementary states $x,y\in\BsB$ to be united by fate it is
necessary and sufficient satisfaction one of the following conditions:

\begin{equation}
y\fff x\quad\textrm{or}\quad x\fff y.\label{eq:CommonLDcondition2}\end{equation}

\EndAs 

\BeginProof 

1) Let $x\in\BsB$. Then, by the definition of time, there exist a
time point $t\in\TmB$ such, that $x\in\psi(t)$. By assertion \ref{As:LdProperties1},
the elementary-time state $\omega_{x}=(t,x)\in\BSB$ must have an
eigen fate line $\cL\in\LdB$. This fate line $\cL$ must be eigen
fate line of elementary state $x$. 

2) 2.a) Let $x,y\in\BsB$, $y\fff x$. Then, by the property \ref{Prop:BMM}(\ref{enu:BMMProp(fff,Bs)})
(see properties \ref{Prop:BMM}), there exist elementary-time states
$\omega_{1}$,$\omega_{2}\in\BSP$ such, that $\bs{\omega_{1}}=x$,
$\bs{\omega_{2}}=y$ and $\omega_{2}\fff\omega_{1}$. By assertion
\ref{As:LdProperties1}, there exist a common fate line $\cL\in\LdB$
for the elementary-time states $\omega_{1}$,$\omega_{2}$ (ie $\omega_{1},\omega_{2}\in\cL$).
By definition \ref{Def:LinDoli}, this fate line $\cL$ must be eigen
fate line of both elementary states $x$ and $y$. 

2.b) Suppose, that for the elementary states $x,y\in\BsB$ there exist
a common eigen fate line $\cL\in\LdB$. Then, there exist elementary-time
states $\omega_{1},\omega_{2}\in\BSB$, such, that $\bs{\omega_{1}}=x$,
$\bs{\omega_{2}}=y$ and $\omega_{1},\omega_{2}\in\cL$. Hence, by
assertion \ref{As:LdProperties1}, statement 2), one of the conditions
$\omega_{2}\fff\omega_{1}$ or $\omega_{1}\fff\omega_{2}$ must be
satisfied. Then, by the property \ref{Prop:BMM}(\ref{enu:BMMProp(fff,bs,tm)}),
at least one of the conditions (\ref{eq:CommonLDcondition2}) must
be fulfilled. ~ ~ ~ ~ \EndProof 

As it was shown in the theorem \ref{Thm:AtDefProperties}, any system
of abstract trajectories, defined on some linearly ordered set $\TT=\left(\T,\leq\right)$,
generates the basic changeable set $\At(\TT,\cR)$. The next aim is
to show, that any basic changeable set $\BBB$ can be represented
in the form $\BBB=\At(\TT,\cR)$, where $\cR$ is some system of abstract
trajectories, defined on some linearly ordered set $\TT$.

\BeginDef 

Let $\cR$ be a system of abstract trajectories from $\TT=\left(\T,\leq\right)$
to $M$.

\begin{enumerate}
\item Trajectory $r\in\cR$ will be named a \textbf{maximum trajectory}
(relatively the $\cR$) if and only if there not exist any trajectory
$\rho\in\cR$ ($\rho\neq r$) such, that $\Dm\left(r\right)\subset\Dm\left(\rho\right)$
and $r(t)=\rho(t)$ $t\in\Dm\left(r\right)$ (that is such, that $r\subset\rho$). 
\item The system of abstract trajectories $\cR$ will be referred to as
the \textbf{system of maximum trajectories} if and only if any trajectory
$r\in\cR$ is maximum trajectory (relatively the $\cR$). 
\end{enumerate}
\EndDef 

Further, for any basic changeable set $\BBB$ we will use the denotation:\[
\TMB:=\left(\TmB,\leq_{\BBB}\right).\]

\BeginAs \label{As:SysAbstrTraekt1}

Let $\BBB$ be a basic changeable set. Then: 

1) Any chain $\cL\subseteq\BSB$ of the oriented set $\left(\BSB,\fff\right)$
is an abstract trajectory from $\TMB$ to $\BsB$. 

2) The set $\LlB$ of all chains of the oriented set $\left(\BSB,\fff\right)$
is a system of abstract trajectories from $\TMB$ to $\BsB$. 

3) Any fate line $\cL\in\LdB\subseteq\LlB$ is a maximum trajectory
(relatively the system of abstract trajectories $\LlB$). 

4) The set $\LdB$ is a system of maximum trajectories (from $\TMB$
to $\BsB$). 

\EndAs 

\BeginProof 

1) Let $\cL\subseteq\BSB$ be a chain of the oriented set $\left(\BSB,\fff\right)$.
Since $\BSB\subseteq\TmB\times\BsB$ and $\cL\subseteq\BSB$, then
$\cL$ is a binary relation from the set $\TmB$ to the set $\BsB$.
Thus, to make sure that $\cL$ is an abstract trajectory from $\TMB$
to $\BsB$, it is sufficient to prove, that this relation $\cL$ is
a function from $\TmB$ to $\BsB$. Suppose contrary. Then there exist
elementary-time states $\omega_{1},\omega_{2}\in\cL$ of kind $\omega_{1}=\left(t,x_{1}\right)$,
$\omega_{2}=\left(t,x_{2}\right)$, where $x_{1}\neq x_{2}$. Since
$\cL$ is a chain, one of the conditions $\omega_{2}\fff\omega_{1}$
or $\omega_{1}\fff\omega_{2}$ must be satisfied. Assume, that $\omega_{2}\fff\omega_{1}$.
Then, by the property \ref{Prop:BMM}(\ref{enu:BMMProp(fff,ffff)})
(see properties \ref{Prop:BMM}), $\omega_{2}\ffff\omega_{1}$. Hence,
taking into account, that $\omega_{2}\neq\omega_{1}$, by assertion
\ref{As:EChsProperties1} (item 1), we obtain $t<t$, which is impossible.
Similarly the assumption $\omega_{1}\fff\omega_{2}$ also leads to
contradiction. The obtained contradiction proves that the chain $\cL$
is a function. 

Taking into account, that, according to the proved above, any chain
$\cL$ of the oriented set $\left(\BSB,\fff\right)$ is an abstract
trajectory, we may use the notations $\Dm(\cL)$ for the domain of
$\cL$ and $x=\cL(t)$ (where $t\in\Dm(\cL)$) to indicate the fact
that $(t,x)\in\cL$. 

2) Let $\LlB$ be the set of all chains of the oriented set $\left(\BSB,\fff\right)$.
Chose any elementary state $x\in\BsB$. By the time definition, there
exist a time point $t\in\TmB$ such, that $x\in\psi(t)$. By assertion
\ref{As:2elem-transitive-chain}, item 2, the singleton set $\cL_{x}=\left\{ (t,x)\right\} \subseteq\BSB$
is a chain of the oriented set $\left(\BSB,\fff\right)$. Besides,
$\Rg\left(\cL_{x}\right)=\{x\}\ni x$. Thus, any elementary state
$x\in\BsB$ is contained in the range of some abstract trajectory
$\cL_{x}\in\LlB$. Therefore, $\bigcup_{\cL\in\LlB}\Rg(\cL)=\BsB$.
Thus, taking into account the statement 1) of this assertion we conclude,
that $\LlB$ is the system of abstract trajectories from $\TMB$ to
$\BsB$. 

3) Let $\cL\in\LlB$ be a fate line of $\BBB$ (ie $\cL$ is a maximum
chain of the oriented set $\left(\BSB,\fff\right)$). Then, there
not exist any chain (abstract trajectory) $\cL_{1}\in\LlB$ such,
that $\cL\subset\cL_{1}$. Hence, $\cL$ is a maximum trajectory (relatively
the system of abstract trajectories $\LlB$). 

4) Now, we are going to prove, that $\bigcup_{\cL\in\LdB}\Rg(\cL)=\BsB$.
Since $\bigcup_{\cL\in\LdB}\cL\subseteq\BSB\subseteq\TmB\times\BsB$,
we have $\bigcup_{\cL\in\LdB}\Rg(\cL)\subseteq\BsB$. Thus, it remains
to prove the inverse inclusion. Chose any elementary state $x\in\BsB$.
By assertion \ref{As:LdProperties2} (item 1), the elementary state
$x$ must have an eigen fate line $\cL_{x}\in\LdB$. This (by definition
\ref{Def:LinDoli}) means, that there exist an elementary-time state
$\omega_{x}=\left(t,x\right)\in\BSB$ such, that $\omega_{x}\in\cL_{x}$.
Since $\left(t,x\right)\in\cL_{x}$, then $\cL_{x}(t)=x$. Therefore,
$x\in\Rg\left(\cL_{x}\right)\subseteq\bigcup_{\cL\in\LdB}\Rg(\cL)$.
Thus, $\bigcup_{\cL\in\LdB}\Rg(\cL)=\BsB$. Hence, $\LdB$ is a system
of abstract trajectories from $\TMB$ to $\BsB$. Since (by item 3
of this assertion) any fate line $\cL\in\LdB\subseteq\LlB$ is a maximum
trajectory relatively the system of abstract trajectories $\LlB$,
it is the maximum trajectory relatively the narrower system of abstract
trajectories $\LdB$. ~ ~ ~ \EndProof 

The next theorem shows, that any basic changeable set can be generated
by some system of maximum trajectories.

\BeginThm \label{Thm:AtLinDoli}

For any basic changeable set $\BBB$ the following equality is true: 

\[
\At\left(\TMB,\,\LdB\right)=\BBB.\]

\EndThm 

\BeginProof 

Denote: $\cR:=\LdB$. We need to prove, that $\At(\cR)=\BBB$. 

1) By the assertion \ref{As:SysAbstrTraekt1}, ~ $\cR=\LdB$ is the
system of abstract trajectories from $\TMB=\left(\TmB,\leq_{\BBB}\right)$
to $\BsB$. Hence, by the first item of the theorem \ref{Thm:AtDefProperties},
\[
\Tm\left(\At(\cR)\right)=\TmB,\quad\leq_{\At(\cR)}=\leq_{\BBB}.\]

2) By the second item of the theorem \ref{Thm:AtDefProperties}: \begin{gather}
\BS\left(\At(\cR)\right)=\bigcup_{r\in\cR}r=\bigcup_{\cL\in\LdB}\cL\:\subseteq\:\BSB.\label{eq:BSAtR_BSB_embed}\end{gather}
On the other hand, by the assertion \ref{As:LdProperties1}, for any
$\omega\in\BSB$ the fate line $\cL_{\omega}\subseteq\BSB$ exists
such, that $\omega\in\cL_{\omega}$. Threfore, $\BSB\subseteq\bigcup_{\cL\in\LdB}\cL=\BS\left(\At(\cR)\right)$.
And, taking into account (\ref{eq:BSAtR_BSB_embed}) we obtain: \[
\BS\left(\At(\cR)\right)=\BSB.\]

3) Let us consider any elementary-time states $\omega_{1}=\left(t_{1},x_{1}\right)$,
$\omega_{2}=\left(t_{2},x_{2}\right)\in\BSB=\BS\left(\At(\cR)\right)$. 

3.a) Suppose, that $\omega_{2}\fff\limits _{\BBB}\omega_{1}$. By
the property \ref{Prop:BMM}(\ref{enu:BMMProp(fff,bs,tm)}) (see properties
\ref{Prop:BMM}), $\tm{\omega_{1}}\leq\tm{\omega_{2}}$. Moreover,
by the assertion \ref{As:LdProperties1} (item 2) the fate line $\LL\in\LdB$
exists such, that $\omega_{1},\omega_{2}\in\LL$. Thus, by the theorem
\ref{Thm:AtDefProperties} (item 3), $\omega_{2}\fff\limits _{\At(\LdB)}\omega_{1}$,
that is $\omega_{2}\fff\limits _{\At(\cR)}\omega_{1}$. 

3.b) Conversely, suppose, that $\omega_{2}\fff\limits _{\At(\cR)}\omega_{1}$,
scilicet $\omega_{2}\fff\limits _{\At(\LdB)}\omega_{1}$. Then, by
the theorem \ref{Thm:AtDefProperties} (item 3), $\tm{\omega_{1}}\leq\tm{\omega_{2}}$
and there exists the fate line $\LL\in\LdB$ exists such, that $\omega_{1},\omega_{2}\in\LL$.
Since the fate line $\LL$ is a chain, at least one from the correlations~
$\omega_{2}\fff\limits _{\BBB}\omega_{1}$ or $\omega_{1}\fff\limits _{\BBB}\omega_{2}$~
must be true. We shall prove, that $\omega_{2}\fff\limits _{\BBB}\omega_{1}$.
Assume the contrary ($\omega_{2}\nff\limits _{\BBB}\omega_{1}$).
Then, we have $\omega_{1}\fff\limits _{\BBB}\omega_{2}$ and $\omega_{2}\neq\omega_{1}$
(because in the case $\omega_{1}=\omega_{2}$ we have $\omega_{2}\fff\limits _{\BBB}\omega_{1}$).
Hence, by the property \ref{Prop:BMM}(\ref{enu:BMMProp(fff,ffff)})
$\omega_{1}\ffff\limits _{\BBB}\omega_{2}$. Since $\omega_{1}\ffff\limits _{\BBB}\omega_{2}$
and $\omega_{2}\neq\omega_{1}$, by the definition \ref{Def:EChS_Synh,fff}
we obtain, $\tm{\omega_{2}}<\tm{\omega_{1}}$. The last inequality
is impossible, because we have proved, that $\tm{\omega_{1}}\leq\tm{\omega_{2}}$.
Therefore, $\omega_{2}\fff\limits _{\BBB}\omega_{1}$. 

From the items 3.a) and 3.b) it follows, that $\fff\limits _{\BBB}=\fff\limits _{\At(\cR)}$
(for the bases of elementary processes on $\BS\left(\At(\cR)\right)=\BSB$). 

Thus, the basic changeable set $\BBB$ satisfies the conditions 1)-3)
of the theorem \ref{Thm:AtDefProperties} (for the system of abstract
trajectories $\cR=\LdB$), and, by this theorem, $\At(\cR)=\BBB$.
~ ~ ~ \EndProof

\section{Multi-figurativeness and Unification of Perception. General Definition
of Changeable Set }

\subsection{Changeable Systems and Processes }

\BeginDef 

Let $\BBB$ be a basic changeable set. Any subset $S\subseteq\BSB$
we will name a \textbf{changeable system} of the basic changeable
set $\BBB$. 

\EndDef 

In the mechanics the elementary states can be interpreted as the states
or positions of material point in various moments of time. That is
why, the concept of changeable system may be considered as the abstract
generalization of the notion of physical body, which, in the general
case, has not constant composition. 

\BeginDef \label{Def:Process}

Let $\BBB$ be a basic changeable set. Any mapping $s:\TmB:\mapsto2^{\BsB}$
such, that $s(t)\subseteq\psi(t)$, $t\in\TmB$ will be referred to
as a \textbf{process} of the basic changeable set $\BBB$. 

\EndDef 

Since primitive changeable sets can be interpreted as basic changeable
set with the base of elementary processes $\ffff$, the chronometric
processes, introduced in the definition \ref{Def:ChronoProc} can
be considered as the particular cases of the processes, introduced
in the definition \ref{Def:Process}. 

Let $S\subseteq\BSB$ be an arbitrary changeable system of any basic
changeable set $\BBB$. Denote: 

\begin{equation}
\Sxv(t):=\left\{ x\in\BsB\,|\:(t,x)\in S\right\} ,\quad t\in\TmB.\label{eq:SxvDef}\end{equation}
It is easy to see, that $\Sxv(t)\subseteq\psi(t)$, $t\in\TmB$. Thus,
by definition \ref{Def:Process}, $\Sxv$ is a process on the basic
changeable set $\BBB$. 

\BeginDef  

The process $\Sxv$ will be named the \textbf{process of transformations}
of the changeable system $S$. 

\EndDef 

\BeginAs \label{As:SysProcEquval}

Let $\BBB$ be a basic changeable set.

1. For any changeable systems $S_{1},S_{2}\in\BSB$ the equality $\Sxv_{1}=\Sxv_{2}$
holds if and only if $S_{1}=S_{2}$. 

2. For an arbitrary process~ $s$~ of the basic changeable set $\BBB$
a unique changeable system $S\subseteq\BSB$ exists such, that $s=\Sxv$.

\EndAs 

\BeginProof 

1. To prove the first statement, it is enough to verify that for any
$S_{1},S_{2}\in\BSB$ the equality $\Sxv_{1}=\Sxv_{2}$ implies the
equality $S_{1}=S_{2}$. Hence, suppose, that $\Sxv_{1}=\Sxv_{2}$.
Then for any $t\in\TmB$ we have $\Sxv_{1}(t)=\Sxv_{2}(t)$. Therefore,
by (\ref{eq:SxvDef}), for an arbitrary $t\in\TmB$ the condition
$(t,x)\in S_{1}$ is equivalent to the condition $(t,x)\in S_{2}$.
But, this means, that $S_{1}=S_{2}$. 

2. Let $s$ be a process of a basic changeable set $\BBB$. Denote: 

\[
S:=\left\{ (t,x)\,|\: t\in\TmB,\; x\in s(t)\right\} =\bigcup_{t\in\TmB}(\{t\}\times s(t)),\]
where the symbol $\times$ denotes Cartesian product of sets. Since
for any pair $(t,x)\in S$ it is true $x\in s(t)\subseteq\psi(t)$,
we have $S\subseteq\BSB$. Therefore, $S$ is a changeable system
of $\BBB$. Moreover, for any $t\in\TmB$ we obtain:

\begin{gather*}
\Sxv(t)=\left\{ x\in\BsB\,|\:(t,x)\in S\right\} =\left\{ x\in\BsB\,|\: x\in s(t)\right\} =s(t).\end{gather*}
 Consequently, $\Sxv=s$. Suppose, an other changeable system $S_{1}$
exists such, that $\Sxv_{1}=s$. Then, $\Sxv=\Sxv_{1}$, and, by the
statement 1, $S=S_{1}$. Thus, changeable system $S$, satisfying
$\Sxv=s$ is unique. ~ ~ ~ \EndProof 

Therefore, the mapping $(\cdot)^{\sim}$ provides one-to-one correspondence
between changeable systems and processes of any basic changeable set.
Taking into account this fact, further we will {}``identify'' changeable
systems and processes of any basic changeable set, and for denotation
of processes of a basic changeable set we will use letters with tilde,
keeping in mind, that any process is the process of transformations
of some changeable system. 

\emph{We say, that a changeable system $U\subseteq\BSB$ in a basic
changeable set $\BBB$ is a }\textbf{\emph{subsystem}}\emph{ of a
changeable system $S\subseteq\BSB$ if and only if $U\subseteq S$}.
The following assertion is true: 

\BeginAs \label{As:SubsystVarSyst}

Changeable system $U\subseteq\BsB$ is a subsystem of a changeable
system $S\subseteq\BSB$ if and only if: 

\[
\forall\, t\in\TmB\quad\Uxv(t)\subseteq\Sxv(t).\]

\EndAs 

\BeginProof  

1. Let $S,U\subseteq\BSB$ and $U\subseteq S$. Then, by (\ref{eq:SxvDef}),
for any $t\in\TmB$ we obtain:  \begin{gather*}
\Uxv(t)=\left\{ x\in\BsB\,|\:(t,x)\in U\right\} \subseteq\left\{ x\in\BsB\,|\:(t,x)\in S\right\} =\Sxv(t).\end{gather*}

2. Conversely, suppose, that $\Uxv(t)\subseteq\Sxv(t)$ for any $t\in\TmB$.
Denote: 

\[
S_{1}:=\bigcup_{t\in\TmB}\{t\}\times\Sxv(t);\quad U_{1}(t):=\bigcup_{t\in\TmB}\{t\}\times\Uxv(t).\]
As it had been shown in the proof of statement 2 of the assertion
\ref{As:SysProcEquval}, $\Sxv_{1}=\Sxv$, $\Uxv_{1}=\Uxv$. Therefore,
by the first item of the assertion \ref{As:SysProcEquval}, $S_{1}=S$,
$U_{1}=U$. Thus: \begin{gather*}
U=U_{1}=\bigcup_{t\in\TmB}\{t\}\times\Uxv(t)\subseteq\bigcup_{t\in\TmB}\{t\}\times\Sxv(t)=S_{1}=S.\end{gather*}
\EndProof 

\BeginDef  

We say, that the elementary state $x\in\BsB$ of a basic changeable
set $\BBB$ \textbf{belongs} to a changeable system $S\subseteq\BSB$
in a time point $t\in\TmB$ if and only if $x\in\Sxv(t)$. 

\EndDef 

The fact, that elementary state of a basic changeable set $\BBB$
belongs to a changeable system $S$ in a time point $t$, will be
denoted by: 

\[
x\inn{t,\BBB}S,\]
 and in the case, when the basic changeable set is clear, we will
use the denotation: \[
x\inn{t}S.\]
 By the assertion \ref{As:SubsystVarSyst}, for any changeable systems
$U,S\subseteq\BSB$ the correlation $U\subseteq S$ holds if and only
if for any $t\in\TmB$ and $x\in\BsB$ the condition $x\inn{t}U$
assures $x\inn{t}S$. 

The last remark indicates that a changeable system of any basic changeable
set can be interpreted as analog of the subset notion in the classic
set theory, and the relation $\inn{\cdot}$ can be interpreted as
analog of the belonging relation of the classic set theory. However,
the elementary-time state is not the complete analogue of the notion
of element in the classic set theory, because knowing all the elementary-time
states of a basic changeable set, we can not fully recover this basic
changeable set. 

It is evident, that any fate line $\cL\in\LdB$ of a basic changeable
set $\BBB$ is the changeable system of $\BBB$. 

\BeginDef  

The process $\cL^{\sim}$, generated by a fate line $\cL\in\LdB$
of a basic changeable set $\BBB$ we name the \textbf{elementary process}
of $\BBB$. 

\EndDef 

The concept of elementary process can be considered as the complete
analogue of the notion of element in the classic set theory, because
knowing all the elementary process of a basic changeable set, we can
fully recover this basic changeable set, using the theorem \ref{Thm:AtLinDoli}.

\subsection{General Definition of Changeable Set }

Basic changeable set can be treated as mathematical abstraction of
physical processes models (in macro level) in the case, when the observations
are conducted from one, fixed point (one, fixed frame of reference).
But, real, physical nature is multi-figurative, because in physics
(in particular in special relative theory) {}``picture of the world''
can significantly vary, according to the frame of reference. Therefore,
we obtain not one but many basic changeable sets (connected with everyone
frame of reference, of the physical model under consideration). Any
of these basic changeable sets can be interpreted as individual image
(or area of perception) of the physical reality. Also it can be naturally
assumed, that there is a natural unification between any two areas
of perception, this means, that it must be defined some rule, which
specifies how the object or process from one area of perception will
be looked out in other area. More precisely, we equate, using certain
rules, some object or process from one area of perception with the
other object or process from other area of perception, saying that
it is the same object, but visible from another area of perception.
In the classical mechanics such {}``unification of perception''
is defined by the Galilean group of transformations, and in the special
relative theory this unification is determined by the group of the
Lorentz-Poincare. It should be noted that in the both cases the unification
of perception is made not at the level of objects and processes, but
at the level of elementary-time states (points 4-dimensional space-time).
This means that in the both cases there is assumed, that any elementary-time
state, {}``visible'' from some area of perception is {}``visible''
from other areas. On author opinion, this assumption is too strong
to construct our abstract theory. That is why, in the definition below
the unification of perception is made on the level of objects and
processes. We recall, that in the previous subsection it had been
introduced the concept of changeable system (subset of the set $\BSB$,
generated by basic changeable set $\BBB$) as an abstract analog of
the notion of physical object or process. 

\BeginDef \label{Def:GlobalUni,ChSets}

Let $\vcB=\left(\BBB_{\alpha}\,|\:\alpha\in\AAA\right)$ be an indexed
family of basic changeable sets (where $\AAA$ is the some set of
indexes). The system of mappings $\vfU=\left(\fU_{\beta\alpha}\,|\:\alpha,\beta\in\AAA\right)$
of kind: 

\[
\fU_{\beta\alpha}:2^{\BSBB{\alpha}}\longmapsto2^{\BSBB{\beta}}\qquad(\alpha,\beta\in\AAA)\]
 is referred to as \textbf{unification of perception} on $\vcB$ if
and only if the following conditions are satisfied: 

\begin{enumerate}
\item $\fU_{\alpha\alpha}A\equiv A$ for any $\alpha\in\AAA$ and $A\subseteq\BSBB{\alpha}$.
\\
{\em(Here and further we denote by $\fU_{\beta\alpha}A$ the action
of the mapping $\fU_{\beta\alpha}$ to the set $A\subseteq\BSBB{\alpha}$,
that is $\fU_{\beta\alpha}A:=\fU_{\beta\alpha}(A)$.)}
\item Any mapping $\fU_{\beta\alpha}$ is a monotonous mapping of sets,
ie for any $\alpha,\beta\in\AAA$ and $A,B\subseteq\BSBB{\alpha}$
the condition $A\subseteq B$ assures $\fU_{\beta\alpha}A\subseteq\fU_{\beta\alpha}B$. 
\item For any $\alpha,\beta,\gamma\in\AAA$ and $A\subseteq\BSBB{\alpha}$
the following inclusion holds:  \begin{equation}
\fU_{\gamma\beta}\fU_{\beta\alpha}A\subseteq\fU_{\gamma\alpha}A.\label{eq:MainUniEmbedding}\end{equation}

\end{enumerate}
In this case the mappings $\fU_{\alpha\beta}$ ($\alpha,\beta\in\AAA$)
we name \textbf{unification mappings}, and the triple of kind: 

\[
\cZ=\left(\AAA,\vcB,\vfU\right)\]
 will be named \textbf{changeable set}. 

\EndDef 

The first condition of the definition \ref{Def:GlobalUni,ChSets}
is quite obvious. The second condition is dictated by the natural
desire {}``to see'' a subsystem of a given changeable system in
a given area of perception as the subsystem of {}``the same'' changeable
system in other area of perception. In the case of classical mechanics
or special relativity theory the third condition of the definition
\ref{Def:GlobalUni,ChSets} may be transformed to the following (stronger)
condition: 

\begin{equation}
\fU_{\gamma\beta}\fU_{\beta\alpha}A=\fU_{\gamma\alpha}A\quad(\alpha,\beta,\gamma\in\AAA,\: A\subseteq\BSBB{\alpha})\label{eq:GrpMaps1}\end{equation}
The replacement of the equal sign by the sign inclusion is caused
by the permission to {}``distort the picture of reality'' during
{}``transition'' to other area of perception in the case of the
our abstract theory. We suppose, that during this {}``transition''
some elementary-time states may turn out to be {}``invisible'' in
other area of perception. Further this idea will be explained more
detailed (see the section \ref{sec:Visibility}, in particular, theorem
\ref{Thm:FviCriterion01}).

\subsection{Remarks on the Terminology and Denotations \label{sub:CsSets:RmkTermDef} }

Let $\cZ=\left(\AAA,\vcB,\vfU\right)$ be a changeable set, where
$\vcB=\left(\BBB_{\alpha}\,|\:\alpha\in\AAA\right)$ is an indexed
family of basic changeable sets and $\vfU=\left(\fU_{\beta\alpha}\,|\:\alpha,\beta\in\AAA\right)$
is an unification of perception on $\vcB$. Later we will use the
following terms and notations: 

\medskip{}

1) The set $\AAA$ will be named the \textbf{\emph{index set}} of
the changeable set $\cZ$, and it will be denoted by $\IndZ$. 

2) For any index $\alpha\in\IndZ$ the pair $\left(\alpha,\BBB_{\alpha}\right)$
will be referred to as \textbf{\emph{area of perception}} or \textbf{\emph{frame
of reference}} or \textbf{\emph{lik}} of the changeable set $\cZ$. 

3) The set of all areas of perception $\cZ$ will be denoted by $\LkZ$: 

\[
\LkZ:=\left\{ \left(\alpha,\BBB_{\alpha}\right)\,|\:\alpha\in\IndZ\right\} .\]
 Areas of perception will typically be denoted by small Latin letters
($l,m,k,p$ and so on). 

4) For $l=\left(\alpha,\BBB_{\alpha}\right)\in\LkZ$ we introduce
the following denotations: 

\[
\ind{l}:=\alpha;\quad\oll:=\BBB_{\alpha}.\]
Thus, for any area of perception $l\in\LkZ$ the object $\ol{l}$
is a basic changeable set. 

Further, when it does not cause confusion, for any area of perception
$l\in\LkZ$ in denotations: 

\begin{gather}
\Bs\left(\oll\right),\:\BS\left(\oll\right),\:\Tm\left(\oll\right),\:\leq_{\oll},\:<_{\oll},\nonumber \\
\geq_{\oll},\:>_{\oll},\:\psi_{\oll},\:\fff\limits _{\oll},\:\Ld\left(\oll\right)\label{eq:LkDenotations01}\end{gather}
the symbol {}``$\ol{\,}$'' will be omitted, and the following denotations
will be used instead: 

\begin{gather}
\Bs\left(l\right),\:\BS\left(l\right),\:\Tm\left(l\right),\:\leq_{l},\:<_{l},\nonumber \\
\geq_{l},\:>_{l},\:\psi_{l},\:\fff\limits _{l},\:\Ld\left(l\right).\label{eq:LkDenotations02}\end{gather}

5) For any areas of perception $l,m\in\LkZ$ the mapping $\fU_{\ind{m},\ind{l}}$
will be denoted by $\un{l,\cZ}{m}$ or by $\uni{l}{m,\cZ}$. Hence: 

\[
\un{l,\cZ}{m}=\uni{l}{m,\cZ}=\fU_{\ind{m},\ind{l}}.\]
In the case, when the basic changeable $\cZ$ set is known, the symbol
$\cZ$ in the above notations will be omitted, and the denotations
{}``$\un{l}{m}$, $\uni{l}{m}$'' will be used instead. Moreover,
in the case, when it does not cause confusion in the notations {}``$\leq_{l}$,
$<_{l}$, $\geq_{l}$, $>_{l}$, $\fff\limits _{l}$'' the symbol
{}``$l$'' will be omitted, and the denotations {}``$\leq$, $<$,
$\geq$, $>$, $\fff$'' will be used instead.

\subsection{Elementary Properties of Changeable Sets \label{sub:BaseMMProperties} }

Using the definition \ref{Def:GlobalUni,ChSets} and notations, introduced
in the subsection \ref{sub:CsSets:RmkTermDef}, we can write the following
\textbf{basic properties of changeable sets}. 

\BeginProp  \label{Prop:ChSets(basic)}

In the properties 1-6 $\cZ$ is any changeable set and $l,m,p\in\Lk{\cZ}$
are any areas of perception of $\cZ$. 

\begin{enumerate}
\item $l=\left(\ind{l},\oll\right)$;
\item $\oll=\left(\left(\left(\Bs(l),\fff\limits _{l}\right),\left(\Tm(l),\leq_{l}\right),\psi_{l}\right),\fff\limits _{l}\right)$
is a basic changeable set. 
\item $\un{l}{l}A=\uni{l}{l}A=A$, $A\subseteq\BS(l)$;
\item $\un{l}{m}A=\uni{l}{m}A$, $A\subseteq\BS(l)$;
\item If $A\subseteq B\;\subseteq\BS(l)$, then $\un{l}{m}A\subseteq\un{l}{m}B$
(or, in other words, $\uni{l}{m}A\subseteq\uni{l}{m}B$);
\item $\un{m}{p}\un{l}{m}A\subseteq\un{l}{p}A$ (or, in other words $\uni{m}{p}\uni{l}{m}A\subseteq\uni{l}{p}A$),
where $A\subseteq\BS(l)$.
\end{enumerate}
\EndProp 

In the future we will not use the definition \ref{Def:GlobalUni,ChSets},
and usually will apply the properties \ref{Prop:ChSets(basic)}. The
following assertions are elementary corollaries of the properties
\ref{Prop:ChSets(basic)}. In these assertions the symbol $\cZ$ denotes
any changeable set. 

\BeginAs \label{As:Zproperties01}

For any $l,m\in\LkZ$ the following equality is true: \[
\un{l}{m}\emptyset=\emptyset.\]

\EndAs 

\BeginProof 

Denote $B:=\un{l}{m}\emptyset\subseteq\BS(m)$. By properties \ref{Prop:ChSets(basic)}~(6
and 3) we obtain: \begin{gather*}
\un{m}{l}B=\un{m}{l}\un{l}{m}\emptyset\subseteq\un{l}{l}\emptyset=\emptyset.\end{gather*}
Therefore, $\un{m}{l}B=\emptyset$. Since $\emptyset\subseteq B$,
then, by property \ref{Prop:ChSets(basic)}(5), \[
\un{m}{l}\emptyset\subseteq\un{m}{l}B=\emptyset,\]
 that is $\un{m}{l}\emptyset=\emptyset$. Hence, by properties \ref{Prop:ChSets(basic)}~(3
and 6), we obtain: \[
\emptyset=\un{m}{m}\emptyset\supseteq\un{l}{m}\un{m}{l}\emptyset=\un{l}{m}\emptyset=B.\]

\EndProof 

\BeginAs \label{As:Zproperties02}

For any $l,m\in\LkZ$ and any family of changeable systems $\left(A_{\alpha}|\alpha\in\AAA\right)$
($A_{\alpha}\subseteq\BS(l)$, $\alpha\in\AAA$) the following inclusions
take place: 

1) $\un{l}{m}\left(\bigcap\limits _{\alpha\in\AAA}A_{\alpha}\right)\subseteq\bigcap\limits _{\alpha\in\AAA}\un{l}{m}A_{\alpha}$;

2) $\bigcap\limits _{\alpha\in\AAA}A_{\alpha}\supseteq\un{m}{l}\left(\bigcap\limits _{\alpha\in\AAA}\un{l}{m}A_{\alpha}\right)$;

3) $\un{l}{m}\left(\bigcup\limits _{\alpha\in\AAA}A_{\alpha}\right)\supseteq\bigcup\limits _{\alpha\in\AAA}\un{l}{m}A_{\alpha}$.

\EndAs 

Note, that the set of indexes $\AAA$ in the last assertion is an
arbitrary, and, in general, it does not coincide with the set of indexes
in the definition \ref{Def:GlobalUni,ChSets}. 

\BeginProof 1) Denote $A:=\bigcap_{\alpha\in\AAA}A_{\alpha}$. Taking
into account, that $A\subseteq A_{\alpha}$, $\alpha\in\AAA$ and
using the property \ref{Prop:ChSets(basic)}(5) we obtain: \begin{gather*}
\un{l}{m}A\subseteq\un{l}{m}A_{\alpha},\quad\alpha\in\AAA.\end{gather*}
Thus, $\un{l}{m}A\subseteq\bigcap_{\alpha\in\AAA}\un{l}{m}A_{\alpha}$. 

2) Denote: $Q:=\bigcap\limits _{\alpha\in\AAA}\un{l}{m}A_{\alpha}$.
Then $Q\subseteq\un{l}{m}A_{\alpha}$, $\alpha\in\AAA$. Hence, by
properties \ref{Prop:ChSets(basic)}(6~and~3) we obtain: \[
\un{m}{l}Q\subseteq\un{m}{l}\un{l}{m}A_{\alpha}\subseteq\un{l}{l}A_{\alpha}=A_{\alpha},\quad\alpha\in\AAA.\]
 Hence, $\un{m}{l}Q\subseteq\bigcap_{\alpha\in\AAA}A_{\alpha}$, and
that it was necessary to prove. 

3) Denote: $A:=\bigcup_{\alpha\in\AAA}A_{\alpha}$. Taking into account,
that $A_{\alpha}\subseteq A$, $\alpha\in\AAA$ and using the property
\ref{Prop:ChSets(basic)}(5) we obtain \begin{gather*}
\un{l}{m}A_{\alpha}\subseteq\un{l}{m}A,\quad\alpha\in\AAA.\end{gather*}
Hence, $\bigcup_{\alpha\in\AAA}\un{l}{m}A_{\alpha}\subseteq\un{l}{m}A$.
~ ~ ~ ~ \EndProof

\section{Examples of Changeable Sets}

~ ~ \BeginEx  \label{Ex:Zpv}

Let $\vcB=\left(\BBB_{\alpha}\,|\:\alpha\in\AAA\right)$ be an non-empty
($\AAA\neq\emptyset$) indexed family of basic changeable sets such,
that $\BSBB{\alpha}$ and $\BSBB{\beta}$ are equipotent for any $\alpha,\beta\in\AAA$,
that is $\card\left(\BSBB{\alpha}\right)=\card\left(\BSBB{\beta}\right),\quad\alpha,\beta\in\AAA$,
where $\card(M)$ is the \emph{cardinality} of the set $M$. Let us
consider any indexed family of bijections (one-to-one correspondences)
$\left(W_{\beta\alpha}|\:\alpha,\beta\in\AAA\right)$ of kind $W_{\beta\alpha}:\BSBB{\alpha}\mapsto\BSBB{\beta}$,
satisfying the following {}``pseudo-group'' conditions: 

\begin{gather}
W_{\alpha\alpha}(\omega)=\omega,\qquad\alpha\in\AAA,\:\omega\in\BSBB{\alpha};\nonumber \\
W_{\gamma\beta}\left(W_{\beta\alpha}\omega\right)=W_{\gamma\alpha}(\omega),\quad\alpha,\beta,\gamma\in\AAA,\:\omega\in\BSBB{\alpha}.\label{eq:PseudoGrpConditions}\end{gather}

\begin{flushright}
{\normalsize%
\begin{minipage}[t][1\totalheight][c]{0.9\columnwidth}%
\BeginRmk 

Such family of bijections can be easily constructed by the following
way. 

Since $\AAA\neq\emptyset$, we can chose any (fixed) index $\alpha_{0}\in\AAA$.
Also chose any family of bijections $\overleftarrow{\mathcal{W}}=\left(\mathcal{W}_{\alpha}\,|\:\alpha\in\AAA\right)$
of kind $\mathcal{W}_{\alpha}:\BSBB{\alpha}\mapsto\BSBB{\alpha_{0}}$
(such family of bijections necessarily must exist, because of $\card\left(\BSBB{\alpha}\right)=\card\left(\BSBB{\beta}\right),\quad\alpha,\beta\in\AAA$).
Denote: 

\[
W_{\beta\alpha}\left(\omega\right):=\arc{\mathcal{W}_{\beta}}\mathcal{W}_{\alpha}\left(\omega\right),\qquad\alpha,\beta\in\AAA,\;\omega\in\BSBB{\alpha}.\]
 where $\arc{\mathcal{W}_{\beta}}$ is the inverse mapping to $\mathcal{W}_{\beta}$.
It is easy to verify, that the family of bijections $\left(W_{\beta\alpha}|\:\alpha,\beta\in\AAA\right)$
satisfies the conditions (\ref{eq:PseudoGrpConditions}). 

\EndRmk ~%
\end{minipage}%
}
\par\end{flushright}

Let us put: \[
\fU_{\beta\alpha}A:=W_{\beta\alpha}(A)=\left\{ W_{\beta\alpha}(\omega)\,|\:\omega\in A\right\} ,\quad A\subseteq\BSBB{\alpha}.\]
It is easy to see, that the family of mappings $\vfU=\left(\fU_{\beta\alpha}\,|\:\alpha,\beta\in\AAA\right)$
satisfies all conditions of the definition \ref{Def:GlobalUni,ChSets},
moreover, the third condition of this definition can be replaced by
more strong condition (\ref{eq:GrpMaps1}). Thus the triple: 

\[
\Zpv{\vcB,\vW}=\left(\AAA,\vcB,\vfU\right)\]
 is a changeable set. The changeable set $\Zpv{\vcB,\vW}$ will be
named a \textbf{\emph{precisely visible changeable set}}, generated
by the system of basic changeable sets $\vcB$ and the system of mappings
$\vW$. 

Note, that for any areas of perception $l,m\in\Zpv{\vcB,\vW}$ the
following equalities are true: 

\[
\oll=\BBB_{\ind{l}};\qquad\un{l}{m}A=W_{\ind{m},\ind{l}}(A),\quad A\subseteq\BS(l).\]

%
{} 

\EndEx 

To construct the next example we need to introduce the concept of
image of basic changeable set during the mapping of the set of it's
elementary-time states. 

Let $T,X$ --- be any sets. For any element $\omega=(t,x)\in T\times X$
we put: \begin{equation}
\tm{\omega}:=t,\quad\bs{\omega}:=x.\label{eq:bs()tm()def}\end{equation}
 Hence, for any $\omega\in T\times X$ we have, $\omega=\left(\tm{\omega},\:\bs{\omega}\right)$.
Note, that the denotations (\ref{eq:bs()tm()def}) have been used
before, but only for elementary-time states $\omega\in\BSP\subseteq\TmP\times\BsP$,
where $\cP$ is a primitive or basic changeable set (see subsection
\ref{sub:EChS!}). 

Let $\BBB$ be a basic changeable set. Consider any mapping $U:\,\TmB\times\BsB\mapsto\TmB\times X$,
where $X$ is some set. The mapping of this type will be named as
\textbf{\emph{transforming mapping}}, for the $\BBB$. 

\BeginThm \label{Thm:U[B]exist}

For an arbitrary transforming mapping $U$ for a basic changeable
set $\BBB$ there exists the unique basic changeable set $U[\BBB]$
satisfying the following conditions: 

\begin{enumerate}
\item $\BS(U[\BBB])=U(\BSB)$;
\item $\TM\left(U[\BBB]\right)=\TMB$; 
\item If $\w_{1},\w_{2}\in\BS(U[\BBB])$ and $\tm{\w_{1}}\neq\tm{\w_{2}}$,
then $\w_{1}$ and $\w_{2}$ are united by fate in $U[\BBB]$ if and
only if there exist united by fate in $\BBB$ elementary-time states
$\omega_{1},\omega_{2}\in\BSB$ such, that $\w_{1}=U\left(\omega_{1}\right)$,
$\w_{2}=U\left(\omega_{2}\right)$. 
\end{enumerate}
\EndThm 

\BeginProof 

\textbf{Proof of existence}. 

1. Let $U:\,\TmB\times\BsB\mapsto\TmB\times X$ be a transforming
mapping for the basic changeable set $\BBB$. Denote: \begin{gather*}
U[\BsB]:=\bs{U(\BSB)}=\left\{ \bs{\w}\,|\:\w\in U\left(\BSB\right)\right\} =\left\{ \bs{U(\omega)}\,|\:\omega\in\BSB\right\} .\end{gather*}

It is evident, that $U[\BsB]\subseteq X$. Let $x_{1},x_{2}\in U[\BsB]$.
We will suppose, that $x_{2}\xxx\limits _{U[\BBB]}x_{1}$, if and
only if one of the following conditions is satisfied: 

\begin{description}
\item [{(C1)}] $x_{1}=x_{2}$.
\item [{(C2)}] Elements $x_{1},x_{2}$ can be represented in the form,
$x_{1}=\bs{U\left(\omega_{1}\right)}$, $x_{2}=\bs{U\left(\omega_{2}\right)}$,
where elementary-time states $\omega_{1},\omega_{2}\in\BSB$ are united
by fate in $\BBB$ and $\tm{U\left(\omega_{1}\right)}<\tm{U\left(\omega_{2}\right)}$. 
\end{description}
By definition, $\xxx\limits _{U[\BBB]}$ is reflexive relation on
$U[\BsP]$. Hence, the pair $\left(U[\BsB],\xxx\limits _{U[\BBB]}\right)$
is an oriented set. 

2. For an arbitrary $t\in\TmB$ we put:

\begin{equation}
\widetilde{\psi}_{U[\BBB]}(t):=\left\{ \bs{U(\omega)}\,|\:\omega\in\BSB,\;\tm{U(\omega)}=t\,\right\} \label{eq:U[P]timeDef}\end{equation}
(in particular, $\widetilde{\psi}_{U[\BBB]}(t)=\emptyset$, in the
case, when there not exist any elementary-time state $\omega\in\BSB$
such, that $\tm{U(\omega)}=t$). 

We are going to prove, that the mapping $\widetilde{\psi}_{U[\BBB]}:U[\TmB]\mapsto2^{U[\BsB]}$
is a time on the oriented set $\left(U[\BsB],\xxx\limits _{U[\BBB]}\right)$. 

2.a) Let, $x\in U[\BsB]$. Then $x=\bs{U(\omega)}$ for some $\omega\in\BSB$.
Denote, $t:=\tm{U(\omega)}$. Then, by the definition of the mapping
$\widetilde{\psi}_{U[\BBB]}$, we have $x\in\widetilde{\psi}_{U[\BBB]}(t)$.
Thus, the first condition of the time definition \ref{Def:ChronoMain}
holds. 

2.b) Suppose, that $x_{1},x_{2}\in U[\BsB]$, $x_{2}\xxx\limits _{U[\cP]}x_{1}$
and $x_{1}\neq x_{2}$. Then, by the definition of the relation $\xxx\limits _{U[\cP]}$,
the elements $x_{1},x_{2}$ can be represented in the form $x_{1}=\bs{U\left(\omega_{1}\right)}$,
$x_{2}=\bs{U\left(\omega_{2}\right)}$, where $\omega_{1},\omega_{2}\in\BSB$
and $\tm{U\left(\omega_{1}\right)}<\tm{U\left(\omega_{2}\right)}$.
Denote, $t_{1}:=\tm{U\left(\omega_{1}\right)}$, $t_{2}:=\tm{U\left(\omega_{2}\right)}$.
Then $t_{1}<t_{2}$, and, by the definition of the mapping $\widetilde{\psi}_{U[\BBB]}$,
we obtain, $x_{1}\in\widetilde{\psi}_{U[\BBB]}(t_{1})$, $x_{2}\in\widetilde{\psi}_{U[\BBB]}(t_{2})$.
Hence, the second condition of the definition \ref{Def:ChronoMain}
also is satisfied. 

Thus, the triple: 

\[
U_{p}[\BBB]=\left(\left(U[\BsB],\xxx\limits _{U[\BBB]}\right),\:(\TmB,\leq),\:\widetilde{\psi}_{U[\BBB]}\right)\]
 is a primitive changeable set, satisfying the following conditions
\begin{flalign}
 & \Bs\left(U_{p}[\BBB]\right)=U[\BsB],\nonumber \\
 & \Tm\left(U_{p}[\BBB]\right)=\TmB,\;\leq_{U_{p}[\BBB]}=\leq=\leq_{\BBB},\;\psi_{U_{p}[\BBB]}=\widetilde{\psi}_{U[\BBB]}.\label{eq:U[B]components01}\end{flalign}

3. Moreover, using the definition \ref{Def:EChS}, correlation (\ref{eq:U[B]components01})
and the definition (\ref{eq:U[P]timeDef}) of the time $\widetilde{\psi}_{U[\BBB]}$
we obtain: 

\begin{multline}
\BS\left(U_{p}[\BBB]\right)=\left\{ \left(t,x\right)\,|\, t\in\Tm\left(U_{p}[\BBB]\right),\, x\in\psi_{U_{p}[\BBB]}(t)\right\} =\\
=\left\{ \left(t,x\right)\,|\, t\in\TmB,\, x\in\widetilde{\psi}_{U[\BBB]}(t)\right\} =\\
=\left\{ \left(t,x\right)\,|\, t\in\TmB,\, x=\bs{U(\omega)},\,\omega\in\BSB,\,\tm{U(\omega)}=t\,\right\} =\\
=\left\{ \left(t,x\right)\,|\, x=\bs{U(\omega)},\, t=\tm{U(\omega)},\,\omega\in\BSB\,\right\} =\\
=\left\{ \left(\tm{U(\omega)},\bs{U(\omega)}\right)\,|\,\omega\in\BSB\,\right\} =\\
=\left\{ U(\omega)\,|\,\omega\in\BSB\,\right\} =U(\BSB).\label{eq:BSUp[B]}\end{multline}

4. Let $\w_{1},\w_{2}\in\BS\left(U_{p}[\BBB]\right)=U(\BSB)$. We
will consider, that $\w_{2}\xxx\limits _{U[\BBB]}\w_{1}$, if and
only if one of the following conditions is satisfied: 

\begin{description}
\item [{(C3)}] $\w_{1}=\w_{2}$;
\item [{(C4)}] $\tm{\w_{1}}<\tm{\w_{2}}$, while $\w_{1}=U\left(\omega_{1}\right)$,
$\w_{2}=U\left(\omega_{2}\right)$, where where elementary-time states
$\omega_{1},\omega_{2}$ are united by fate in $\BBB$.
\end{description}
We are aim to prove, that $\xxx\limits _{U[\BBB]}$ is a base of elementary
processes on the primitive changeable set $U_{p}[\BBB]$. 

4.a) By the condition (C3) for any $\w\in\BS\left(U_{p}[\BBB]\right)$
we have, $\w\xxx\limits _{U[\BBB]}\w$. Consequently, the first condition
of the definition \ref{Def:BazovaMM} is satisfied. 

4.b) Suppose, that $\w_{1}=\left(t_{1},x_{1}\right),\:\w_{2}=\left(t_{2},x_{2}\right)\in\BS\left(U_{p}[\BBB]\right)$
and $\w_{2}\xxx\limits _{U[\BBB]}\w_{1}$. In the case $\w_{1}=\w_{2}$,
by definition \ref{Def:EChS_Synh,fff}, we obtain $\w_{2}\:\ffff\limits _{U_{p}[\BBB]}\,\w_{1}$.
Therefore, it remains to consider the case $\w_{1}\neq\w_{2}$. Since
$\w_{2}\xxx\limits _{U[\BBB]}\w_{1}$ and $\w_{1}\neq\w_{2}$, then,
by the definition of the relation $\xxx\limits _{U[\BBB]}$, we have
$t_{1}<t_{2}$, and $\w_{1}=U\left(\omega_{1}\right)$, $\w_{2}=U\left(\omega_{2}\right)$,
where elementary-time states $\omega_{1},\omega_{2}$ are united by
fate in $\BBB$. Hence, by the definition of the relation $\xxx\limits _{U[\BBB]}$
on $\Bs\left(U_{p}[\BBB]\right)=U[\BsB]$, we obtain $x_{2}\,\xxx\limits _{U[\BBB]}\, x_{1}$.
Therefore, $x_{2}\,\fff\limits _{U_{p}[\BBB]}\, x_{1}$ and $t_{1}<t_{2}$.
Consequently, $\w_{2}\:\ffff\limits _{U_{p}[\BBB]}\,\w_{1}$. Thus,
the second condition of the definition \ref{Def:BazovaMM} also holds. 

4.c) Let, $x_{1},x_{2}\in\Bs\left(U_{p}[\BBB]\right)=U[\BsB]$ and
$x_{2}\fff\limits _{U_{p}[\BBB]}x_{1}$ (that is $x_{2}\xxx\limits _{U[\BBB]}x_{1}$).
By definition of the relation $\xxx\limits _{U[\BBB]}$, the latter
can occur only when at least one of the conditions (C1), (C2) is satisfied. 

\emph{Case (C1)}: $x_{1}=x_{2}$. Since $x_{1}\in U[\BsB]=\bs{U(\BSB)}$,
there exists elementary-time state $\omega\in\BSB$ such, that $x_{1}=\bs{U(\omega)})$.
Denote, $\w=U(\omega)$. Then $x_{1}=x_{2}=\bs{\w}$, where, by item
4.a), $\w\xxx\limits _{U[\BBB]}\w$. Thus, in this case, the third
condition of the definition \ref{Def:BazovaMM} holds. 

\emph{Case (C2)}: $x_{1}=\bs{U\left(\omega_{1}\right)}$, $x_{2}=\bs{U\left(\omega_{2}\right)}$,
where elementary-time states $\omega_{1},\omega_{2}\in\BSB$ are united
by fate in $\BBB$ and $\tm{U\left(\omega_{1}\right)}<\tm{U\left(\omega_{2}\right)}$.
Denote: \begin{gather*}
\w_{1}:=U\left(\omega_{1}\right),\quad\w_{2}:=U\left(\omega_{2}\right).\end{gather*}
 Then, $x_{1}=\bs{\w_{1}}$, $x_{2}=\bs{\w_{2}}$, where, by definition
of the relation $\xxx\limits _{U[\BBB]}$ on $\BS\left(U_{p}[\BBB]\right)$
(condition (C4)), $\w_{2}\xxx\limits _{U[\BBB]}\w_{1}$. 

Thus, the relation $\xxx\limits _{U[\BBB]}$, defined on $\BS\left(U_{p}[\BBB]\right)$,
is the base of elementary processes. Hence, the pair \[
U[\BBB]=\left(U_{p}[\BBB],\,\xxx\limits _{U[\BBB]}\right)\]
 is a basic changeable set. 

5. By the equality (\ref{eq:BSUp[B]}), $\BS\left(U[\BBB]\right)=\BS\left(U_{p}[\BBB]\right)=U(\BSB)$.
Also from (\ref{eq:U[B]components01}) it follows, that $\TM\left(U[\BBB]\right)=\left(\Tm(U[\BBB]),\leq_{U[\BBB]}\right)=\left(\TmB,\leq_{\BBB}\right)=\TMB$.
Thus, the basic changeable set $U[\BBB]$ satisfies the first two
conditions of the Theorem. 

We are aim to prove, that the third condition of the Theorem for the
basic changeable set $U[\BBB]$ also is satisfied. Consider any elementary-time
states $\w_{1},\w_{2}\in\BS(U[\BBB])$ such, that $\tm{\w_{1}}\neq\tm{\w_{2}}$. 

5.a) Suppose, that $\w_{1},\w_{2}$ are united by fate in $U[\BBB]$.
Since $\tm{\w_{1}}\neq\tm{\w_{2}}$, we have 

 $\w_{1}\neq\w_{2}$. Since $\w_{1},\w_{2}$ are united by fate in
$U[\BBB]$, by the assertion \ref{As:LdProperties1}, at least one
of the conditions $\w_{2}\fff\limits _{U[\BBB]}\w_{1}$ or $\w_{1}\fff\limits _{U[\BBB]}\w_{2}$
must be fulfilled. For example we consider the case $\w_{2}\fff\limits _{U[\BBB]}\w_{1}$
(another case is considered similarly). By definition of the base
of elementary processes on $U[\BBB]$ the last relation means, that
$\w_{2}\,\xxx\limits _{U[\BBB]}\,\w_{1}$. Since $\w_{2}\,\xxx\limits _{U[\BBB]}\,\w_{1}$
and $\w_{1}\neq\w_{2}$, the condition (C4) must be satisfied. Thus,
there exist united by fate in $\BBB$ elementary-time states $\omega_{1},\omega_{2}\in\BSB$
such, that $\w_{1}=U\left(\omega_{1}\right)$, $\w_{2}=U\left(\omega_{2}\right)$,
what is needed to prove. 

5.b) Now, suppose, that $\w_{1}=U\left(\omega_{1}\right)$, $\w_{2}=U\left(\omega_{2}\right)$,
where the elementary-time states $\omega_{1},\omega_{2}\in\BSB$ are
united by fate in $\BBB$. Since $\tm{\w_{1}}\neq\tm{\w_{2}}$, by
the condition (C4) in the case $\tm{\w_{1}}<\tm{\w_{2}}$ we obtain,
that $\w_{2}\,\xxx\limits _{U[\BBB]}\,\w_{1}$, that is $\w_{2}\fff\limits _{U[\BBB]}\w_{1}$,
and in the case $\tm{\w_{2}}<\tm{\w_{1}}$ we have $\w_{1}\fff\limits _{U[\BBB]}\w_{2}$.
Thus, by the assertion \ref{As:LdProperties1}, $\w_{1}$ and $\w_{2}$
are united by fate in $U[\BBB]$. 

\textbf{Proof of the uniqueness}.

Let $\Uxv[\BBB]$ be other basic changeable set, which satisfies the
conditions 1,2,3 of this theorem. Then, from first two conditions
of the theorem, it follows, that $\BS\left(U[\BBB]\right)=\BS\left(\Uxv[\BBB]\right)$,
$\Tm\left(U[\BBB]\right)=\Tm\left(\Uxv[\BBB]\right)$, $\leq_{U[\BBB]}=\leq_{\Uxv[\BBB]}$.
Next, we are going to prove, that the bases of elementary processes
$\fff\limits _{U[\BBB]}$ and $\fff\limits _{\Uxv[\BBB]}$ on the
set $\BS\left(U[\BBB]\right)=\BS\left(\Uxv[\BBB]\right)$ are identical. 

Suppose, that $\w_{1},\w_{2}\in\BS\left(U[\BBB]\right)$ and $\w_{2}\,\fff\limits _{U[\BBB]}\,\w_{1}$. 

In the case $\w_{1}=\w_{2}$, by the definition of base of elementary
processes \ref{Def:BazovaMM}, we have $\w_{2}\fff\limits _{\Uxv[\BBB]}\w_{1}$. 

Thus, it remains to consider the case $\w_{1}\neq\w_{2}$. Since $\w_{2}\,\fff\limits _{U[\BBB]}\,\w_{1}$,
then, by the property \ref{Prop:BMM}(\ref{enu:BMMProp(fff,ffff)}),
$\w_{2}\,\ffff\limits _{U[\BBB]}\,\w_{1}$. Since $\w_{2}\,\ffff\limits _{U[\BBB]}\,\w_{1}$
and $\w_{1}\neq\w_{2}$, by definition \ref{Def:EChS_Synh,fff}, $\tm{\w_{1}}<_{U[\BBB]}\,\tm{\w_{2}}$.
Hence, $\tm{\w_{1}}\neq\tm{\w_{2}}$. By the assertion \ref{As:LdProperties1},
$\w_{1}$ and $\w_{2}$ are united by fate in $U[\BBB]$. So, taking
into account, that the basic changeable set $U[\BBB]$ satisfies the
third condition of this theorem, we obtain, that there exists united
by fate in $\BBB$ elementary-time states $\omega_{1},\omega_{2}\in\BSB$
such, that $\w_{1}=U\left(\omega_{1}\right)$, $\w_{2}=U\left(\omega_{2}\right)$.
Hence, since the basic changeable set $\Uxv[\BBB]$ also satisfies
the third condition of this theorem and $\tm{\w_{1}}\neq\tm{\w_{2}}$,
$\w_{1}$ and $\w_{2}$ are united by fate in $\Uxv[\BBB]$. Thus,
one of the conditions $\w_{2}\fff\limits _{\Uxv[\BBB]}\w_{1}$ or
$\w_{1}\fff\limits _{\Uxv[\BBB]}\w_{2}$ must be fulfilled. Since
$\tm{\w_{1}}<_{U[\BBB]}\,\tm{\w_{2}}$ and $\leq_{U[\BBB]}=\leq_{\Uxv[\BBB]}$,
we have $\tm{\w_{1}}<_{\Uxv[\BBB]}\,\tm{\w_{2}}$. Thus, the condition
$\w_{1}\fff\limits _{\Uxv[\BBB]}\w_{2}$ is impossible by the property
\ref{Prop:BMM}(\ref{enu:BMMProp(fff,ffff)}). Consequently, $\w_{2}\fff\limits _{\Uxv[\BBB]}\w_{1}$. 

Thus, we have proved, that for any $\w_{1},\w_{2}\in\BS\left(U[\BBB]\right)=\BS\left(\Uxv[\BBB]\right)$
condition $\w_{2}\,\fff\limits _{U[\BBB]}\,\w_{1}$ involves the condition
$\w_{2}\fff\limits _{\Uxv[\BBB]}\w_{1}$. Similarly, it can be proved,
that condition $\w_{2}\fff\limits _{\Uxv[\BBB]}\w_{1}$ involves the
condition $\w_{2}\,\fff\limits _{U[\BBB]}\,\w_{1}$. This means, that
$\fff\limits _{U[\BBB]}\upharpoonright_{\BS(U[\BBB])}=\fff\limits _{\Uxv[\BBB]}\upharpoonright_{\BS(\Uxv[\BBB])}$,
where $\fff\limits _{U[\BBB]}\upharpoonright_{\BS(U[\BBB])}$ and
$\fff\limits _{\Uxv[\BBB]}\upharpoonright_{\BS(\Uxv[\BBB])}$ are
the bases of elementary processes on $U[\BBB]$ and $\Uxv[\BBB]$
correspondingly. 

Now we have proved the equalities $\BS\left(U[\BBB]\right)=\BS\left(\Uxv[\BBB]\right)$,
$\Tm\left(U[\BBB]\right)=\Tm\left(\Uxv[\BBB]\right)$, $\leq_{U[\BBB]}=\leq_{\Uxv[\BBB]}$
and $\fff\limits _{U[\BBB]}\upharpoonright_{\BS(U[\BBB])}=\fff\limits _{\Uxv[\BBB]}\upharpoonright_{\BS(\Uxv[\BBB])}$.
From these equalities it follows, that other components of the basic
changeable sets $U[\BBB]$ and $\Uxv[\BBB]$ also are identical (according
to the properties \ref{Prop:BMM}(\ref{enu:BMMProp(fff,Bs)},\ref{enu:BMMProp(Bs)})).
Thus, $U[\BBB]=\Uxv[\BBB]$. ~ ~ \EndProof  

\BeginDef 

The basic changeable set $U[\BBB]$, which satisfies the conditions
1,2,3 of the theorem \ref{Thm:U[B]exist} will be named the \textbf{image
of the basic changeable set} $\BBB$ \textbf{during the transforming
mapping} $U:\,\TmB\times\BsB\mapsto\TmB\times X$. 

\EndDef 

\BeginRmk 

According to conditions (C3),(C4) in the proof of theorem \ref{Thm:U[B]exist}
for any elementary-time states $\w_{1},\,\w_{2}\in\BS(U[\BBB])$ the
relation $\w_{2}\fff\limits _{U[\BBB]}\w_{1}$ is true if and only
if $\w_{1}=\w_{2}$ or $\tm{\w_{1}}<\,\tm{\w_{2}}$ and there exists
united by fate in $\BBB$ elementary-time states $\omega_{1},\omega_{2}\in\BSB$
such, that $\w_{1}=U\left(\omega_{1}\right)$, $\w_{2}=U\left(\omega_{2}\right)$. 

\EndRmk 

\BeginEx \label{Ex:Zimg}

Let $\BBB$ be a basic changeable set, and $X$ --- an arbitrary set
such, that $\BsB\subseteq X$. And let $\bbU$ be any set of bijections
(one-to-one correspondences) of kind: 

\[
U:\TmB\times X\longmapsto\TmB\times X\qquad(U\in\bbU)\]

Such set of bijections $\bbU$ will be referred to as \textbf{\emph{transforming
set of mappins}} for the basic changeable set $\BBB$. 

Denote: \begin{gather*}
\AAA:=\bbU;\\
U_{\alpha}:=\alpha,\quad\alpha\in\AAA\end{gather*}

Then we obtain the indexed set of mappings $\vU=\left(U_{\alpha}\,|\:\alpha\in\AAA\right)$
such, that $U_{\alpha}\neq U_{\beta}$, for $\alpha\neq\beta$.

Any mapping $U_{\alpha}$ ($\alpha\in\AAA$) is a transforming mapping,
for the basic changeable set $\BBB$. Thus, we obtain a family of
basic changeable sets: 

\begin{gather*}
\BBB_{\alpha}:=U_{\alpha}[\BBB],\qquad\alpha\in\AAA;\\
\vU[\BBB]=\left(U_{\alpha}[\BBB]\,|\:\alpha\in\AAA\right)=\left(U[\BBB]\,|\: U\in\bbU\right).\end{gather*}
By theorem \ref{Thm:U[B]exist}:\[
\BS\left(U_{\alpha}[\BBB]\right)=U_{\alpha}\left(\BSB\right),\quad\alpha\in\AAA,\]
so any mapping $U_{\alpha}$ is a bijection from the set $\BSB$ to
the set $\BS\left(U_{\alpha}[\BBB]\right)$. Hence, we can consider
the family of mappings: \[
\widetilde{U}_{\beta\alpha}:=U_{\beta}U_{\alpha}^{-1}=U_{\beta}\left(U_{\alpha}^{-1}\right),\qquad\alpha,\beta\in\AAA.\]
 For any $\alpha,\beta\in\AAA$ the mapping $\widetilde{U}_{\beta\alpha}$
is bijection from the set $\BS\left(U_{\alpha}[\BBB]\right)$ to the
set $\BS\left(U_{\beta}[\BBB]\right)$. We are near to prove that
the family of mappings $\overleftarrow{U^{\sim}}=\left(\widetilde{U}_{\beta\alpha}\,|\:\alpha,\beta\in\AAA\right)$
satisfies the conditions (\ref{eq:PseudoGrpConditions}). Indeed:
\begin{gather*}
\widetilde{U}_{\alpha\alpha}(\omega):=U_{\alpha}U_{\alpha}^{-1}(\omega)=\omega;\\
\widetilde{U}_{\gamma\beta}\widetilde{U}_{\beta\alpha}(\omega)=U_{\gamma}U_{\beta}^{-1}U_{\beta}U_{\alpha}^{-1}(\omega)=U_{\gamma}U_{\alpha}^{-1}(\omega)=\widetilde{U}_{\gamma\alpha}(\omega)\\
\hspace{5cm}(\alpha,\beta,\gamma\in\AAA,\:\omega\in\BS\left(U_{\alpha}[\BBB]\right)).\end{gather*}
Thus, by results of example \ref{Ex:Zpv}, we can construct the changeable
set: \[
\Zim{\bbU,\BBB}=\Zpv{\vU[\BBB],\overleftarrow{U^{\sim}}}.\]
The changeable set $\Zim{\bbU,\BBB}$ will be named \textbf{\emph{multi-figurative
image}} of the basic changeable set $\BBB$ relatively the transforming
set of mappins $\bbU$. 

\EndEx  

\BeginEx \label{Ex:ZRelativity}

Let $\BBB$ be a basic changeable set such, that 

\[
\BsB\subseteq\R^{3},\quad\TmB=\R\]
 (for example it may be, that $\BBB=\At\left(\cR\right)$, where $\cR$
is a system of abstract trajectories from $\R$ to $\R^{3}$). The
Poincare group $\bbU=P(1,3)$, defined on the 4-dimensional space-time
$\R^{4}=\R\times\R^{3}\supseteq\TmB\times\BsB$ is transforming set
of mappins for this basic changeable set $\BBB$. Hence, we obtain
the changeable set $\Zim{P(1,3),\BBB}$, which can be applied to formalization
of the cinematics of special relativity theory in the inertial frames
of reference. 

\EndEx 

In the examples \ref{Ex:Zpv}-\ref{Ex:ZRelativity} the unification
mappings $\un{l}{m}$ ($l,m\in\LkZ$) are defined by means of bijections
(one-to-one correspondences) between the sets of elementary-time states
$\BS(l)$ and $\BS(m)$ (that is $\un{l}{m}A=\bigcup_{\omega\in A}\un{l}{m}\left\{ \omega\right\} $
and the third condition of the definition \ref{Def:GlobalUni,ChSets}
may be replaced by more strong condition (\ref{eq:GrpMaps1})). But
really the conditions of the definition \ref{Def:GlobalUni,ChSets}
are enough general. And the next examples show, how far in this definition
was made a departure from the usual for physics {}``pointwise''
comparison between elementary-time states of different areas of perception
(frames of reference). 

\BeginEx \label{Ex:Znv}

Let $\vcB=\left(\BBB_{\alpha}\,|\:\alpha\in\AAA\right)$ be any indexed
family of basic changeable sets. Denote: \[
\fU_{\beta\alpha}A:=\begin{cases}
A, & \alpha=\beta\\
\emptyset, & \alpha\neq\beta\end{cases},\qquad\alpha,\beta\in\AAA,\; A\subseteq\BSBB{\alpha}.\]
It is easy to verify, that for the family of mappings $\vfU=\left(\fU_{\beta\alpha}\,|\:\alpha,\beta\in\AAA\right)$
all conditions of the definition \ref{Def:GlobalUni,ChSets} are satisfied.
Therefore, the triple \[
\Znv{\vcB}=\left(\AAA,\vcB,\vfU\right)\]
is a changeable set. 

\EndEx 

The changeable set $\Znv{\vcB}$ will be named the \textbf{\emph{fully
invisible changeable set}}, generated by the system of basic changeable
sets $\vcB$. 

Note, that any basic changeable set $\BBB$ can be identified with
the changeable set of kind $\Znv{\vcB}$, where $\AAA=\{1\}$, $\BBB_{1}=\BBB$
and $\vcB=\left(\BBB_{\alpha}\,|\,\alpha\in\AAA\right)=\left(\BBB_{1}\right)$. 

\BeginEx \label{Ex:ZVdNeNormVd}

\begin{onehalfspace}
Let, $\vcB=\left(\BBB_{1},\BBB_{2}\right)$ ($\AAA=\{1,2\}$) be a
family of two basic changeable sets. Choose any elementary-time state
$\omega\in\BSBB{2}$. Denote: \begin{gather*}
\fU_{11}A:=A,\quad A\subseteq\BSBB{1};\qquad\fU_{22}A:=A,\quad A\subseteq\BSBB{2};\\
\fU_{21}A:=\begin{cases}
\emptyset, & A\neq\BSBB{1}\\
\{\omega\}, & A=\BSBB{1}\end{cases},\qquad A\subseteq\BSBB{1};\\
\fU_{12}A:=\begin{cases}
\emptyset, & \omega\notin A\\
\BSBB{1}, & \omega\in A\end{cases},\qquad A\subseteq\BSBB{2};\end{gather*}

\end{onehalfspace}

1. Since $\fU_{11},\fU_{22}$ are identity mappings of sets, the first
condition of the definition \ref{Def:GlobalUni,ChSets} is performed
by a trivial way. For the same reason the second condition of this
definition also is satisfied in the case $\alpha=\beta$. 

2. Suppose, that $\alpha,\beta\in\AAA=\{1,2\}$, $A,B\subseteq\BSBB{\alpha}$,
$A\subseteq B$. According to the remark in the end of previous item,
it is enough to consider only the case $\alpha\neq\beta$. Thus, we
have the next two subcases. 

2.a) $\alpha=1$, $\beta=2$. In the case $A\neq\BSBB{1}$ we obtain
$\fU_{\beta\alpha}A=\fU_{21}A=\emptyset\subseteq\fU_{\beta\alpha}B$,
and in the case $A=\BSBB{1}$, since $A\subseteq B$ we have $B=\BSBB{1}$,
and, therefore, $\fU_{\beta\alpha}A=\fU_{\beta\alpha}B$. 

2.b) $\alpha=2$, $\beta=1$. In the case $\omega\notin A$ we obtain
$\fU_{\beta\alpha}A=\fU_{12}A=\emptyset\subseteq\fU_{\beta\alpha}B$.
In the case $\omega\in A$ from the condition $A\subseteq B$ it follows,
that $\omega\in B$, so $\fU_{\beta\alpha}A=\fU_{12}A=\BSBB{1}=\fU_{12}B=\fU_{\beta\alpha}B$. 

3. Let $\alpha,\beta,\gamma\in\AAA=\{1,2\}$, $A\subseteq\BSBB{\alpha}$.
We consider the following cases. 

3.a) $\alpha=\beta$. In this case $\fU_{\beta\alpha}A=A$. Consequently:
\[
\fU_{\gamma\beta}\fU_{\beta\alpha}A=\fU_{\gamma\beta}A=\fU_{\gamma\alpha}A.\]

3.b) $\beta=\gamma$. In this case $\fU_{\gamma\beta}S=S$, $S\subseteq\BSBB{\beta}$.
Hence:\[
\fU_{\gamma\beta}\fU_{\beta\alpha}A=\fU_{\beta\alpha}A=\fU_{\gamma\alpha}A.\]

3.c) $\alpha\neq\beta\neq\gamma$. Since the set $\AAA$ is two-element,
this case can be divided into the following two subcases:

3.c.1) Let $\alpha=1$, $\beta=2$, $\gamma=1$. Then in the case
$A\neq\BSBB{1}$ we obtain: \[
\fU_{\gamma\beta}\fU_{\beta\alpha}A=\fU_{12}\fU_{21}A=\fU_{12}\emptyset=\emptyset\subseteq\fU_{\gamma\alpha}A,\]
 and in the case $A=\BSBB{1}$ we calculate: \[
\fU_{\gamma\beta}\fU_{\beta\alpha}A=\fU_{12}\fU_{21}A=\fU_{12}\{\omega\}=\BSBB{1}=A=\fU_{\gamma\alpha}A.\]

3.c.2) Let, $\alpha=2$, $\beta=1$, $\gamma=2$. Then in the case
$\omega\notin A$ we have:\[
\fU_{\gamma\beta}\fU_{\beta\alpha}A=\fU_{21}\fU_{12}A=\fU_{21}\emptyset=\emptyset\subseteq\fU_{\gamma\alpha}A,\]
 and in the case $\omega\in A$ we obtain: \[
\fU_{\gamma\beta}\fU_{\beta\alpha}A=\fU_{21}\fU_{12}A=\fU_{21}\BSBB{1}=\{\omega\}\subseteq A=\fU_{\gamma\alpha}A.\]
Consequently, the triple: \[
\cZ_{1}=\left(\AAA,\vcB,\vfU\right),\]
where $\vfU=\left(\fU_{\beta\alpha}\,|\:\alpha,\beta\in\AAA\right)$
is a changeable set. 

\EndEx 

\BeginEx  \label{Ex:ZNormVdmNeChitka}

Let $\AAA$, $\BBB_{1}$, $\BBB_{2}$, $\omega$ be the same as in
the example \ref{Ex:ZVdNeNormVd}. As well as in the previous example
\ref{Ex:ZVdNeNormVd}, $\fU_{11}$ and $\fU_{22}$ are the identical
mappings of the sets. Also, we denote: \begin{gather*}
\fU_{21}A:=\begin{cases}
\emptyset, & A=\emptyset\\
\{\omega\}, & A\neq\emptyset\end{cases},\qquad A\subseteq\BSBB{1};\\
\fU_{12}A:=\emptyset,\qquad A\subseteq\BSBB{2}.\end{gather*}

1,2. Since, $\fU_{11}$ and $\fU_{22}$, are the identical mappings
of the sets, the first condition of the definition \ref{Def:GlobalUni,ChSets}
is satisfied by a trivial way. The second condition of this definition
also is easy to verify. 

3. In the cases $\alpha=\beta=\gamma$, $\alpha\neq\beta=\gamma$,
$\alpha=\beta\neq\gamma$ verification of the third condition of the
definition \ref{Def:GlobalUni,ChSets} is the same, as in the example
\ref{Ex:ZVdNeNormVd}. Thus it remains to consider the case $\alpha\ne\beta\neq\gamma$.
Like the previous example we divide this case into the following two
subcases: 

3.1) Let, $\alpha=1$, $\beta=2$, $\gamma=1$. Then: \[
\fU_{\gamma\beta}\fU_{\beta\alpha}A=\fU_{12}\fU_{21}A=\emptyset\subseteq\fU_{\gamma\alpha}A.\]

3.2) Let, $\alpha=2$, $\beta=1$, $\gamma=2$. Then:\[
\fU_{\gamma\beta}\fU_{\beta\alpha}A=\fU_{21}\fU_{12}A=\fU_{21}\emptyset=\emptyset\subseteq\fU_{\gamma\alpha}A.\]
Thus, the triple: \[
\cZ_{2}=\left(\AAA,\vcB,\vfU\right),\]
is a changeable set. 

\EndEx

\section{Visibility in Changeable Sets \label{sec:Visibility}}

\subsection{Gradations of Visibility}

\BeginDef \label{Def:Vi:Systems}

Let $\cZ$ be any changeable set, and $l,m\in\LkZ$ be any areas of
perception of $\cZ$. We say, that a changeable system $A\subseteq\BS(l)$
of the area of perception $l$ is:

\begin{enumerate}
\item \textbf{visible }(partially visible) from the area of perception $m$,
if and only if $\un{l}{m}A\neq\emptyset$;
\item \textbf{normally visible} from the area of perception $m$, if and
only if an arbitrary nonempty subsystem $B\subseteq A$ of the changeable
system $A$ is visible from $m$ (that is $\forall\, B:\emptyset\neq B\subseteq A$
$\un{l}{m}B\neq\emptyset$); 
\item \textbf{precisely visible} from $m$, if and only if:

\begin{enumerate}
\item $A$ is normally visible from $m$; 
\item for any family $\left\{ A_{\alpha}\,|\,\alpha\in\AAA\right\} \subseteq2^{A}$
of changeable subsystems $A$ such, that $\bigsqcup_{\alpha\in\AAA}A_{\alpha}=A$
the following equality holds \[
\un{l}{m}A=\bigsqcup_{\alpha\in\AAA}\un{l}{m}A_{\alpha},\]
 where $\bigsqcup_{\alpha\in\AAA}A_{\alpha}$ denotes the disjoint
union of the family of sets $\left\{ A_{\alpha}\,|\,\alpha\in\AAA\right\} $,
that is the union $\bigcup_{\alpha\in\AAA}A_{\alpha}$, with additional
condition $A_{\alpha}\cap A_{\beta}=\emptyset$, $\alpha\neq\beta$.
\textbf{}
\end{enumerate}
\item \textbf{invisible }from the area of perception $m$, if and only if
$\un{l}{m}A=\emptyset$;
\end{enumerate}
\EndDef 

\BeginRmk \label{Rmk:vi01}

It is apparently, that the precise visibility of the changeable system
$A\subseteq\BS(l)$ ($l\in\LkZ$) from the area of perception $m\in\LkZ$
involves the normal visibility of $\AAA$ from $m$, and the normal
visibility of any changeable system $A\subseteq\BS(l)$ from $m$
involves it's visibility (partial visibility) from $m$. 

\EndRmk 

\BeginAs \label{As:ViProperties01}

For any changeable set $\cZ$ the following properties of visibility
of changeable systems are true: 

\begin{enumerate}
\item Empty changeable system $\emptyset\subseteq\BS(l)$ always is invisible
from any area of perception $m\in\LkZ$. 
\item Any nonempty changeable system $A\subseteq\BS(l)$, $A\neq\emptyset$
always is precisely visible from its own area of perception $l$. 
\item If a changeable system $A\subseteq\BS(l)$ (where $l\in\LkZ$) includes
a subsystem $B\subseteq A$, which is visible from area of perception
$m\in\LkZ$, then the changeable system $A$ also is visible from
$m$.
\item If a changeable system $A\subseteq\BS(l)$ is normally visible (precisely
visible) from area of perception $m$, then any nonempty subsystem
$B\subseteq A$, $B\neq\emptyset$ of changeable system $A$ also
is normally visible (precisely visible) from $m$. 
\end{enumerate}
\EndAs 

\BeginProof 

Statements 1,2,3 of this assertion follow from the assertion \ref{As:Zproperties01}
and properties \ref{Prop:ChSets(basic)} of changeable sets. Statement
4 for the case of normal visibility is trivial. Thus, it remains to
prove the statement 4 for the case of precise visibility. Let a changeable
system $A\subseteq\BS(l)$ be precisely visible from the area of perception
$m$. Consider any changeable system $B$ such, that $\emptyset\neq B\subseteq A$.
Since precise visibility involves the normal visibility, $B$ is normally
visible from $m$. Suppose, that $B=\bigsqcup_{\alpha\in\AAA}B_{\alpha}$.
Using the equalities: 

\begin{gather*}
A=B\sqcup\left(A\setminus B\right);\qquad A=\bigsqcup_{\alpha\in\AAA}B_{\alpha}\sqcup\left(A\setminus B\right),\end{gather*}
 and taking into account precise visibility of the changeable system
$A$ from $m$, we obtain: \begin{gather*}
\un{l}{m}A=\un{l}{m}B\sqcup\un{l}{m}\left(A\setminus B\right);\qquad\\
\un{l}{m}A=\bigsqcup_{\alpha\in\AAA}\un{l}{m}B_{\alpha}\sqcup\un{l}{m}\left(A\setminus B\right).\end{gather*}
Consequently, $\un{l}{m}B\sqcup\un{l}{m}\left(A\setminus B\right)=\bigsqcup_{\alpha\in\AAA}\un{l}{m}B_{\alpha}\sqcup\un{l}{m}\left(A\setminus B\right)$.
Hence: \[
\un{l}{m}B=\bigsqcup_{\alpha\in\AAA}\un{l}{m}B_{\alpha}.\]
Thus, \textbf{$B$ }is precisely visible from $m$. ~ ~ ~ ~ ~
\EndProof  

\BeginDef \label{Def:Vi:Lk}

We say, that an area of perception $l\in\LkZ$ is:

\begin{enumerate}
\item \textbf{visible }(partially visible) from the area of perception $m\in\LkZ$
(denotation is $l\vi m\:(\cZ)$), if and only there exists at least
one visible from the $m$ changeable system $A\subseteq\BS(l)$ (that
is $\exists\, A\subseteq\BS(l)$ $\un{l}{m}A\neq\emptyset$). 
\item \textbf{normally visible} from the area of perception $m\in\LkZ$
(denotation is $l\nvi m\:(\cZ)$), if and only if any nonempty changeable
system $A\subseteq\BS(l)$ ($A\neq\emptyset$) is normally visible
from the $m$. 
\item \textbf{precisely visible} from $m$ (denotation is $l\fvi m\:(\cZ)$),
if and only if any nonempty changeable system $A\subseteq\BS(l)$
($A\neq\emptyset$) is precisely visible from the area of perception
$m$. 
\item \textbf{invisible }from the area of perception $m$, if and only if
any changeable system $A\subseteq\BS(l)$ is invisible from the $m$. 
\end{enumerate}
\EndDef  

In the case, when the changeable set $\cZ$ is known in advance in
the denotations $l\vi m\:(\cZ)$, $l\nvi m\:(\cZ)$, $l\fvi m\:(\cZ)$
the sequence of symbols {}``$(\cZ)$'' will be omitted, and the
denotations $l\vi m$, $l\nvi m$, $l\fvi m$ will be used instead. 

\BeginRmk  \label{Rmk:vi02}

From the remark \ref{Rmk:vi01} it follows, that for the areas of
perception $l,m\in\LkZ$ the next propositions are true

$\bullet$ if $l\fvi m$, then $l\nvi m$; 

$\bullet$ if $l\nvi m$, then $l\vi m$. 

Thus, precise visibility involves the normal visibility and normal
visibility involves visibility (partial visibility). The example \ref{Ex:ZVdNeNormVd}
shows, that visibility do not involve the normal visibility. Indeed,
we may consider the case, when in this example $\card\left(\BSBB{1}\right)\geq2$.
In this case for the areas of perception $l_{1}=\left(1,\BBB_{1}\right)$,
$l_{2}=\left(2,\BBB_{2}\right)$ we have, that the changeable system
$\BS\left(l_{1}\right)=\BS\left(\BBB_{1}\right)$ is visible from
$l_{2}$, but it is not normally visible from $l_{2}$, because any
subset $A\subset\BS\left(l_{1}\right)=\BSBB{1}$ ($A\neq\BSBB{1}$)
is invisible from $l_{2}$. Thus, in the case $\card\left(\BSBB{1}\right)\geq2$
we obtain $l_{1}\vi l_{2}$, but \textbf{not} $l_{1}\nvi l_{2}$. 

The example \ref{Ex:ZNormVdmNeChitka} shows, that normal visibility
do not involve the precise visibility. In this example any nonempty
changeable system $A\subseteq\BS\left(l_{1}\right)$ ($l_{1}=\left(1,\BBB_{1}\right)$)
is normally visible from the area of perception $l_{2}=\left(2,\BBB_{2}\right)$.
But, in the case $\card\left(A\right)\geq2$ the changeable system
$A$ is not precisely visible from $l_{2}$, because in this case
there exist nonempty sets $A_{1},A_{2}\subseteq A$ such, that $A_{1}\sqcup A_{2}=A$,
but the images of these sets ($\un{l_{1}}{l_{2}}A_{1}=\fU_{21}A_{1}=\left\{ \omega\right\} $,
$\un{l_{1}}{l_{2}}A_{2}=\fU_{21}A_{2}=\left\{ \omega\right\} $) are
not disjoint. Thus, in the case $\card\left(\BSBB{1}\right)\geq2$
we have $l_{1}\nvi l_{2}$, but \textbf{not} $l_{1}\fvi l_{2}$. 

In the examples \ref{Ex:Zpv}, \ref{Ex:Zimg} and \ref{Ex:ZRelativity}
any area of perception of the changeable sets $\Zpv{\vcB,\vW}$ and
$\Zim{\bbU,\BBB}$ is precisely visible from another. 

\EndRmk 

\medskip{}

The next three assertions immediately follow from the definitions
\ref{Def:Vi:Lk},\ref{Def:Vi:Systems} and the assertion \ref{As:ViProperties01}. 

\BeginAs \label{As:ViProperties02} 

For any changeable set $\cZ$ the following propositions are equivalent:

\begin{description}
\item [{(Vi1)}] Area of perception $l\in\LkZ$ is visible from area of
perception $m\in\LkZ$ ($l\vi m$).
\item [{(Vi2)}] The set $\BS(l)$ of all elementary-time states of $l$
is visible from $m$. 
\end{description}
\EndAs 

\BeginAs \label{As:ViProperties03} 

For an arbitrary changeable set $\cZ$ the following propositions
are equivalent:

\begin{description}
\item [{(nVi1)}] Area of perception $l\in\LkZ$ is normally visible from
area of perception $m\in\LkZ$ ($l\nvi m$).
\item [{(nVi2)}] The set $\BS(l)$ of all elementary-time states of $l$
is normally visible from $m$. 
\item [{(nVi3)}] Any nonempty changeable system $A\subseteq\BS(l)$ is
visible from $m$ ($\forall\subseteq\BS(l)$ $\left(A\neq\emptyset\Rightarrow\un{l}{m}A\neq\emptyset\right)$). 
\end{description}
\EndAs  

\BeginAs \label{As:ViProperties04} 

Let $\cZ$ --- be an arbitrary changeable set. Then: \smallskip{}

1. Any area of perception $l\in\LkZ$ is precisely visible from itself
(that is $\forall l\in\LkZ\quad l\fvi l$).\smallskip{}

2. The following propositions are equivalent:

\begin{description}
\item [{(pVi1)}] Area of perception $l\in\LkZ$ is precisely visible from
area of perception $m\in\LkZ$ ($l\fvi m$). 
\item [{(pVi2)}] The set $\BS(l)$ of all elementary-time states of $l$
is precisely visible from $m$. 
\end{description}
\EndAs  \smallskip{}

\BeginAs \label{As:NviQuasiOrder}

For any changeable set $\cZ$ the binary relation $\nvi$ quasi order
on the set $\LkZ$ of all areas of perception $\cZ$. 

\EndAs 

\BeginProof 

Reflexivity of the relation $\nvi$ follows from the first item of
the assertion \ref{As:ViProperties04} and from the remark \ref{Rmk:vi02}.
Thus, we need to prove the transitivity of the relation $\nvi$. 

Suppose, that $l\nvi m$ and $m\nvi p$, where $l,m,p\in\LkZ$. Then,
using the assertion \ref{As:ViProperties03} (equivalence between
(nVi1) and (nVi3)), for any nonempty changeable system $A\subseteq\BS(l)$
we obtain, $\un{l}{p}A\supseteq\un{m}{p}\un{l}{m}A\neq\emptyset$,
thus, by the assertion \ref{As:ViProperties03}, $l\nvi p$. ~ ~
~ \EndProof 

\BeginRmk 

First item of the assertion \ref{As:ViProperties04} and remark \ref{Rmk:vi02}
also bring about the reflexivity of the relations $\fvi$ and $\vi$
on the set $\LkZ$ (for any changeable set $\cZ$). But these relations,
in general, are not transitive. And the next examples explain the
last statement. 

\EndRmk 

\BeginEx \label{Ex:FviNoTransitive}

\begin{onehalfspace}
Let $\BBB$ be any basic changeable set. We consider the family $\vcB=\left(\BBB_{\alpha}\,|\:\alpha\in\N\right)$
of basic changeable sets, which is defined as follows: 
\end{onehalfspace}

\[
\BBB_{\alpha}:=\BBB,\qquad\alpha\in\N.\]
 For $\alpha,\beta\in\N$ we define the mappings $\fU_{\beta\alpha}:\BSBB{\alpha}\mapsto\BSBB{\beta}$
by the following way: \begin{equation}
\fU_{\beta\alpha}A:=\begin{cases}
A, & \beta\in\left\{ \alpha,\alpha+1\right\} ;\\
\BSB, & \beta>\alpha+1,\: A\neq\emptyset;\\
\emptyset, & \beta>\alpha+1,\: A=\emptyset;\\
\emptyset, & \beta<\alpha,\end{cases}\qquad(A\in\BSBB{\alpha}=\BSB,\; n\in\N)\label{eq:UDef01}\end{equation}
(where the symbols $<,>$ denote the usual order on the set of natural
numbers). 

We shell prove, that the system of mappings $\vfU=\left(\fU_{\beta\alpha}\,|\:\alpha,\beta\in\N\right)$
is unification of perception. 

The first two conditions of the definition \ref{Def:GlobalUni,ChSets}
for the system of mappings $\vfU$ are performed by a trivial way.
Thus, we need to verify the third condition of this definition. Let
$\alpha,\beta,\gamma\in\N$ and $A\subseteq\BSBB{\alpha}=\BSB$. Then
in the case $\alpha\leq\beta\leq\gamma$, by (\ref{eq:UDef01}), we
obtain: 

\begin{equation}
\fU_{\gamma\beta}\fU_{\beta\alpha}A=\begin{cases}
\emptyset, & A=\emptyset;\\
A, & A\neq\emptyset,\:\beta\in\left\{ \alpha,\alpha+1\right\} ,\,\gamma\in\left\{ \beta,\beta+1\right\} ;\\
\BSB, & A\neq\emptyset,\;\textrm{and}\;(\beta>\alpha+1\:\textrm{or}\:\gamma>\beta+1).\end{cases}\label{eq:UgbUba01}\end{equation}
Since $\fU_{\gamma\alpha}A\in\left\{ A,\BSB\right\} $ for $\alpha\leq\gamma$,
in the first two cases of the formula (\ref{eq:UgbUba01}) the inclusion
$\fU_{\gamma\beta}\fU_{\beta\alpha}A\subseteq\fU_{\gamma\alpha}A$
holds. In the third case of the formula (\ref{eq:UgbUba01}) we have
$\gamma>\alpha+1$, and hence, $\fU_{\gamma\alpha}A=\BSB$. Thus,
in this case, the last inclusion also is performed. If the condition
$\alpha\leq\beta\leq\gamma$ is not satisfied, we have $\alpha>\beta$
or $\beta>\gamma$. Therefore, by the formula (\ref{eq:UDef01}),
we have, $\fU_{\gamma\beta}\fU_{\beta\alpha}A=\emptyset$. Consequently,
in this case we also have the inclusion $\fU_{\gamma\beta}\fU_{\beta\alpha}A\subseteq\fU_{\gamma\alpha}A$.
Thus, all conditions of the definition \ref{Def:GlobalUni,ChSets}
are satisfied. 

Hence, the triple $\cZ=\left(\N,\vcB,\vfU\right)$ is a changeable
set. For this changeable set $\cZ$ we have: 

\begin{gather*}
\LkZ=\left\{ l_{n}\,|\: n\in\N\right\} ,\;\textrm{where}\\
l_{n}=\left(n,\BBB_{n}\right)=\left(n,\BBB\right),\; n\in\N,\end{gather*}
and for $l_{n},l_{m}\in\LkZ$ the equality $\un{l_{m}}{l_{n}}=\fU_{nm}$
holds. Thus, by (\ref{eq:UDef01}):  \begin{gather*}
\un{l_{n}}{l_{n+1}}A=A,\qquad A\subseteq\BS\left(l_{n}\right)=\BSB,\; n\in\N;\\
\un{l_{n}}{l_{n+2}}A=\begin{cases}
\BSB, & A\neq\emptyset\\
\emptyset, & A=\emptyset\end{cases},\quad A\subseteq\BS\left(l_{n}\right)=\BSB,\; n\in\N;\end{gather*}
The last equality shows, that $l_{n}\fvi l_{n+1}$ ($n\in\N$). But,
in the case $\card(\BSB)\geq2$, $l_{n}$ is normally visible, but
not precisely visible from $l_{n+2}$. Thus, in the case $\card(\BSB)\geq2$
for any $n\in\N$ we have $l_{n}\fvi l_{n+1}$,~ $l_{n+1}\fvi l_{n+2}$,
although the correlation $l_{n}\fvi l_{n+2}$ is not true. 

\EndEx 

\BeginEx \label{Ex:ViNoTransitive} 

\begin{onehalfspace}
Let basic changeable set $\BBB$ be such, that the set $\BSB$ is
infinite. Then there exists the sequence $\left(\omega_{n}\right)_{n=1}^{\infty}\subseteq\BSB$
of elementary-time states such, that $\omega_{n}\neq\omega_{m}$,
$m\neq n$. Denote: 
\end{onehalfspace}

\begin{gather}
\BBB_{\alpha}:=\BBB,\quad\alpha\in\N;\qquad\vcB:=\left(\BBB_{\alpha}\,|\:\alpha\in\N\right);\nonumber \\
\fU_{\beta\alpha}A:=\begin{cases}
A, & \beta=\alpha\\
\left\{ \omega_{\beta}\right\} , & \beta=\alpha+1,\:\omega_{\beta}\in A\\
\emptyset, & \beta=\alpha+1,\:\omega_{\beta}\notin A\\
\emptyset, & \beta\notin\left\{ \alpha,\alpha+1\right\} .\end{cases}\qquad(A\in\BSBB{\alpha}=\BSB,\; n\in\N)\label{eq:UDef02}\end{gather}
We shell prove, that the system of mappings $\vfU=\left(\fU_{\beta\alpha}\,|\:\alpha,\beta\in\N\right)$
is unification of perception. The first two conditions of the definition
\ref{Def:GlobalUni,ChSets} for the system of mappings $\vfU$ are
performed by a trivial way. Thus, we need to verify the third condition
of this definition. Let $\alpha,\beta,\gamma\in\N$. It should be
noted, that from (\ref{eq:UDef02}) it follows, that $\fU_{\beta\alpha}\emptyset=\emptyset$
for any $\alpha,\beta\in\N$. Thus, according to (\ref{eq:UDef02}),
if one of the conditions $\alpha\leq\beta$ or $\beta\leq\gamma$,
are not performed, then we have $\fU_{\gamma\beta}\fU_{\beta\alpha}A=\emptyset\subseteq\fU_{\gamma\alpha}A$,
$A\in\BSB$. Hence, we shell consider the case $\alpha\leq\beta\leq\gamma$.
In the case, when $\alpha=\beta$ or $\beta=\gamma$, similarly to
the example \ref{Def:GlobalUni,ChSets}, we obtain $\fU_{\gamma\beta}\fU_{\beta\alpha}A=\fU_{\gamma\alpha}A$.
Thus, it remains to consider only the case $\alpha<\beta<\gamma$.
In the cases $\beta>\alpha+1$ or $\gamma>\beta+1$, by (\ref{eq:UDef02}),
we obtain $\fU_{\gamma\beta}\fU_{\beta\alpha}A=\emptyset\subseteq\fU_{\gamma\alpha}A$,
$A\in\BSB$. Hence, it remains only the case $\beta=\alpha+1$ and
$\gamma=\beta+1$. If $\omega_{\beta}\notin A$, then, by (\ref{eq:UDef02}),
$\fU_{\beta\alpha}A=\emptyset$, and we have, $\fU_{\gamma\beta}\fU_{\beta\alpha}A=\emptyset\subseteq\fU_{\gamma\alpha}A$.
And in the case $\omega_{\beta}\in A$, we obtain $\omega_{\gamma}=\omega_{\beta+1}\notin\left\{ \omega_{\beta}\right\} $.
Thus, in this case:  \[
\fU_{\gamma\beta}\fU_{\beta\alpha}A=\fU_{\gamma\beta}\fU_{\beta\alpha}A=\fU_{\gamma\beta}\left\{ \omega_{\beta}\right\} =\emptyset\subseteq\fU_{\gamma\alpha}A.\]

Consequently, the triple $\cZ=\left(\N,\vcB,\vfU\right)$ is a changeable
set, satisfying: \begin{gather*}
\LkZ=\left\{ l_{n}\,|\: n\in\N\right\} ,\;\textrm{where}\; l_{n}=\left(n,\BBB_{n}\right)=\left(n,\BBB\right),\; n\in\N,\\
\un{l_{m}}{l_{n}}=\fU_{nm},\quad m,n\in\N\;(l_{n},l_{m}\in\LkZ).\end{gather*}
From (\ref{eq:UDef02}) it follows, that any $n\in\N$ $\un{l_{n}}{l_{n+2}}A=\fU_{n+2,n}A=\emptyset$,
~ $A\subseteq\BSB=\BS\left(l_{n}\right)$, but, under the condition,
$\omega_{n+1},\omega_{n+2}\in A$ we have $\un{l_{n}}{l_{n+1}}A=\left\{ \omega_{n+1}\right\} \neq\emptyset$,
$\un{l_{n+1}}{l_{n+2}}A=\left\{ \omega_{n+2}\right\} \neq\emptyset$.
Therefore, $l_{n}\vi l_{n+1}$, $l_{n+1}\vi l_{n+2}$, although the
area of perception $l_{n}$ invisible from $l_{n+2}$ ($l_{n}\not\vi l_{n+2}$). 

\EndEx  

\BeginDef 

We say, that a changeable set $\cZ$ is \textbf{visible (normally
visible, precisely visible)} if and only if for any $l,m\in\LkZ$
it satisfied the condition $l\vi m$ ($l\nvi m$, $l\fvi m$) correspondingly. 

\EndDef 

From remark \ref{Rmk:vi02} it follows, that any normally visible
changeable set is visible. The example \ref{Ex:ZVdNeNormVd} shows,
that the inverse assertion is not true. Indeed, we may consider the
case, when in this example $\card\left(\BSBB{1}\right)\geq2$. As
it has been shown in the remark \ref{Rmk:vi02}, in this case for
the areas of perception $l_{1}=\left(1,\BBB_{1}\right)$, $l_{2}=\left(2,\BBB_{2}\right)$
we have, $l_{1}\vi l_{2}$, but \textbf{not} $l_{1}\nvi l_{2}$. Since
in this example $\omega\in\BSBB{2}$, we obtain $\un{l_{2}}{l_{1}}\BSBB{2}=\fU_{12}\BSBB{2}=\BSBB{1}\neq\emptyset$.
Hence, $l_{2}\vi l_{1}$. Thus $l_{1}\vi l_{2}$, $l_{2}\vi l_{1}$,
but \textbf{not} $l_{1}\nvi l_{2}$. And, taking into account, that
$\Lk{\cZ_{1}}=\left\{ l_{1},l_{2}\right\} $, we obtain, that the
changeable set $\cZ_{1}$ in the example \ref{Ex:ZVdNeNormVd} is
visible, but not normally visible. In the subsection \ref{sub:Vi-classes}
(corollary \ref{Nasl:ChSet:FviEqNvi}) it will be shown, that the
changeable set $\cZ$ is precisely visible if and only if it is normally
visible.

\subsection{Visibility Classes \label{sub:Vi-classes}}

\BeginAs \label{As:FviEq}

For any areas of perception $l,m\in\LkZ$ of any changeable set $\cZ$
the following propositions are equivalent: 

\begin{description}
\item [{(I)}] $l\nvi m$ and $m\nvi l$; 
\item [{(II)}] $l\fvi m$ and $m\fvi l$. 
\end{description}
\EndAs 

\BeginProof 

Since precise visibility always involves normal visibility, it is
enough only to prove the implication (I)$\Rightarrow$(II). Hence,
suppose, that $l,m\in\LkZ$, $l\nvi m$, $m\nvi l$. 

1) First we shall prove, that for any $A,B\subseteq\BS(l)$, the equality
$A\cap B=\emptyset$ is true if and only if $\un{l}{m}A\cap\un{l}{m}B=\emptyset$.
Suppose, that $A\cap B=\emptyset$. Then, according to second item
of the assertion \ref{As:Zproperties02}, $\emptyset=A\cap B\supseteq\un{m}{l}\left(\un{l}{m}A\cap\un{l}{m}B\right)$.
Since $m\nvi l$ and $\un{m}{l}\left(\un{l}{m}A\cap\un{l}{m}B\right)=\emptyset$,
then, by the definition of normal visibility, $\un{l}{m}A\cap\un{l}{m}B=\emptyset$,
what is necessary to prove. Conversely, let $\un{l}{m}A\cap\un{l}{m}B=\emptyset$.
Then, by first item of the assertion \ref{As:Zproperties02}, $\un{l}{m}\left(A\cap B\right)\subseteq\un{l}{m}A\cap\un{l}{m}B=\emptyset$.
Since $\un{l}{m}\left(A\cap B\right)=\emptyset$ and $l\nvi m$, then,
by the definition of normal visibility, $A\cap B=\emptyset$. 

2) Let, $A\in\BS(l)$ and $A=\bigsqcup_{\alpha\in\AAA}A_{\alpha}$
(where $A_{\alpha}\subseteq A$, $\alpha\in\AAA$; $A_{\alpha}\cap A_{\beta}=\emptyset$,
$\alpha\neq\beta$). By the item 3) of the assertion \ref{As:Zproperties02},
$\un{l}{m}A\supseteq\bigcup_{\alpha\in\AAA}\un{l}{m}A_{\alpha}$.
Since the family of sets $\left(A_{\alpha}\,|\:\alpha\in\AAA\right)$
is disjoint, by the first item of this proof, the family of sets $\left(\un{l}{m}A_{\alpha}\,|\:\alpha\in\AAA\right)$
also is disjoint, that is $\un{l}{m}A\supseteq\bigsqcup_{\alpha\in\AAA}\un{l}{m}A_{\alpha}$.
Assume, that the last inclusion is strict (ie $\un{l}{m}A\neq\bigsqcup_{\alpha\in\AAA}\un{l}{m}A_{\alpha}$).
Then the set $\Bxv=(\un{l}{m}A)\setminus\left(\bigsqcup_{\alpha\in\AAA}\un{l}{m}A_{\alpha}\right)$
is nonempty. Hence, by the definition of normal visibility, the set
$B=\un{m}{l}\Bxv$ also is nonempty. Since $\Bxv\subseteq\un{l}{m}A$,
by the properties \ref{Prop:ChSets(basic)}, $B=\un{m}{l}\Bxv\subseteq\un{m}{l}\un{l}{m}A\subseteq\un{l}{l}A=A$.
Since the set $\Bxv=(\un{l}{m}A)\setminus\left(\bigsqcup_{\alpha\in\AAA}\un{l}{m}A_{\alpha}\right)$
is disjoint with with any of the sets $\un{l}{m}A_{\alpha}$ ($\alpha\in\AAA$),
the set $\un{l}{m}B=\un{l}{m}\un{m}{l}\Bxv\subseteq\un{m}{m}\Bxv=\Bxv$
also is disjoint with with any of $\un{l}{m}A_{\alpha}$ ($\alpha\in\AAA$)
(ie $\un{l}{m}B\cap\un{l}{m}A_{\alpha}=\emptyset$, $\alpha\in\AAA$).
Hence, by the first item of this proof, $B\cap A_{\alpha}=\emptyset$,
$\alpha\in\AAA$. Thus, we can conclude, that there exist the \textbf{nonempty}
set $B\subseteq A$ such, that $B\cap A_{\alpha}=\emptyset$, $\alpha\in\AAA$,
which contradicts the equality $A=\bigsqcup_{\alpha\in\AAA}A_{\alpha}$.
Thus, the assumption above is wrong, and, consequently, we obtain
$\un{l}{m}A=\bigsqcup_{\alpha\in\AAA}\un{l}{m}A_{\alpha}$. 

Thus, any set $A\subseteq\BS(l)$ is precisely visible from the area
of perception $m$, ie $l\fvi m$. Similarly, we obtain, that $m\fvi l$.
~ ~ ~ ~ \EndProof  

The next corollary immediately follows from the assertion \ref{As:FviEq}. 

\BeginNasl  \label{Nasl:ChSet:FviEqNvi}

Changeable set $\cZ$ is precisely visible if and only if it is normally
visible. 

\EndNasl 

Taking into account the corollary \ref{Nasl:ChSet:FviEqNvi}, the
notion {}``normally visible changeable set'' will be not used henceforth. 

\BeginDef 

We say, that areas of perception $l,m\in\LkZ$ are \textbf{equivalent
respectively the precise visibility} (or, abbreviated, \textbf{precisely-equivalent})
if and only if it is satisfied the condition (II) (or, equivalently,
the condition (I)) of the assertion \ref{As:FviEq}. 

\EndDef 

The fact, that areas of perception $l,m\in\LkZ$ are precisely-equivalent
will be denoted by the following way: \[
l\fvieq m\,(\cZ).\]
 And in the case, when changeable set $\cZ$ known in advance we shall
use the denotation $l\fvieq m$ instead. 

\BeginAs 

Relation $\fvieq$ is relation of equivalence on the set $\LkZ$. 

\EndAs 

\BeginProof 

For $l,m\in\LkZ$ condition $l\fvieq m$ is equivalent to the condition
(I) of the assertion \ref{As:FviEq}. Thus, since (by the assertion
\ref{As:NviQuasiOrder}) the relation $\nvi$ is quasi order on $\LkZ$,
the desired result follows from \cite[page. 21]{Birkhoff}. ~ ~
~ ~ ~ \EndProof 

\BeginDef 

Equivalence classes, generated by the relation $\fvieq$ will be referred
to as \textbf{precise visibility classes} of the changeable set $\cZ$. 

\EndDef 

Thus, for any changeable set, the set of all its areas of perception
can be splited on the precise visibility classes. Within an arbitrary
precise visibility class any area of perception is precisely visible
from other. It is evident, that changeable set $\cZ$ is precisely
visible if and only if $\LkZ$ contains only one precise visibility
class.   \medskip{}

It turns out, that, using the relation of visibility {}``$\vi$'',
we can also divide the set $\LkZ$ by equivalence classes. 

\BeginDef 

Let $\cZ$ be a changeable set. 

\begin{description}
\item [{(a)}] We say, that areas of perception $l,m\in\LkZ$ are \textbf{directly
connected by visibility} (denotation is $l\vicnt m\:(\cZ)$, or $l\vicnt m$
in the case, when changeable set $\cZ$ known in advance) if and only
if at least one of the following conditions is satisfied: \[
l\vi m\quad\textrm{or}\quad m\vi l.\]

\item [{(b)}] We say, that areas of perception $l,m\in\LkZ$ are\textbf{
connected by visibility} (denotation is $l\vieq m\:(\cZ)$, or $l\vieq m$
in the case, when changeable set $\cZ$ known in advance) if and only
if there exists a sequence $l_{0},l_{1},\cdots,l_{\nu}\in\LkZ$ ($\nu\in\N$)
such, that: \begin{gather*}
l_{0}=l,\quad l_{\nu}=m,\;\textrm{and}\quad l_{i}\vicnt l_{i-1}\;(i\in\overline{1,\nu}).\end{gather*}

\end{description}
\EndDef 

\BeginAs 

Relation $\vieq$ is relation of equivalence on the set $\LkZ$. 

\EndAs 

\BeginProof 

Since the relation of visibility is reflexive, the relation $\vicnt$
is reflexive and symmetric on $\LkZ$. The relation $\vieq$ is transitive
closure of the relation $\vicnt$ in the sense of \cite[page 69]{Karnaych},
\cite[page. 32]{Cuznetsov}.  Thus, by \cite[assertion 5.8, 5.9; theorem 5.8]{Karnaych},
$\vieq$ is equivalence relation on \textbf{\emph{$\LkZ$}}. ~ ~
~ ~ \EndProof 

\BeginDef 

Equivalence classes in the set $\LkZ$, generated by the relation
$\vieq$ will be referred to as\textbf{ visibility classes} of the
changeable set $\cZ$. 

\EndDef 

But it may occur, that in the changeable set only one visibility class
exist. 

\BeginDef 

We say, that a changeable set $\cZ$ is \textbf{connected visible}
if and only if for any $l,m\in\LkZ$ it is true the correlation $l\vieq m$. 

\EndDef 

It is evident, that any visible changeable set is connected visible.
Analyzing the examples \ref{Ex:FviNoTransitive} and \ref{Ex:ViNoTransitive}
it is easy to verify that the inverse proposition, in general, is
false. 

So, we see, that in the case, when a changeable set $\cZ$ is not
connected visible the set of all it's areas of perception is splitted
by {}``parallel worlds'' (visibility classes) and any visibility
class is {}``fully invisible'' from other visibility classes. As
formal example of changeable set with many visibility classes it can
be considered the changeable set $\Znv{\vcB}$ (see example \ref{Ex:Znv})
with $\card(\vcB)\geq2$. In the changeable set $\Znv{\vcB}$ any
area of perception forms the separated visibility class. 

Precise visibility classes also can be interpreted as {}``parallel
worlds''. But these {}``parallel worlds'' may be partially visible
from other {}``parallel worlds''.

\subsection{Precisely Visible Changeable Sets }

In the classical mechanics and special relativity theory it is supposed,
that any elementary-time state (or {}``physical event'') is visible
in any frame of reference. Hence, the precisely visible changeable
sets are to be important for physics. In this subsection we investigate
precisely visible changeable sets in more details. The changeable
sets $\Zpv{\vcB,\vW}$ and $\Zim{\bbU,\BBB}$, introduced in the examples
\ref{Ex:Zpv},\ref{Ex:Zimg} and \ref{Ex:ZRelativity}, evidently
are precisely visible. 

\BeginRmk  \label{Rmk:fvi01}

It should be noted, that by the assertion \ref{As:FviEq} and definition
of the relation $\fvieq$, \emph{for any changeable set $\cZ$ the
following propositions are equivalent: }

\vspace{-2mm}

\begin{description}
\item [{(I)}] \emph{$\cZ$ is precisely visible changeable set};\vspace{-2mm}

\item [{(II)}] \emph{for any $l,m\in\LkZ$ it is performed the condition
$l\fvi m$};\vspace{-2mm}

\item [{(III)}] \emph{for any $l,m\in\LkZ$ it is performed the condition
$l\nvi m$;}
\item [{(IV)}] \emph{for any $l,m\in\LkZ$ it is performed the condition
$l\fvieq m$.}
\end{description}
\EndRmk 

Note also that in the first item of the proof of assertion \ref{As:FviEq}
it was proved, the following lemma. 

\BeginLem \label{Lem:FviMapDisjunkt}

Let $\cZ$ be a precisely visible changeable set. Then for any $l,m\in\LkZ$
and $A,B\subseteq\BS(l)$ the equality $\un{l}{m}A\cap\un{l}{m}B=\emptyset$
is true if and only if $A\cap B=\emptyset$. 

\EndLem 

\BeginThm \label{Thm:FviCriterion01}

Changeable set $\cZ$ is precisely visible if and only if for any
$l,m,p\in\LkZ$ the followind equality is true: \begin{equation}
\un{m}{p}\un{l}{m}=\un{l}{p}.\label{eq:QusiGroupUni}\end{equation}

\EndThm 

\BeginProof 

\textbf{Sufficiency}. Suppose, that for any $l,m,p\in\LkZ$ the equality
(\ref{eq:QusiGroupUni}) holds. Chose any areas of perception $l,m\in\LkZ$
and any changeable system $A\subseteq\BS(l)$ such, that $A\neq\emptyset$.
Then, by (\ref{eq:QusiGroupUni}), \[
A=\un{l}{l}A=\un{m}{l}\un{l}{m}A.\]
 Therefore, by the assertion \ref{As:Zproperties01}, $\un{l}{m}A\neq\emptyset$.
Thus, by the assertion \ref{As:ViProperties03}, $l\nvi m$ (for any
areas of perception $l,m\in\LkZ$). Hence, by the remark \ref{Rmk:fvi01},
the changeable set $\cZ$ is precisely visible.

\textbf{Necessity}. Conversely, suppose, that the changeable set $\cZ$
is precisely visible. Consider any areas of perception $l,m,p\in\LkZ$
and any changeable system $A\subseteq\BS(l)$. By the properties \ref{Prop:ChSets(basic)}
$\un{m}{p}\un{l}{m}A\subseteq\un{l}{p}A$. Denote: \[
B_{1}:=\un{l}{p}A\setminus\un{m}{p}\un{l}{m}A.\]
 Then, $B_{1}\subseteq\un{l}{p}A$ and $B_{1}\cap\un{m}{p}\un{l}{m}A=\emptyset$.
Denote $B:=\un{m}{l}\un{p}{m}B_{1}$. Using the properties \ref{Prop:ChSets(basic)}
we obtain: \begin{gather*}
B=\un{m}{l}\un{p}{m}B_{1}\subseteq\un{m}{l}\un{p}{m}\un{l}{p}A\subseteq\\
\subseteq\un{l}{l}A=A;\\
\un{m}{p}\un{l}{m}B=\un{m}{p}\un{l}{m}\un{m}{l}\un{p}{m}B_{1}\subseteq\\
\subseteq\un{p}{p}B_{1}=B_{1}.\end{gather*}
 Hence, since $B_{1}\cap\un{l}{p}\un{l}{m}A=\emptyset$, we have $\un{m}{p}\un{l}{m}B\cap\un{m}{p}\un{l}{m}A=\emptyset$.
Consequently, using lemma \ref{Lem:FviMapDisjunkt}, we obtain $B\cap A=\emptyset$.
Since $B\subseteq A$ and $B\cap A=\emptyset$, we obtain $B=\emptyset$.
Since $\un{m}{l}\un{p}{m}B_{1}=B=\emptyset$, taking into account,
that, by remark \ref{Rmk:fvi01} $p\nvi m$ and $m\nvi l$, we obtain
(by definition of normal visibility) $B_{1}=\emptyset$. ~ ~ ~
\EndProof 

Note, that, for the changeable set $\cZ=\left(\AAA,\vcB,\vfU\right)$
from the definition \ref{Def:GlobalUni,ChSets}, the condition (\ref{eq:QusiGroupUni})
is equivalent to the condition (\ref{eq:GrpMaps1}). 

\BeginAs \label{As:ZFviproperties}

Let $\cZ$ be a precisely visible changeable set. Then for any areas
of perception $l,m\in\LkZ$, any family of changeable systems $\left(A_{\alpha}|\alpha\in\AAA\right)$
($A_{\alpha}\subseteq\BS(l)$, $\alpha\in\AAA$) and any changeable
systems $A,B\in\BS(l)$ the following assertions are true:

\begin{enumerate}
\item $\un{l}{m}\left(\bigcap\limits _{\alpha\in\AAA}A_{\alpha}\right)=\bigcap\limits _{\alpha\in\AAA}\un{l}{m}A_{\alpha}$;
 
\item $\un{l}{m}(A\setminus B)=\un{l}{m}A\setminus\un{l}{m}B$;
\item $\un{l}{m}\BS(l)=\BS(m)$;
\item $\un{l}{m}\left(\bigcup\limits _{\alpha\in\AAA}A_{\alpha}\right)=\bigcup\limits _{\alpha\in\AAA}\un{l}{m}A_{\alpha}$;

\item If a changeable system $A\subseteq\BS(l)$ is a singleton ($\card(A)=1$),
then the changeable system $\un{l}{m}A$ also is a singleton. 
\end{enumerate}
\EndAs 

\BeginProof 

1) Using the assertion \ref{As:Zproperties02}, item 2), properties
\ref{Prop:ChSets(basic)} and theorem \ref{Thm:FviCriterion01} (equality
(\ref{eq:QusiGroupUni})) we obtain: \begin{gather*}
\un{l}{m}\left(\bigcap\limits _{\alpha\in\AAA}A_{\alpha}\right)\supseteq\un{l}{m}\un{m}{l}\left(\bigcap\limits _{\alpha\in\AAA}\un{l}{m}A_{\alpha}\right)=\\
=\un{m}{m}\left(\bigcap\limits _{\alpha\in\AAA}\un{l}{m}A_{\alpha}\right)=\bigcap\limits _{\alpha\in\AAA}\un{l}{m}A_{\alpha}.\end{gather*}
Hence, $\un{l}{m}\left(\bigcap_{\alpha\in\AAA}A_{\alpha}\right)\supseteq\bigcap_{\alpha\in\AAA}\un{l}{m}A_{\alpha}$.
The inverse inclusion had been proved in the assertion \ref{As:Zproperties02},
item 1). 

2) Since $A\setminus B\subseteq A$, then by the property \ref{Prop:ChSets(basic)}(5)
we have, $\un{l}{m}(A\setminus B)\subseteq\un{l}{m}A$. Since $(A\setminus B)\cap B=\emptyset$,
then, by lemma \ref{Lem:FviMapDisjunkt}, $\un{l}{m}(A\setminus B)\cap\un{l}{m}B=\emptyset$.
Hence: \begin{equation}
\un{l}{m}(A\setminus B)\subseteq\un{l}{m}A\setminus\un{l}{m}B.\label{eq:UnSetminusEmbed}\end{equation}
Using the correlation (\ref{eq:UnSetminusEmbed}) to the sets $\un{l}{m}A$,
$\un{l}{m}B$, with unification mapping $\un{m}{l}$, applying the
formula (\ref{eq:QusiGroupUni}) and properties \ref{Prop:ChSets(basic)}
we obtain: \begin{gather*}
\un{m}{l}\left(\un{l}{m}A\setminus\un{l}{m}B\right)\subseteq\\
\subseteq\un{m}{l}\un{l}{m}A\setminus\un{m}{l}\un{l}{m}B=\un{l}{l}A\setminus\un{l}{l}B=A\setminus B.\end{gather*}
 Hence, by the property \ref{Prop:ChSets(basic)}(5) $\un{l}{m}\un{m}{l}\left(\un{l}{m}A\setminus\un{l}{m}B\right)\subseteq\un{l}{m}(A\setminus B)$.
And applying the formula (\ref{eq:QusiGroupUni}), we obtain the inverse
inclusion to (\ref{eq:UnSetminusEmbed}). 

3) By definition of unification mapping, \begin{equation}
\un{l}{m}\BS(l)\subseteq\BS(m).\label{eq:UnBSembed}\end{equation}
Similarly, $\un{m}{l}\BS(m)\subseteq\BS(l)$. Applying to the last
inclusion unification mapping $\un{l}{m}$, and using properties \ref{Prop:ChSets(basic)}
as well as correlation (\ref{eq:QusiGroupUni}) we obtain the inverse
inclusion to (\ref{eq:UnBSembed}). 

4) Note, that: $\bigcup_{\alpha\in\AAA}A_{\alpha}=\BS(l)\setminus\left(\bigcap_{\alpha\in\AAA}\left(\BS(l)\setminus A_{\alpha}\right)\right)$.
Hence, using items 1, 2 and 3 of this assertion we obtain: \begin{gather*}
\un{l}{m}\left(\bigcup_{\alpha\in\AAA}A_{\alpha}\right)=\un{l}{m}\BS(l)\setminus\left(\bigcap_{\alpha\in\AAA}\left(\un{l}{m}\BS(l)\setminus\un{l}{m}A_{\alpha}\right)\right)=\\
=\BS(m)\setminus\left(\bigcap_{\alpha\in\AAA}\left(\BS(m)\setminus\un{l}{m}A_{\alpha}\right)\right)=\bigcup_{\alpha\in\AAA}\un{l}{m}A_{\alpha}.\end{gather*}

5) Let $A\subseteq\BS(l)$, and $A=\left\{ \omega\right\} $ is a
singleton. By remark \ref{Rmk:fvi01}, $l\nvi m$ and, since $A\neq\emptyset$,
by definition of normal visibility, we have $\un{l}{m}A\neq\emptyset$.
Suppose, that the set $B=\un{l}{m}A$ contains more, than one element.
Then, there exist sets $B_{1},B_{2}\subseteq B$ such, that $B_{1},B_{2}\neq\emptyset$
and $B=B_{1}\sqcup B_{2}$. Denote: $A_{1}:=\un{m}{l}B_{1}$, $A_{2}:=\un{m}{l}B_{2}$.
Since $B_{1},B_{2}\neq\emptyset$, then, by the definition of normal
visibility, $A_{1},A_{2}\neq\emptyset$. Since $B=B_{1}\sqcup B_{2}$,
then, by the definition of precise visibility, $\un{m}{l}B=\un{m}{l}B_{1}\sqcup\un{m}{l}B_{2}=A_{1}\sqcup A_{2}$.
Hence, taking into account, that $B=\un{l}{m}A$ and using the equality
(\ref{eq:QusiGroupUni}), we obtain: \[
A_{1}\sqcup A_{2}=\un{m}{l}B=\un{m}{l}\un{l}{m}A=A.\]
Thus, we see, that the set $A$ can be divided into two nonempty disjoint
sets, which contradicts the fact, that the set $A$ is a singleton.
Therefore, the set $\un{l}{m}A$ is nonempty, and can not contain
more, than one element, hence, it is a singleton. ~ ~ ~ \EndProof 

\BeginDef \label{Def:FviUni!}

Let $\cZ$ be a precisely visible changeable set, $l,m\in\LkZ$ and
$\omega\in\BS(l)$. Elementary-time state $\omega'\in\BS(m)$ such,
that $\left\{ \omega'\right\} =\un{l}{m}\left\{ \omega\right\} $
will be referred to as \textbf{visible image }of elementary-time state
$\omega\in\BS(l)$ in the area of perception $m$ and it will be denoted
by $\unn{l}{m}{\omega}$: \[
\omega'=\unn{l}{m}{\omega}.\]

\EndDef 

By the assertion \ref{As:ZFviproperties}, item 5, any elementary-time
state $\omega\in\BS(l)$ always has a visible image $\omega'=\unn{l}{m}{\omega}$.
Hence, by definition \ref{Def:FviUni!}, for any elementary-time state
$\omega\in\BS(l)$ in the area of perception $l\in\LkZ$ of precisely
visible changeable set $\cZ$ the following equality holds: \begin{equation}
\un{l}{m}\left\{ \omega\right\} =\left\{ \unn{l}{m}{\omega}\right\} \quad(m\in\LkZ)\label{eq:MainFviUni01}\end{equation}
Using the equality $A=\bigsqcup_{\,\omega\in A}\left\{ \omega\right\} $,
definition of precise visibility and equality (\ref{eq:MainFviUni01})
we obtain the following theorem. 

\BeginThm \label{Thm:MainFviUni}

For any nonempty changeable system $A\subseteq\BS(l)$ in area of
perception $l\in\LkZ$ of precisely visible changeable set $\cZ$
the following equality is true: \begin{equation}
\un{l}{m}A=\bigsqcup_{\omega\in A}\left\{ \unn{l}{m}{\omega}\right\} =\left\{ \unn{l}{m}{\omega}\,|\:\omega\in A\right\} \quad(m\in\LkZ).\label{eq:MainFviUni02}\end{equation}

\EndThm  

\BeginNasl 

Let $\cZ$ be a precisely visible changeable set and $l,m\in\LkZ$
any it's areas of perception. 

\begin{onehalfspace}
Then for any changeable system $A\subseteq\BS(l)$ the sets $A$ and
$\un{l}{m}A$ are equipotent, in particular the sets $\BS(l)$ and
$\BS(m)$ are equipotent . In the case $A\neq\emptyset$ the mapping:\begin{equation}
f(\omega)=\unn{l}{m}{\omega},\qquad\omega\in\BS(l)\label{eq:FviEquipotentMapDef}\end{equation}
 is bijection between the sets $A$ and $\un{l}{m}A$. 
\end{onehalfspace}

\EndNasl 

\BeginProof 

In the case $A=\emptyset$ the statement of the corollary follows
from the assertion \ref{As:Zproperties01}. In the case $A\neq\emptyset$
from the theorem \ref{Thm:MainFviUni} (equality (\ref{eq:MainFviUni02}))
it follows, that the mapping (\ref{eq:FviEquipotentMapDef}) is bijection
between the sets $A$ and $\un{l}{m}A$. And from the assertion \ref{As:ZFviproperties}
(item 3)) it follows, that the sets $\BS(l)$ and $\BS(m)$ are equipotent.
~ ~ ~ \EndProof

\end{document}